\documentclass[letterpaper,11pt]{article}
\usepackage[unknownkeysallowed]{graphicx}  % Old versions of LaTeX throw error without this: ! Package keyval Error: pdfa undefined.

\usepackage{jheppub} % for details on the use of the package, please
 \usepackage{psfrag}
 \usepackage{array}

\newcommand{\cross}{\times}

\newcommand{\ra}[1]{ \stackrel{\scriptscriptstyle #1}{ \rightarrow  }}  % Put the argument above = as in \ra{gauge}
\newcommand{\eS}{ \varepsilon_{S}}
\newcommand{\eB}{ \varepsilon_{B}}

\newcommand{\DZero}{{D\O}}
\newcommand{\zh}{ZH}

\def\lsim{\mathrel{\rlap{\lower4pt\hbox{\hskip1pt$\sim$}}
    \raise1pt\hbox{$<$}}}
\def\gsim{\mathrel{\rlap{\lower4pt\hbox{\hskip1pt$\sim$}}
    \raise1pt\hbox{$>$}}}

\title{Quark and Gluon Jet Substructure}
\author[a]{Jason Gallicchio,}
\author[b]{Matthew D. Schwartz}

\affiliation[a]{Department of Physics, University of California, Davis, CA 95616, USA}
\affiliation[b]{Jefferson Physical Laboratory, Harvard University, Cambridge, MA 02138, USA}

\emailAdd{jason@frank.harvard.edu}
\emailAdd{schwartz@physics.harvard.edu}

\date{\today}

\abstract{
Distinguishing quark-initiated jets from gluon-initiated jets has the potential to significantly improve
the reach of many beyond-the-standard model searches at the Large Hadron Collider and to provide additional
tests of QCD. To explore whether quark and gluon jets could possibly be distinguished on an event-by-event
basis, we perform a comprehensive simulation-based study. We explore a variety of motivated and unmotivated variables
with a semi-automated multivariate approach. General conclusions are that at 50\% quark jet acceptance efficiency,
around 80\%-90\% of gluon jets can be rejected. Some benefit is gained by combining variables.
Different event generators are compared, as are the effects of using only charged tracks to avoid pileup.
Additional information,
including interactive distributions of most variables and their cut efficiencies, can be
found at \url{http://jets.physics.harvard.edu/qvg}.
}
\begin{document}
\maketitle
\flushbottom

%\begin{fmffile}{higgsopt_feyn}  % For Feynman Diagrams -- must be matched below

\section{Introduction}
\begin{figure}
\label{fig:gluino_chain}
\begin{center}
\includegraphics[width=0.5\textwidth]{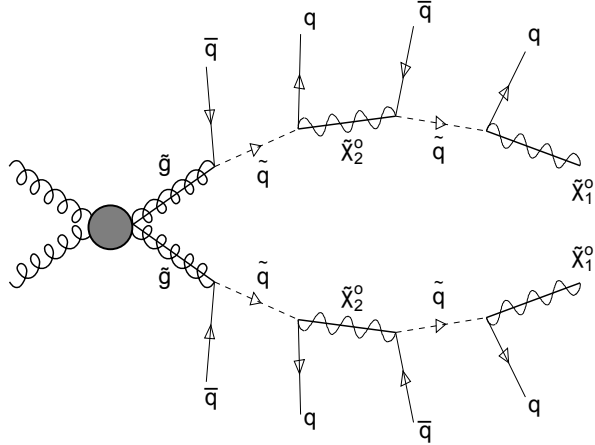}
\caption{
Gluino decay as an of of a quark-heavy signal, in this case with 8 quark jets and no gluon jets produced.
%Produced in pairs, there can be 8 quark jets.
%, and even if both heavy neutralinos %sides' ${X^~}^0_2$ neutralinos
%decay leptonically, there are still 4 quark jets.
Multi-jet events in standard model backgrounds are extremely unlikely to have so many quark jets.
}
\end{center}
\end{figure}
%The ability to distinguish between
%quarks and gluons on a jet-by-jet basis has the potential to dramatically
%improve many searches involving jets.
%For example,
%In beyond-the-standard model physics, 
Being able to distinguish quark-initiated
from gluon-initiated jets reliably at the LHC could be very useful, since signatures of beyond-the-standard-model physics
are often quark heavy. For example, a typical gluino-pair production
topology is pictured in Figure~\ref{fig:gluino_chain}.
Produced in pairs, each gluino's
cascade decay can produce four quarks and missing transverse
momentum due to the escape of the lightest supersymmetric
partner.
Backgrounds to this process have events with many jets produced from QCD. These jets are predominately gluonic.
Additionally, many $R$-parity violating SUSY models
produce quark jets without the missing transverse momentum. To constrain these models, being able to filter
out background QCD events containing gluon jets would be
helpful.
Leptophobic $Z'$ or $W'$ particles provide other
obvious examples where quark/gluon discrimination would be useful.

% Alternatively, any potential $Z'$ would
% decay to quark rather than gluon jets, and this mode might be
% provide a useful handle on a leptophobic $Z'$.
% %$B$-tagging could be useful, but in other situations $b$ quarks
% %are not produced.
% An example is the hadronic decays of a $W$,
% including the fully-hadronic $t \bar t$ decays, which produce
% 2 $b$ jets and 4 light-quark jets.
% Similarly, a charged Higgs boson decays primarily to $c \bar s$
% if it is too light to decay to $t \bar b$.

Gluon-heavy backgrounds are especially problematic for signals
without leptons, gauge bosons, $B$-jets, tops, or missing energy.
Quark/gluon tagging might be one of the few ways to improve these searches.
Another application is to reduce combinatorial ambiguity
within a single event. If jets in a given event could be
identified as quark or gluon, their place in a proposed decay
topology could be constrained, or they could be classified as initial-state
radiation. Examining the quark/gluon
tagging scores of jets produced by a new particle might be the
only way to measure QCD quantum numbers directly.
Alternatively, some signals consist of gluon jets, like coloron
models~\cite{Hill:1993hs}
or buried-Higgs, where $h \rightarrow 2a \rightarrow 4g$ and
$a$ is CP odd scalar \cite{Bellazzini:2009xt}.
The same observables and techniques apply to gluon tagging,
though here we will treat the quark jets as the signal and
the gluon jets as background for concreteness.

\begin{figure}
\begin{center}
%\vspace{-0.25in}
\includegraphics[width=0.6\textwidth]{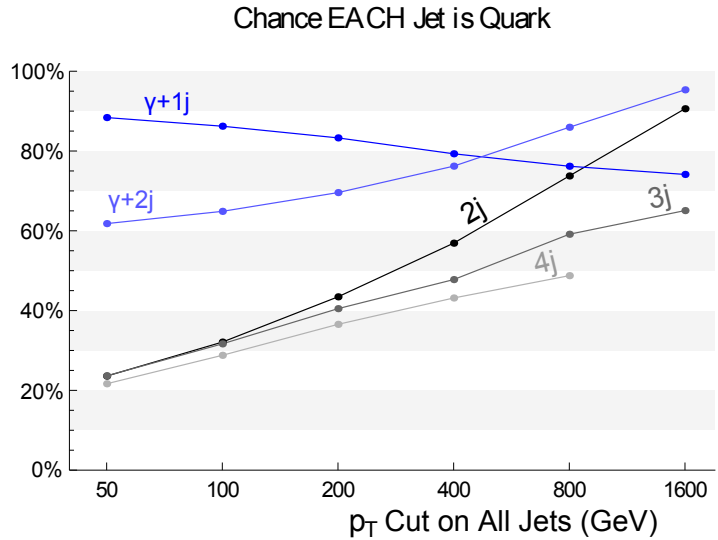}
%\vspace{-0.15in}
 \caption{
 Fraction of jets which are a
 light quark jet (up, down or strange) rather than a gluon jet
 %(bottom and charm aren't included in this ratio.)
 Here all jets have the minimum $p_T$ cut indicated, but
 photons have a minimum $p_T$ of only 20\,GeV.
 \label{fig:Chance_EACH_Jet_is_Quark1}
 }
\end{center}
\end{figure}

Practical quark/gluon discrimination would also be useful for some standard model
studies.
%On the other hand, quark/gluon tagging may be less useful
%for signals involving quark jets produced together with gauge bosons,
%since the standard model backgrounds are already quark-rich.
For example, in vector
boson fusion (VBF), the forward jets are always quark jets
whereas in non-electroweak backgrounds to VBF, the jets
near the beams are often gluonic.
In the standard model, as the $p_T$ of jets increases, or
if they are produced along with an electroweak boson, the fraction
becomes more quark-heavy.  This is shown in Figure \ref{fig:Chance_EACH_Jet_is_Quark1}. Thus,
knowing the quark-to-gluon jet fraction of an event can help determine what are the underlying
hard partons, with applications even in the standard model.
%
% Some effort has already been made to predict and measure differences between
% light quark and gluon jets, and this is discussed in the next section.
% But to our knowledge, this knowledge has not been used as a
% generic tool to improve new physics searches.

Differences between quark and gluon jets were measured in great detail
in LEP 3-jet events, where the flavor
could be known to high accuracy.
Such measurements are described well by perturbative QCD calculations and leading-log parton
showers combined with hadronization models.
In the LHC era, we propose using this accumulated wisdom as a tool to
find new physics.
The small differences between generators do not invalidate
the \emph{use} of these tools to find observables
that can distinguish between quark- and gluon-initiated jets on
a jet-by-jet basis.
Experimental effort can then be focused on the small set of
the most powerful discriminators.

The goal of this hadron-level Monte Carlo study is to find properties of jets
that best distinguish ones initiated by a quark from those initiated by a gluon.
Charged particle count and jet mass
are well known examples, but new observables like
$p_T$-weighted moments as measured from the jet center
and subjet properties provide additional
handles.  Each observable is examined for many jet $p_T$s,
and the best set is combined
into a quark/gluon tagger.
Given a jet of a particular
$p_T$, our tagger assigns a quark/gluon likelihood score. This
can be cut on to purify the flavor content, combined with prior
quark/gluon fraction into a true probability, or used in
conjunction with $b$ and $\tau$ tagging scores to more fully
classify the jet's flavor.
We do not expect quark/gluon tagging to reach the same power as
$B$-tagging or $\tau$-tagging, but our quark-efficiency vs gluon rejection curve
can serve as a first approximation of what is
achievable.

The main result of this paper is that a
small set of 1-3 input observables capture nearly all of the
quark/gluon differences. The most useful variables could become the focus of
theoretical study, experimental measurement, and Monte Carlo
validation.
The multidimensional distributions of quark/gluon discriminating variables
can be experimentally verified, for example,
by looking at many samples with different known
quark/gluon compositions,
especially ones that are relatively pure \cite{Gallicchio:2011xc}.

%The tagger can be used immediately to improve sensitivity in SUSY and $Z'$
%searches or estimate the quark/gluon content of new physics.

Since jet properties depend strongly on $p_T$,
We examine jets in narrow $p_T$ windows
around six central values between 50\,GeV and 1.6\,TeV, in
powers of 2.  As a result of our examination of so many
observables, we can make general statements about some $p_T$ trends.
For example, track counts are more useful at high
$p_T$, whereas geometric moments (which measure the width/girth
of jets) are more useful at low $p_T$. In addition,
some observables are more powerful
discriminants when the operating point of the tagger is chosen
at high quark efficiency, and others are useful when a stronger
cut is used to achieve high quark purity.

In the next section, we
%give some examples of LHC
%analyses that can benefit from quark/gluon discrimination.
%Then we
review past calculations and collider measurements at
LEP and the Tevatron.
After that, we define our observables, show hadron-level
distributions, and quantify their performance.
Finally we combine observables using boosted decision
trees to form a multivariate discriminant.
The final sections include comments on how one might use a quark/gluon
tagger in situations where the signal or background contains
both quark and gluon jets.

%\section{Motivating Searches and Other Applications}

\section{History and Future of Quark/Gluon Measurements}
There are several differences between quarks and gluons that prove
useful in motivating observables that can distinguish between
the jets initiated by quarks as compared to gluons.
Below is a list of properties and observables they motivate:
%\begin{tabular}{|c|c|c|c|}
%\hline
%Property
%\hline
% Color Charge & $C_F$ vs. $C_A$ & jet mass, jet size, track count
%\hline
%\end{tabular}
\begin{itemize}
  \item Color Charge: $C_F$ vs $C_A$ $\rightarrow$ jet
      mass, girth/width, track count
  \item Color Connections: 1 vs 2 $\rightarrow$
      eccentricity, planar flow, and pull
  \item Electrical Charge $\rightarrow$ charge-weighted
      track $p_T$
  \item Spin: 1/2 vs 1 $\rightarrow$ correlations in the
      location of subjets
\end{itemize}

An excellent review of theoretical and experimental results
as of 2003 are presented
% the `Differences between quark and gluon jets' chapter of the book
in~\cite{Dissertori:2003}, some of which we now summarize.
LEP studied the
difference between quark and gluon jets by looking at 3-jet events.
These correspond to $e^+ e^- \rightarrow q \bar q g$ at the parton level. At high center-of-mass
energy, the two hardest jets are quark-initiated 99\% of the time,
thus one can use energy to select a pure sample.
In another selection method, the highest
energy jet is assumed to be a quark jet and one of the other
jets is tagged for heavy flavor, which indicates the third should be gluonic.
 Alternatively, for three jets
of similar energy, two $B$-tags gave a clean sample of
$\sim$30\,GeV gluon jets.

LEP measured the ratio of the
number of particles in gluon vs quark jets. The average
multiplicity of \emph{any} type of particle, along with its
variance are given by the semi-classical approximation
\begin{equation}
    \frac{\langle N_g \rangle}{\langle N_q \rangle} = \frac{C_A}{C_F}
    \qquad
    \frac{\sigma_g^2}{\sigma_q^2} = \frac{C_A}{C_F}
\end{equation}
where $C_A/C_F = 9/4$.
The angular width of the jet, using Sterman-Weinberg definition,
is to leading order
\begin{equation}
    \delta_g  = \delta_q^{C_F/C_A} \ .
\end{equation}
An intuitive explanation for these results
is that a quark jet is dominated by the first gluon emission,
at which point it continues to shower like a gluon jet.
%Leading order analytic calculations treat each splitting independently.
Since gluon jets have more particles, for a given energy
they will have correspondingly fewer hard particles.

In cases where QCD estimates do not agree with full simulation or with data, the reason is often attributed to energy
conservation not being taken into account in each splitting.
Since shower Monte Carlos enforce this energy conservation, they often have better agreement with data
than the analytic estimates.
Multiplicities have been calculated, including energy-momentum
conservation, at N$^3$LO~\cite{Capella:1999ms,Bolzoni:2012ii}.
At LEP I energies, the result was  $\langle N_g
\rangle / \langle N_q \rangle \approx 1.7$.
OPAL~\cite{Ackerstaff:1997xg} studied the charged particle
multiplicity in light quark jets of average energy 45.6\,GeV and gluon jets of 41.8\,GeV.
Agreement in the moments (mean, width, skewness, kurtosis) of the particle-count
distributions was found to agree with the Monte Carlo event generators
and with analytic predictions.

%Some rough analytical predictions for the statistics
%of counting particles are:
%
%\vspace{1em}
%\begin{tabular}{lll}
%Mean:      &  $\mu = \langle N \rangle$  &  $\mu_g/\mu_q \approx \sqrt{C_A/C_F}$  \\
%Width:     &  $\sigma=\sqrt{ \langle (N-\mu)^2 \rangle }$ & $\sigma_g/\sigma_q \approx \sqrt{C_A/C_F} = 3/2$\\
%Skewness:  &  $\gamma = \langle (N-\mu)^3 \rangle / \sigma^3$    & \\
%Kurtosis:  &  $\kappa = \langle (N-\mu)^4 \rangle / \sigma^4 - 3$   & \\
%\end{tabular}
%\vspace{1em}

Subjet multiplicities were also examined at LEP for various subjet sizes~\cite{Barate:1998cp,Abreu:1998ve}. Extremely small subjets
($k_T$=0.1\,GeV) approach the limit of particles, and therefore probed
hadronization. But larger subjets ($k_T$=5\,GeV) probed the
better modeled, perturbative physics \emph{and} gave the
largest ratio between quark and gluon subjet multiplicities.
For the first study cited, the average energy of the quark jets was 32\,GeV,
while that for gluon jets was 28\,GeV.  Later in this paper,
we show that smaller subjets always improved quark/gluon
discrimination at the LHC, down to the smallest subjets we probed with a
resolution-limited size of $R_\textrm{sub} \approx 0.1$.

The particle types identified within jets also
differ between quarks and gluons. For example, the numbers of $K^0$,
$\Lambda$, $\pi^\pm$, $K^\pm$, $p$, $\eta$, $\eta'$, and
$\pi^0$ particles have been studied.
LEP found an increase in baryons (protons and Lambdas) for 
gluon jets and an increase in kaons for quark jets,
though $Z$ decays to $s \bar s$ or $c \bar c$ will have
a higher strange-quark content than LHC's $u$ and $d$-dominated
quark jets.
Hadron identification within jets is reviewed in~\cite{Dissertori:2003},
where table 12.1 lists results of many LEP experiments.
Some of the most relevant include DELPHI
\cite{Abreu:1997cn, Abreu:2000nw}, % DELPHI 1997c -- Lambdas, DELPHI 2000c  -- pions, kaons and protons
and OPAL
\cite{Ackerstaff:1998ev, Abbiendi:2000cv}  % Opal 1999b for Lambdas, OPAL 2000b
measurements of identified particle ratios.
We do not consider variables based on particle type, since their use
depends strongly on how well these can be experimentally measured.

Many of the LEP measurements were performed on the $Z$ pole, where the jets studied come from the decay
of an on-shell particle (the $Z$ boson). These jets may not be typical of jets produced from QCD at hadron colliders. For example,
the particle multiplicity can be computed in the $Z$ rest frame and is therefore independent of jet energy. In contrast, 
the particle multiplicity for a QCD jet at a hadron collider grows with the jet's energy.

LEP studies found $B$-jets to be more similar to gluon jets than
to light quark jets \cite{Buskulic:1995sw,Biebel:1996mc}. The
number of particles was higher in $B$-jets than in light quark jets, as was the angular spread.
Both of these effects are due to the longer decay chain of $B$-hadrons, which
overwhelms the effect of perturbative parton shower.
%Since the $b$
%quark's mass only becomes relevant at $m_b$ rather than $p_T^{jet}$.
The peculiarity of $B$-jets should be
lessened in the LHC, which has higher $p_T$ jets and more boosted
$B$-hadrons: for higher $p_T$, the QCD shower produces more
particles, whereas the particle multiplicity is relatively
fixed in the $B$-hadron decay.
When the decay products of a $B$-hadron are
specifically removed from consideration, the
properties of $B$-jets will again look more quark-like.

Importantly, $B$-taggers rely largely on impact parameters or a secondary vertex,
and so their efficiency should be largely uncorrelated with the observables we consider,
which are constructed from the momenta of the particles.
Thus, $B$-mistag rates are not significantly
different for quarks as compared to gluons.
One can imagine a 3-dimensional tagger based on the probability that a jet is either $b$, quark or gluon
initiated. Since $B$-tagging is very dependent on experimental issues, we do not attempt to consider
such a tagger here. We note that gluon splittings to heavy flavor, $g \rightarrow b \bar b$,
are included in our simulation of gluon jets.

Compared to LEP studies using 25 to 45\,GeV jets, the LHC will typically have
higher energy jets. Since high-energy jets are of particular interest to new physics searches, we
consider jets with energies up to 1.6\,TeV in this study.
At high energy, it is helpful to use longitudinally
boost-invariant measures like transverse momentum and rapidity as opposed to
energy and angle.  Sometimes this motivates new variables
appropriate to hadron colliders by replacing the LEP variable
$E$ by $p_T$ and $\theta$ by $r=\sqrt{\Delta y^2 + \Delta \phi^2}$.

Measures of the angular width of jets were used in 
\DZero \, \cite{Abbott:1999mr} and later CDF \cite{Aaltonen:2010pe} to
reject gluons and purify fully-hadronic top-quark samples. This may have been the first experimental
application and proof that a separation
exists in a complex hadronic collider environment. The CDF study showed that quark
and gluon jets can be calibrated in a ``naturally pure'' quark sample
(semileptonic tops without any explicit quark/gluon tagging).

Quark/gluon tagging should be even more useful at the LHC than
it has been at LEP or the Tevatron. Compared to CDF and \DZero, ATLAS and CMS have
better tracking and calorimeters, with spatial resolutions up
to 10x as high. CMS's particle flow and ATLAS's
individually calibrated TopoClusters give jet substructure
techniques new power (especially if associated with the primary
vertex and corrected for magnetic field bending). Also the
LHC's proton-proton initial state, higher energy, and higher
luminosity make gluon jets more common and
more new physics signals are buried under multi-jet
events.  In addition, we find that higher $p_T$ jets of the LHC are
more taggable than lower $p_T$ jets of previous colliders. For
example, the charged track count becomes a better indicator of
flavor as the jet $p_T$ increases.

\section{Theoretical Considerations}

Before cataloging and evaluating jet observables, it is worth
%but the next section discusses what we really mean by quark
%jets and gluon jets and the extent to which
commenting on the extent to which jet flavor is well-defined.
We will argue that in the case of well-separated jets, appropriate
for kinematic reconstruction, each jet can be assigned an unambiguous flavor.
%Does it make sense to even talk about quark or gluon jets?
%How do we define quark/gluon jets in perturbative QCD, in
%a parton shower event record, or in data?
% There are some subtleties,
%but the different definitions will agree.
In other words,
%The reason that one can talk sensibly about quark and gluon jets is that
any situation which is problematic
for quark/gluon tagging is \emph{also} problematic for
kinematic reconstruction.
Thus, quark/gluon tagging is no more poorly defined than reconstructing a decay chain or
other short-distance interpretation of an event with jets.

% and \emph{always} the result of jets not meaningfully
%corresponding to some hard partons.

\begin{figure}[t]
\begin{center}
\includegraphics[width=0.35\textwidth]{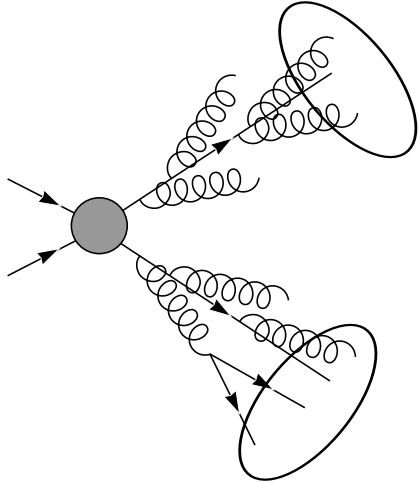}
\caption{
Jets are formed by grouping together collinear radiation.
} \label{fig:dijet_shower}
\end{center}
\end{figure}

In the parton shower picture (which is in excellent agreement with data) a hard
parton, well-separated from other hard partons in the event, 
undergoes showering that produces nearly collinear
radiation. An example is illustrated in Figure \ref{fig:dijet_shower}.
With reasonable assumptions about hadronization, any
infrared and collinear-safe jet algorithm returns jets
whose momenta correspond in some way to the initial hard partons.  This
parton/jet correspondence is implicitly assumed in all searches
which use the resulting 4-vectors to reconstruct heavy objects
like $W$ bosons, Higgs bosons, and tops.  Violating any of these assumptions erodes the
parton/jet correspondence.
%For well-separated jets, asking whether a given jet is quark or gluon is as valid
%as asking whether jets come from the decays of hard particles.

For example, the shower products from two nearby hard partons
could significantly overlap. Depending on the jet algorithm used, the jets might merge or have
strange shapes.
In such a case, the resulting jet
momenta might not be useful for kinematic reconstruction
and the
jet properties (charged particle count or mass) might not be distributed
in a way that corresponds to isolated quark or gluon jets. 
%Even
%given a Monte Carlo event record containing the full shower
%history, there will be plenty of jets that fall into this
%`other' or `mixed' or `ambiguous' category and can't be
%assigned a truth flavor.  
%Even if two quarks from different
%parts of an event happen to merge into one jet, it should not
%be considered a `a quark jet,' since the properties of such a jet will be atypical.
%  Same with gluons
% or $q \bar q$.

\begin{figure}[t]
\begin{center}
\includegraphics[width=0.80\textwidth]{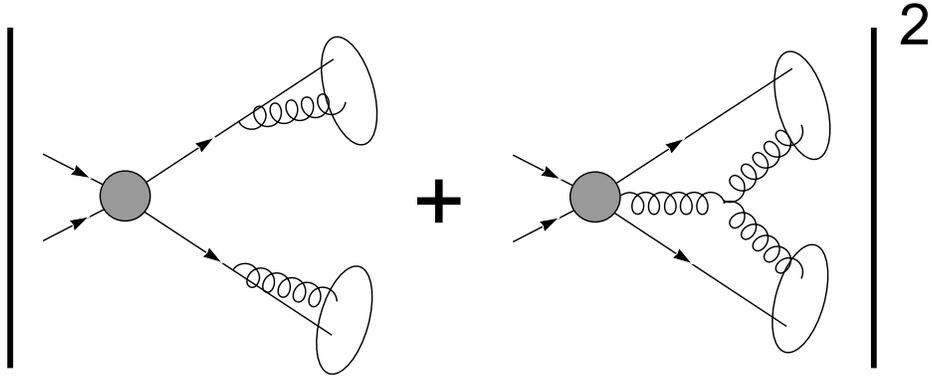}
\caption{
Parton showers produce
%case on the left produces
quark jets whose properties are largely determined by the emitted gluons, as indicated
in the left diagram.
On the right, the same configuration is produced when a third hard parton, in this
case a gluon, splits into two gluons with momenta equal to the
showered gluons.  Since the two amplitudes interfere, it might not make sense to describe this final
state configuration as having two quark jets.
%This is not just an ambiguity
%in determining which shower history is `correct,' since they
%quantum mechanically interfere.
In this case, however, the
amplitude for the shower diagram is \emph{much} larger than the
hard-gluon-splitting diagram for the same final-state kinematics.
In fact, as the gluons become more collinear with the quarks, the
first amplitude is divergent.
%Note that similar ambiguities occur in $B$-tagging: if the hard quarks were $b$ quarks, it might not be
%particularly useful to call them $B$-jets either if the hard
%event was $gZ$ with $Z \rightarrow b \bar b$ since the jets
%no longer reconstruct the $Z$ mass.
}
 \label{fig:dijet_feynman_sum}
\end{center}
\end{figure}

One cause of unease is a sense that
NLO quantum effects invalidate the semiclassical
parton-shower picture.
Much of the NLO corrections comes from real emission diagrams.
At the quantum level, there is indeed
interference between diagrams with the same final particle
flavor and momenta.  This is illustrated in Figure \ref{fig:dijet_feynman_sum} where
in one diagram a collinear gluon emission affects the properties of
unambiguously quark-initiated jets, whereas the other diagram is
a quantum mechanically indistinguishable correction where the gluons
come from a completely unrelated additional hard parton.  Looking only
at the flavor and momenta of the final state, one might be uncomfortable claiming the configuration
corresponds to two quark jets.
However, the parton-shower-like, nearly collinear diagram
has a much larger amplitude and therefore the uncertainty on labeling the configuration as having quark jets
is small.

Up to an overall normalization, much of the NLO effects are reproduced by including matrix element corrections
merged with a parton shower. In a fully-matched sample (using CKKW\cite{Catani:2001cc} or MLM\cite{Hoche:2006ph} for example), each jet comes unambiguously from exactly
one hard parton, and the flavor of this parton is known. The matching procedures have some merging scale, on which the final
distributions depend only weakly. Thus one can make the same conclusion about matching for quark and gluon discrimination
as for almost any other application (such as kinematic reconstruction): it gives unambiguous answers when the final state contains
clearly separated jets. In ambiguous final states, which can be explicitly avoided, there is no well-defined underlying parton topology
relevant for any analysis.

%  When precision studies are
% required and a proper NLO calculation needs to be done.

Ambiguities are always present in event-reconstruction from jets: fully
hadronic $t \bar t$ decay doesn't always produce six clean, well-separated jets
with unambiguous correspondence to $b$ and $W$ decay products. Thus, the problem is no
worse for quark/gluon tagging than for top-reconstruction. Secondly, the mixing effect is numerically
quite small. It is of course suppressed by a factor of $\alpha_s$. But also, hard splittings
which change quarks to gluons or vice versa are power suppressed, for example by $m_{\mathrm jet}/E_{\mathrm jet}$.
NLO ambiguities are important for
measurements like the inclusive jet cross sections, but
the bottom line is that NLO effects do
not prevent a quark/gluon tagger from being a practical tool for many new physics
searches.

For additional justification, we point out that $B$-tagging is at least as ambiguous as quark/gluon-tagging,
and has been well-proven to be useful. For $B$-tagging, there are many simple, leading-order $B$-jet
definitions. For example, $B$-jets can be defined as jets that contain a $B$-hadron among its decay products.
This does not mean that a perfectly accurately $B$-tagged jet has a
momentum that corresponds to the initiating $b$ quark. For example, an event
with a $Z$ decaying to $b$ and $\bar b$ quarks doesn't
necessarily produce a pair of $B$-tagged jets that have an
invariant mass that corresponds to the $Z$.  Away from the mass
peak, either the wrong jets contained the $B$-hadron or there
is simply no single jet that corresponds to each $b$ quark. At
the Monte Carlo level, the hard $b$-quarks won't correspond
to $B$-tagged jets any better than a $W$ boson's direct quark decay
products will correspond to two jets that should obviously be
labeled \emph{the} quark jets.
%
%
% Loops don't create any ambiguity.  They might change the
% relative probability to see a particular final state, but there
% is no interference between diagrams with different final
% flavors.

An alternative to using the event record in a Monte-Carlo to extract truth information
about the hard parton initiating a jet would be to cluster the partons in the jet such that
flavor information is retained. An algorithm for doing so was
proposed by Banfi, Salam, and Zanderighi in~\cite{Banfi:2006hf}.
Their idea was to count the number of quarks minus anti-quarks in a jet.
By itself, this would not be a good infrared and collinear-safe definition.
But they modified the $k_T$ jet-clustering algorithm to combine only partons that
preserve flavor in an infrared and collinear-safe way.
Gluons can be combined with $u$-quarks to make a a $u$-jet,
$u$ and $\bar u$ can be combined into a flavorless gluon-jet,
but $u$ and $d$-quarks cannot be combined.
The focus of~\cite{Banfi:2006hf} was on precision calculations in  perturbative QCD
with a small number of partons involved.
The applications of quark/gluon tagging at hadron colliders are somewhat different. 
Since the observables at colliders are tracks and
calorimeter deposits from color-neutral hadrons, a parton-level jet-flavor algorithm like the one in~\cite{Banfi:2006hf} is
not directly applicable.
%Many of the most interesting such jets arise from the decay of new
%particles. Finding these events, rather than
%measuring QCD cross sections, is what interests us here.
The exact quark minus anti-quark count is not reliably observable,
nor does it directly capture the useful but vague
notion that a particular jet `was initiated by' a particular quark or gluon.
The algorithm in~\cite{Banfi:2006hf} could be used on the pre-hadronization partons in a
Monte Carlo event record to assign a truth-flavor in a non-matched sample. But if 
the relevant hard partons are available in the event record, one might as well use them for
the truth information, since this corresponds exactly to what one is trying to extract from the event.

%
% Is flavor meaningful beyond leading order? Flavor is
% well-defined to to \emph{all} orders in QCD perturbation theory
% in the sense that at each vertex in a collinear parton shower,
% the number of quarks minus anti-quarks is conserved. Ambiguity
% arises only when further radiation (hard QCD and soft
% showering) doesn't match jet grouping. These situations are
% described by power corrections that affect any collinear and IR
% safe jet algorithm's parton correspondence and involve
% $\Lambda_{QCD}/E$, jet size $R$, jet's mass-to-energy ratio
% $m/E$, etc.

To verify the distributions of the variables discussed below,
samples of known flavor-composition can be used.
For example, in a $\gamma jj$
event when the softer jet is near enough to the photon, it
is over 98\% likely to be a quark jet. (This can be understood from the
simple observation that quarks radiate photons but gluons do not.)
A catalog of high cross-section processes and kinematic cuts which can
be used to purify samples was given in
\cite{Gallicchio:2011xc}. 
The fraction of a cross section consisting of quark or gluon jets
is in fact well-defined beyond the leading order in perturbation theory
as long as the jets are hard and well-separated. This follows
essentially because helicity is conserved in the collinear limit, as discussed in detail in
~\cite{Gallicchio:2011xc}. 

Jets from the pure samples discussed above might not be representative of jets in a
signal or background of interest.  
To predict the jet properties of a new signal, simulations must
be employed at some level, and there are problems assigning truth-level
flavor to these jets.
One popular method is to `match' the jets to the hard process using
their  $\Delta R$.
While this is common, it doesn't
take into account how well the energies match. It is also not
guaranteed, for example, that the 4-momenta of hard partons from MadGraph
are preserved when {\sc Pythia} adds initial state
radiation and has to rebalance the event.
This procedure can only be trusted in a matched sample, where
the hardest jets have explicit matrix-level counterparts.
Another method is to examine the shower history, which can
be used to assign a jet a truth tag if a large enough fraction
of its energy is `descended' from a single hard parton. 
Changing the definition of `large enough' might alter the 
distribution of the jet properties of interest. 
But a larger problem is that soft radiation is really a
dipole effect, sourced by two hard partons, whereas
{\sc Pythia} randomly assigns this soft radiation to either
one or the other parent.
%Contaminated or split/merged jets will either not
%be assigned a truth flavor or they will be assigned to an
%`other' category. Because the hadronization step leaves too many hadrons
%related to each other, an alternative procedure is to
%trace this history using post-showered, but pre-hadronized
%partons. This would involve matching
%fully-simulated jets (or post-hadronization particle jets) to
%pre-hadronization parton jets. Our preliminary studies have found
%that all of these methods agree for reasonably separated jets, as expected.

\section{Event Generation \label{sec:generation}}
Most of the results in this paper pertain to generated with {\sc MadGraph
v4.4.26}~\cite{Alwall:2007st} and showered through {\sc Pythia
v8.140}~\cite{Sjostrand:2007gs} with most recent default tune.
We also compare to the same events showered with {\sc  Herwig++ 2.5.2}~\cite{Bahr:2008pv}.
Jets are reconstructed using {\sc FastJet
v2.4.2}~\cite{Cacciari:2005hq}. The multivariate analysis is
done using the {\sc TMVA v4.0.4} package~\cite{tmva} that comes
with {\sc ROOT v5.27.02}~\cite{root}.

No detector simulation was done. Instead, we discard charged particles with momenta less than 500\,MeV.
These particles are not allowed to
% are allowed ``out of the magnetic field'' to
contribute to either the construction of jets or
the observables involving charged tracks.  This 500\,MeV cut is identical to early ATLAS
studies~\cite{Feng:2010wh} (later studies have raised  the cutoff to 1\,GeV~\cite{Aad:2011he}).
With data and better tunes, a full detector simulation (not publicly available) will
become necessary to validate the various variables.
%Getting this right is difficult for us.

Since experiments also compare
to Monte Carlo truth-hadrons, our study provides a useful rendezvous point.
The goal of this paper is to point out potentially
interesting observables, some new, which might either be used
right away or studied in greater detail. To that end, we have made an effort
to find observables that depend more on the perturbative parton
shower than on hadronization. No effort has been made to
explicitly consider multiple interactions, though they are
included in the underlying event model.  Pileup, however, is not explicitly included
since removing it is best studied with a full detector simulation.

We start from a dijet sample $pp \rightarrow jj$ with the jets
in each sample having their $p_T$ in windows centered around values spaced
by factors of 2 in GeV: $(50, 100, 200, 400, 800, 1600)$.
We also considered a back-to-back $\gamma+jet$ sample for the same jet $p_T$s,
and the results were nearly identical.

The shape of the quickly falling $p_T$ distribution within each window affects the efficiencies of various variables.
For example, the
$p_T$ of a jet depends on the jet algorithm and jet size;
this dependence is \emph{precisely} one of the variables
studied here that usefully distinguishes quarks from gluons.
That the cross sections fall sharply with $p_T$ makes the initial sample selections quite delicate.
%For
%example, we might like to have a 200\,GeV anti-$k_T$ $R=0.5$ jet and compare
%it to quark and gluon jets of the same $p_T$ from that same
%algorithm.
Ideally we would simulate a `natural' dijet
$p_T$ distribution and select jets only within an infinitesimal window
around each $p_T$.  With a narrow enough window, the falling
distribution can have a negligible effect.
But the shower, hadronization, and jet algorithms must all be
run before this determination can be made, so an extremely narrow window is computationally inefficient.

To deal with the rapidly falling distributions, we chose parton-level MadGraph cuts to reproduce
samples with `natural' anti-$k_T$ $R=0.5$ jet $p_T$
distributions within a $\pm$10\% window, starting with the most
narrow parton $p_T$ window that was possible.
Even when quark and gluon partons start with an identical
initial $p_T$, after showering, quark jets had to higher average
$p_T$ than gluon jets. This was compensated by shifting and widening
the initial parton-level $p_T$ windows to efficiently generate
a representative distribution of jet $p_T$s within the narrower jet window.
The gluon and quark parton $p_T$
windows were chosen to have a $\pm$20\% width and were shifted relative to each
other to align the center of their anti-$k_T$ $R=0.5$ jet $p_T$ distributions on the nominal values.
The resulting $p_T$
distributions still aren't exactly identical: the gluon has
more of a lower tail, and the quark distribution has an
upper-tail. So only jets within $\pm
10\%$ of the nominal $p_T$ were kept.
Additionally, for dijets, the $p_T$ distribution of the gluons falls faster
than that for quarks (as can be inferred from the changing fractions in Figure \ref{fig:Chance_EACH_Jet_is_Quark1}),
but for our narrow final jet $p_T$ window, this slope difference is negligible.
Our shifting and spreading successfully decoupled jet $p_T$ from the jet properties
while maintaining somewhat efficient event generation.
This prevents our tagger from picking up on the difference in $p_T$
distribution of the input samples rather than the jet properties.

\begin{figure}[t]
\begin{center}
\psfrag{arbitrary}{\scriptsize{(arbitrary)}}
\psfrag{Pt GeV}{\scriptsize{$p_T$ (GeV)}}
\psfrag{ak05_Pt}{\scriptsize{\quad $p_T$ \quad \ \  50\,GeV anti-$k_T$ R=0.5 Jets}}
\includegraphics[width=0.48\textwidth]{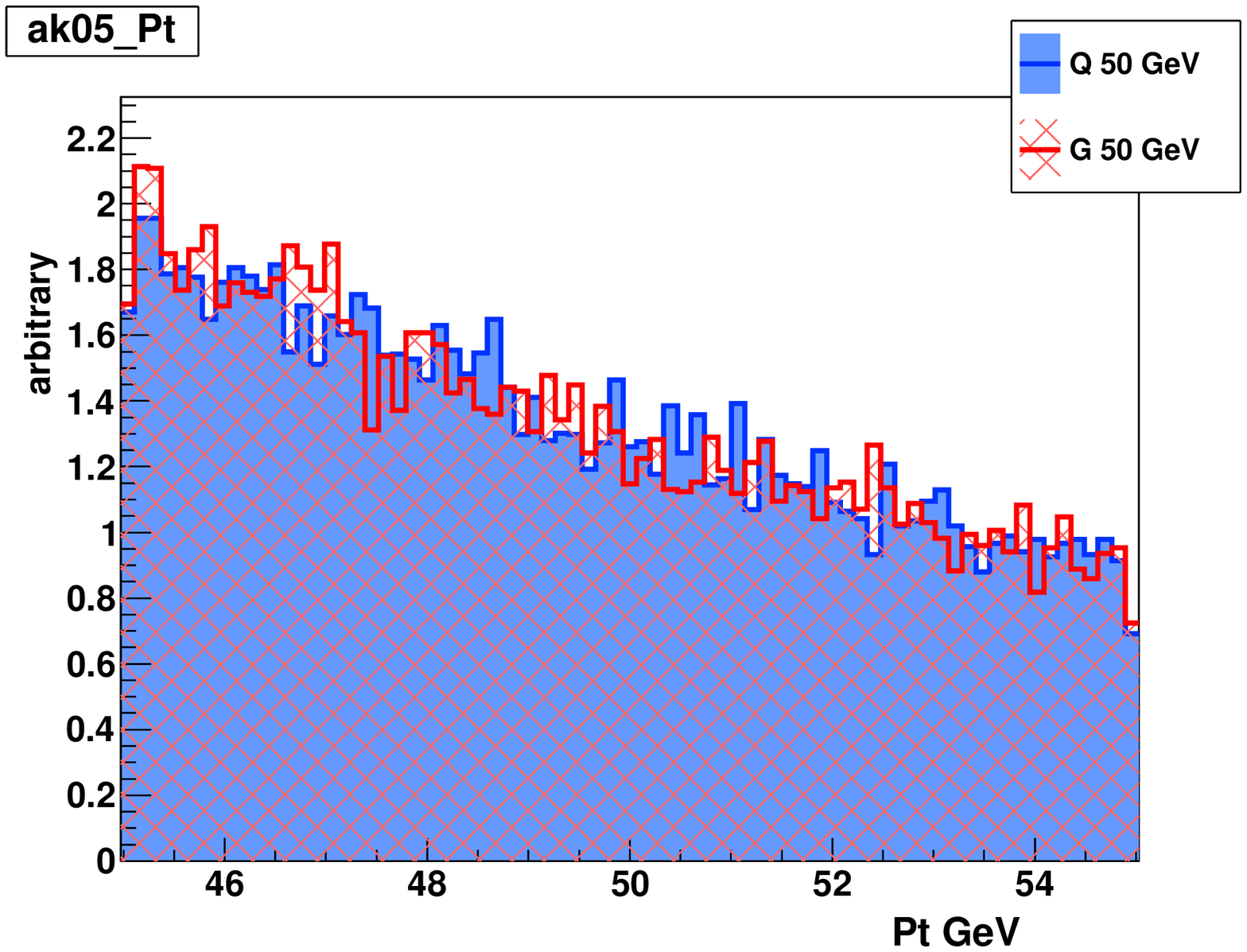}
\psfrag{ak05_Pt}{\scriptsize{\quad $p_T$ \quad \ \ 800\,GeV anti-$k_T$ R=0.5 Jets}}
\includegraphics[width=0.48\textwidth]{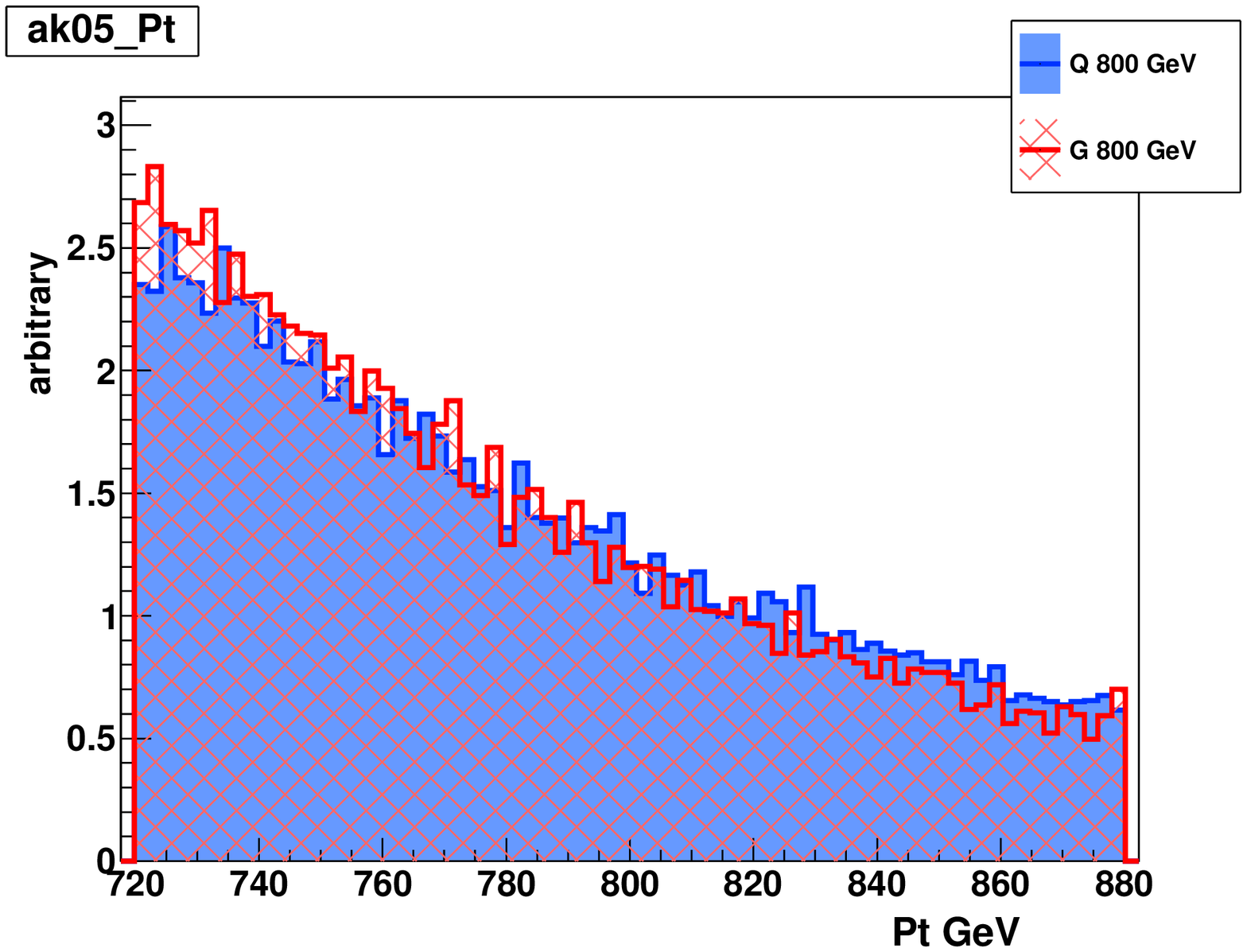}
\caption{
The $p_T$ distributions for two quark and gluon jet samples from {\sc Pythia8} with arbitrary normalization.
Our samples at each $p_T$ were chosen such that anti-$k_T$ R=0.5 jets had $p_T$ values within 10\% of the nominal value.
For all QCD jets, this distribution is falling, but within our window, the
$p_T$ itself cannot be used distinguish quark from gluon jets.
On the left is the 50\,GeV sample, and on the right is the 800\,GeV sample.
}
\label{fig:Pts}
\end{center}
\end{figure}

This pre-shift and post-window procedure above slightly biases
the sample for a finite width. Any `real' set of jets at a
particular $p_T$ will include some whose underlying parton was
much softer, and others where it was much harder. A `natural'
$p_T$ distribution, especially for a QCD background, is
exponentially falling, which means that jets at a particular
$p_T$ will more likely come from softer partons that
showered-up rather than harder partons that showered-down.
With these caveats, the best advice is to take our scores as a rough
guide, focus on ones that don't change drastically with jet
$p_T$, and train any multivariate discriminant either bin-by-bin in $p_T$, or on the actual
underlying $p_T$ distribution of your signal and background samples.
To construct a general gluon tagger, the experiments will need
a canonical jet definition so they can train it on a set quark
and gluon jets with identical $p_T$ distributions with respect
to that jet definition.  Some anti-$k_T$ R=0.5 $p_T$
distributions within our windows are shown in Figure~\ref{fig:Pts}.

%\section{Explosion of Dimensionality and Infrared Safety}
\section{Overview of Observables \label{sec:observables}}

For the purposes of quark/gluon tagging, a jet can be thought
of a set of particles, tracks, or calorimeter deposits.
Each constituent has a 4-momentum and possibly a charge or
particle ID, though this is difficult to determine.
%Thus we assume all constituents are massless.
% (The main exception being for $B$-hadrons, which can decay.)
Given this huge set of constituent
data, the goal is to estimate the likelihood that the jet was
initiated from a quark rather than a gluon.
%Any tagging
%score that is a one-to-one function of the likelihood is
%equally optimal when cut on, and any approximation of such a
%one-to-one function is still useful.

If a jet is made of 100 constituents, each with a
massless 4-momentum, the problem is 300-dimensional.  Ideally we'd
have a fully differential cross section: a 300-dimensional
probability density for quark jets, and one for gluon jets.
This would take into account important correlations when
reducing the list of particles to a one dimensional quark/gluon likelihood,
but of course this is completely unrealistic.
Familiar multivariate classifiers like Neural Networks or
Boosted Decision Trees are designed to estimate this likelihood if properly trained.
Unfortunately, they aren't designed to deal with a long,
variable-length list of inputs. In other words, we can't just give them
the 4-momenta for each particle in the jet and hope for
the best.
The challenge is to find simple observables that allow us to
get as close to this ideal likelihood as possible.
Since the particles are not independent, a good observable
must extract the most important correlations.

As in jet algorithms, we'd like to do this in a way that is
infrared and collinear-safe.  This means that a jet's score
shouldn't change if one particle in the jet is replaced by two
where either (1) both are traveling in the same direction as the
original or (2) one has very soft momentum. Raw particle
count is not infrared safe, nor are things that depend directly on particle count like
average particle $p_T$. Charged particle count with a minimum $p_T$ is safer
with respect to soft emission and also safer with respect to collinear emissions, since these must conserve charge.

Infrared and collinear safety is usually framed as a strict yes/no
requirement in the limit of exactly collinear splitting or zero
momentum soft emission.  In reality, by the time individual tracks or
calorimeter deposits are observed, they would never be exactly
collinear nor infinitely soft. Thus it is more meaningful to
envision spectrum of safety. For example, while all of the
popular iterative jet algorithms are infrared safe,
counting the number of small anti-$k_T$ subjets of size $R$=0.1
is less safe than counting larger $R$=0.3 subjets.  Unfortunately, the smaller
subjets turn out to be more useful, and the charged particles
even more so.
%We've been careful to only consider variables
%that are theoretically safe and experimentally reasonable, but
%each observable will have to be understood and validated by
%each detector through full simulation and data.

%\section{Overview of Observables \label{sec:observables}}

Thus we consider two main types of observables: ones that try to
distinguish individual particles, tracks, or subjets, and ones that
treat the energy or $p_T$ within the jet as a function of
$(\delta y, \delta \phi)$ away from the jet axis.

The first category includes things like count,
average $p_T$, and spread (standard deviation)
in $p_T$ for these discrete objects.
Subjets can be obtained by explicit $k_T$ declustering into $N$ jets, 
or by running a jet algorithm again with a much smaller $R$.
These have been studied and confirmed extensively at
LEP, and provide better discrimination at high quark efficiency
and high $p_T$, but can be more difficult to measure at hadron
collides in crowded jets.
Here CMS particle flow or Atlas
TopoClusters can extract the most information out of each jet.
The identities of particles were not explored by us,
but as mentioned above, LEP found an increase in baryons (protons and Lambdas) for
gluon jets and an increase in kaons for quark jets.

The discrete-category observables that we evaluated are listed below.
In later sections,
the useful ones are described
in more detail, distributions are shown, and gluon rejection
scores are compared for different parameters (like jet size).

\vspace{1em}

\noindent Distinguishable Objects (particles/tracks/subjets)
(Section~\ref{sec:Discrete Results}):
\begin{itemize}
\item Particle/track/subjet multiplicity, with different
    subjet algorithms and sizes $R_\mathrm{sub}$
\item $\langle p_T \rangle$: Average $p_T$ of
    particles/tracks/subjets within the jet
\item Higher statistical moments like $\langle p_T^2
    \rangle$ and $\sigma_{p_T}$
\item Average distance from jet axis $\langle r \rangle$
    and higher moments like $\langle r^2 \rangle$
\item $\langle k_T \rangle$: Average $k_T$, the momenta transverse to the
    jet axis.
\item The $p_T$ fraction or $\Delta R$ of subjets, 
      explicitly $k_T$ declustered into $N$ jets
%\item The $p_T$ of $1^{\text{st}}$ hardest, $2^{\text{nd}}$
%    hardest, and $3^{\text{rd}}$ hardest subjets or tracks
%\item Ratio of leading subjet or track $p_T$ to jet $p_T$
\item Subjet count above a particular $p_T$ or a fraction
    of jet $p_T$
\item Subjet splitting scale
%\item $\Delta R$ between subjets
\item Masses of subjets
\item Charge-weighted $p_T$ sum of tracks
\end{itemize}

The second, more continuous, category includes jet mass, jet
broadening, and the family of radial moments.
These tend to be better for lower $p_T$ jets and better
at achieving high-purity at the cost of
low quark-efficiency.
Some observables we evaluated are listed below with
more detail left to later sections.

\vspace{1em}

\noindent Continuous Shapes  (Section~\ref{sec:Continuous Results}):
\begin{itemize}
\item Jet Mass and $m/p_T$ ratio
    (Section~\ref{sec:Jet Mass})
\item Jet Shapes: integrated and differential, to some
    distance from the jet axis
    (Section~\ref{sec:Traditional Jet Shape})
\item Radial moments
    (Section~\ref{sec:radial_geometric_moments})
\item ``Girth'' of each jet (Section~\ref{sec:girth})
\item Jet Broadening (Section~\ref{sec:girth})
\item Jet Angularities (Section~\ref{sec:jet_angularities})
\item Optimal Radial Moment (Section~\ref{sec:optimal_kernel})
%\item Triangle $p_T$ distance version of Optimal Moment
\item N-Subjettiness (Section~\ref{sec:N-subjettiness})
\item Two-Point Moment (Section~\ref{sec:Two-Point_Moment})
\item Higher geometric moments like eccentricity or planar
    flow (Section~\ref{sec:other_2d_moments})
\item Pull as a measure of color-connections (Section~\ref{sec:pull})
\end{itemize}

%Properties of isolated jets don't depend on $\phi$, and by
%virtue of boost invariance, they and their underlying-event
%contamination don't depend significantly on $\eta$, at least
%for our focus on $|\eta| < 2.5$.  This seems true even for
%things like pull, where higher $\eta$ jets are closer to the
%beam, which is their color-connected partner.  Other jets in
%the event, color connected or not, along with the underlying
%event and multiple interactions complicate matters. A gluon
%tagger should be robust, and applicable to a variety of
%situations. Most of the energy is concentrated close to the
%center of the jet...

% Does the actual shape of the distributions stay constant for
% different $\eta$, or just their `discrimination power'?

When we describe these variables in more detail, it will
be useful to have a way to score or rank them.
In the next section, we describe our scoring methods.

\section{Evaluation of Discrimination Power}

In this section we describe ways of quantifying our
jet observables' quark/gluon tagging power.
We will then use gluon rejection at a fixed quark acceptance
to rank our variables to find the most powerful discriminants.
We will also look at how the discrimination power depends on things like
jet $p_T$ or jet size.

% The choice of a measure is ultimately somewhat
%subjective, and ours will be motivated by the tagger's use
%in new physics searches.
%We'll describe several popular methods, but ultimately settle on
%gluon rejection at a fixed quark acceptance.

\noindent
One method of evaluating an observable's discriminating power is
the \emph{separation} \cite{tmva}
\begin{equation}
\langle S^2 \rangle =
\frac{1}{2}
\int
\frac{\left[ \hat p_S(x) - \hat p_B(x) \right]^2}
{\hat p_S(x) + \hat p_B(x)}
dx \ ,
\end{equation}
where $\hat p_S(x)$ and $\hat p_B(x)$ are the signal
and background probability density functions (PDFs) of some
observable $x$.  Identical distributions give zero separation,
while ones with no overlap give a separation of one. This
definition is invariant under any one-to-one change of variables.
This invariance this fixes the exponent on the PDFs used in the
numerator and denominator.  This measure does not directly
tell us how the observable will perform as a tagger, so we
do not use it.

\begin{figure}
\begin{center}
\includegraphics[width=0.45\textwidth]{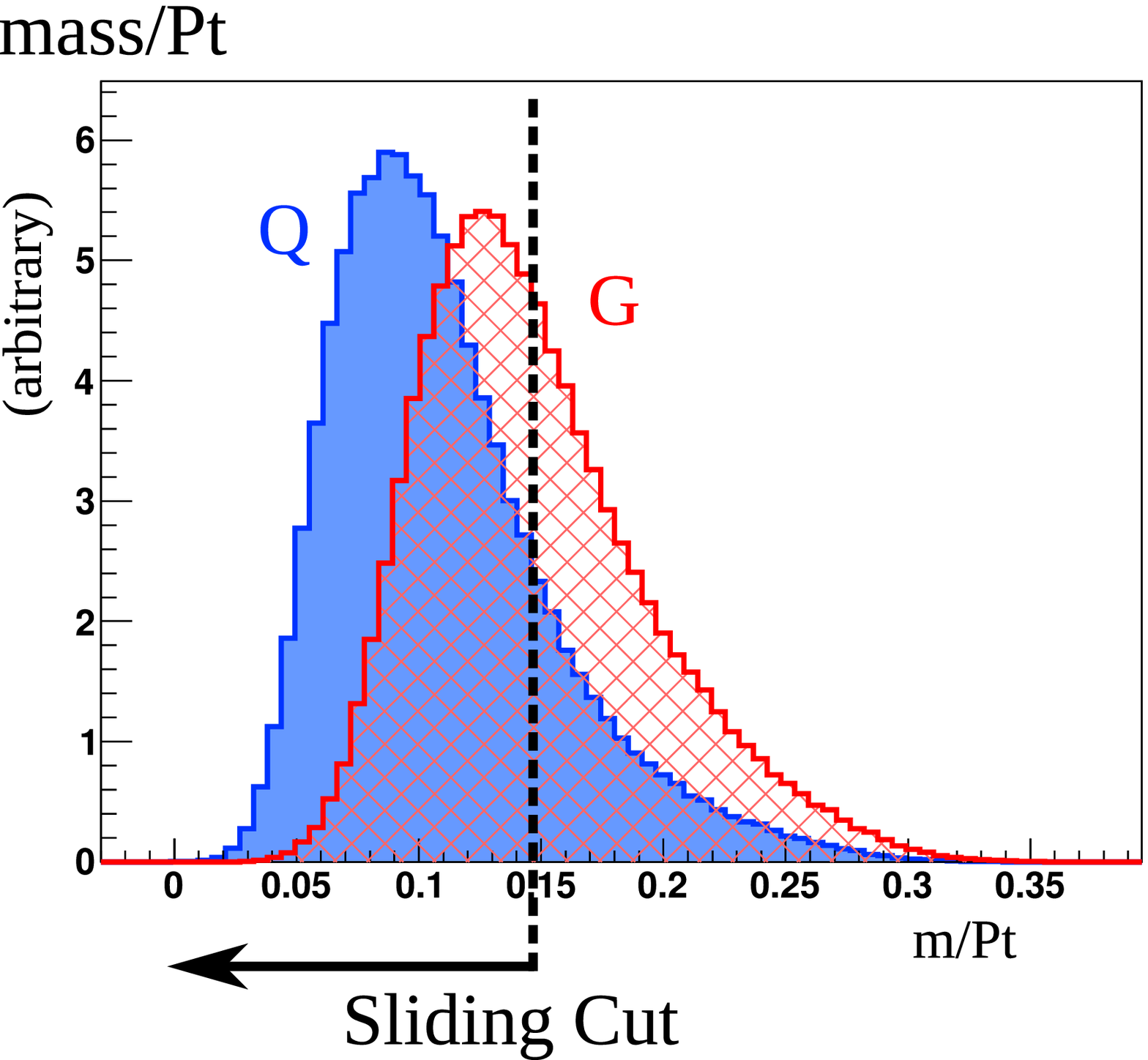}
\ \ 
\includegraphics[width=0.50\textwidth]{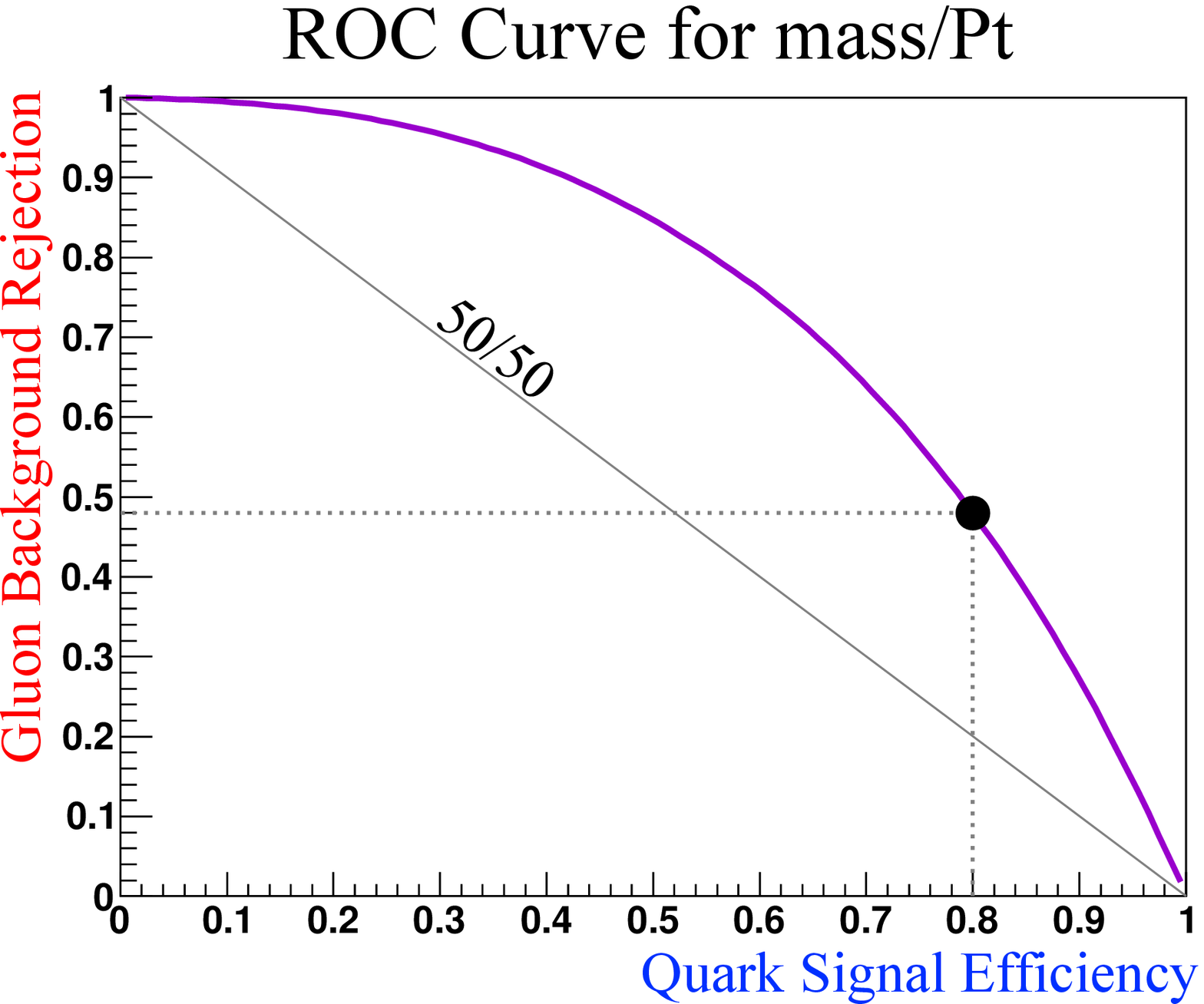}
\end{center}
\caption{
On the left is the distribution of {\sc Pythia8} particle jet mass divided by $p_T$ for
quark and gluon jets of $p_T \approx 200$\,GeV.
In all plots that follow, quarks are always solid blue and gluons hashed red,
and the distributions are arbitrarily normalized to equal area.
Every cut leads to a particular efficiency for keeping quarks
and a (hopefully lower) efficiency for keeping gluons.
One minus this gluon efficiency is the background rejection.
The curve formed by all possible cuts
is called the ROC curve, and is shown on the right.
The particular point shown corresponds to keeping 80\% of the quark
signal while rejecting 50\% of the gluon background.
%Adding more variables and finding more optimal cut contours or surfaces can improve this
%rejection for every quark efficiency.
}
\label{fig:mass_over_pt_roc}
\end{figure}

Other measures of discrimination power involve the so-called
ROC curve\footnote{The name comes from radar. It stands
for Receiver Operating Characteristic.}
shown in Figure~\ref{fig:mass_over_pt_roc}. The ROC
curve is constructed by sliding a cut across the variable and
plotting the gluon `background' rejection against the quark
`signal' acceptance. If the distributions completely
overlapped, the result would be the diagonal line.  Height
above the diagonal represents discrimination power. One way to
quantify this power is to choose a reference signal efficiency
(80\% is shown) and measure background rejection there.
Variables can be ranked by rejection power at this chosen signal efficiency.

Variables with more complicated distributions might require a
two-sided cut with the signal either inside or outside the cut.  For
two-sided cuts there is no unique background rejection for a
given signal efficiency, so the best value is used. Even more
complicated distributions, including multivariate ones, require
more general boundaries (a contour in 2D, a surface in 3D, etc.)
Transforming the (possibly
multidimensional) quark and gluon observable distributions into a single
likelihood $q/(q+g)$ distribution always allows for a
single-sided cut on this likelihood.
If the multidimensional distributions were known a priori,
a sliding cut on this likelihood would form the best possible
ROC curve.  Any one-to-one map of this likelihood like
$\log(q/g)$ can also be cut on to produce an identical ROC
curve.  A good multivariate discriminator
(i.e. neural net or boosted decision tree)
estimates one such
one-to-one map given limited training data.

Sometimes ROC curves for different variables cross.
We will find, for example,
that at high signal efficiency (a loose cut), counting the charged tracks is
the best observable, while for low signal efficiency (a tight cut),
observables like jet broadening are best.
In these situations, there is no unique way
to rank the relative discrimination power since
different variables are superior for different signal efficiencies.
The area under the ROC curve, which is
equivalent to averaging the rejection power over all signal
efficiencies, provides another measure of discrimination power
without having to pick a reference signal efficiency.
%Area is not particularly useful
%when you want a very pure or very inclusive sample. These cuts
%correspond to the start or end of the ROC curve, and the
%distances to the diagonal tend to zero at the ends, so their relative
%power at these extremes contributes very little to the average.
Ranking the variables by area turns out to be very similar to ranking
them by their background rejection at around 50\% signal
efficiency, where the distance to the diagonal is greatest.

Any measure of the discrimination power of a variable is going
to depend on the $p_T$ of the jet, along with
other parameters like the jet algorithm or jet size.  It will
also depend on the source ({\sc Pythia}, {\sc Herwig}, data), although
robust and well simulated variables should minimize this dependence.
For these reasons, we'll do our best to display and summarize our
findings, but the final recommendation
will always be subject to validation on real data.

%For jet mass, different ranking methods are
%shown for different jet $p_T$s and different jet algorithms in
%Figure~\ref{fig:mass_for_diff_pt_and_algs}.

\section{Discrete (Particle/Track/Subjet) Variable Results}
\label{sec:Discrete Results}
\begin{figure}
\begin{center}
\begin{tabular}{cc}
\psfrag{count}{\footnotesize{count} }
\psfrag{arbitrary}{\footnotesize{(arbitrary)} }
\psfrag{components_jet0_ak05_charged_count}{\footnotesize{ \!\!\!Charged Track Count, 100\,GeV} }
\includegraphics[width=0.48\textwidth]{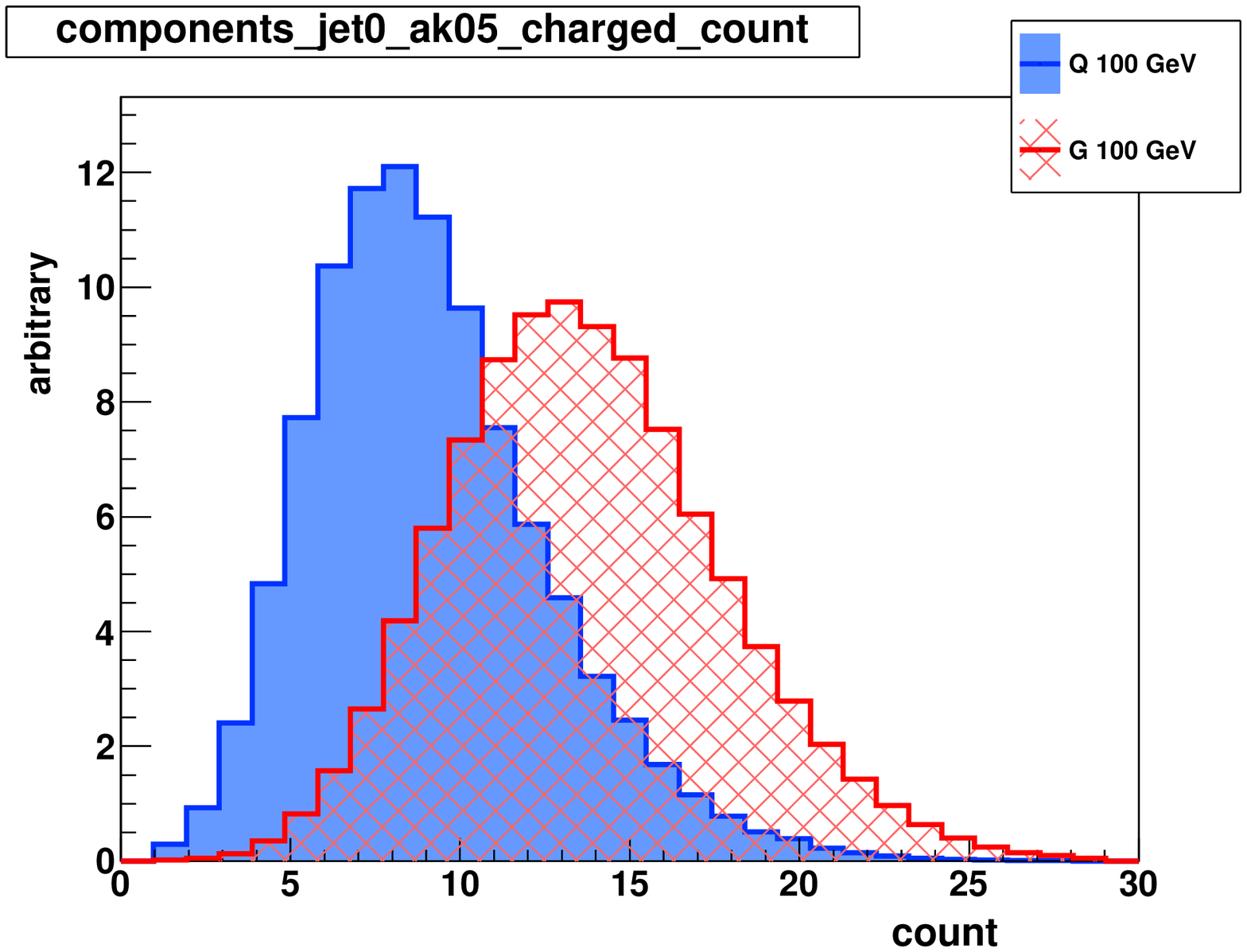}  &
\psfrag{count}{\footnotesize{count} }
\psfrag{arbitrary}{\footnotesize{(arbitrary)} }
\psfrag{components_jet0_ak05_charged_count}{\footnotesize{ Charged Track Count} }
\includegraphics[width=0.48\textwidth]{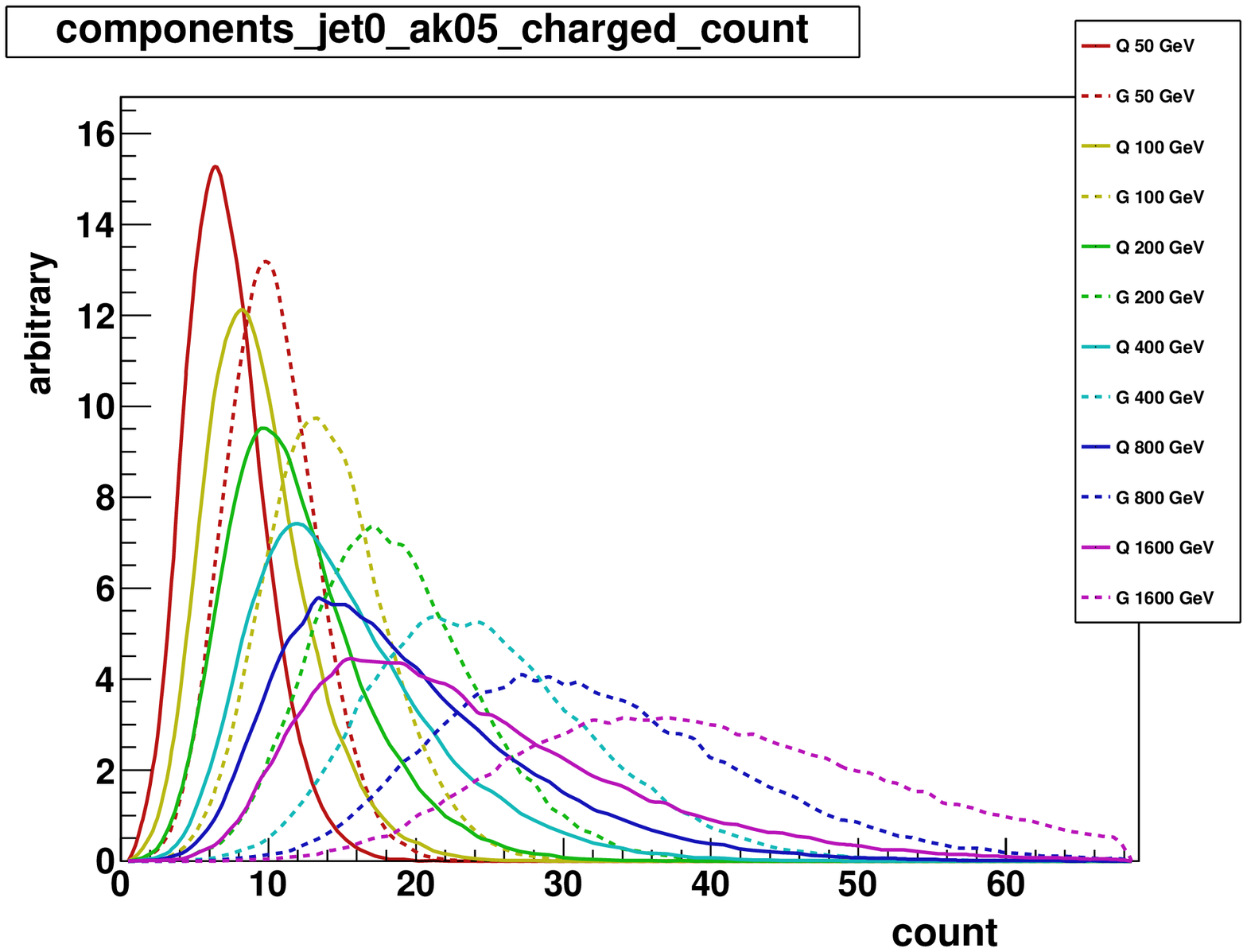}
\end{tabular}
\caption{
Charged track count for {\sc Pythia8} anti-$k_T$ R=0.5 jets, normalized to equal area.
\textbf{Left}: 100\,GeV jets with quark jets in solid blue and gluon jets in hashed red.
\textbf{Right}: All $p_T$ samples where solid is quark and dotted is gluon.
For a variable that is less sensitive to the jet $p_T$,
the count could be divided by the log of the jet $p_T$.
}  \label{fig:components_jet0_ak05_charged_count}
\end{center}
\end{figure}

\begin{figure}
\begin{center}
\begin{tabular}{ccc}
\psfrag{components_jet0}{\small{\qquad Charged Track Count} }
\psfrag{_ak}{ }
\psfrag{_charged}{ }
\psfrag{_count}{ }
\includegraphics[width=0.3\textwidth]{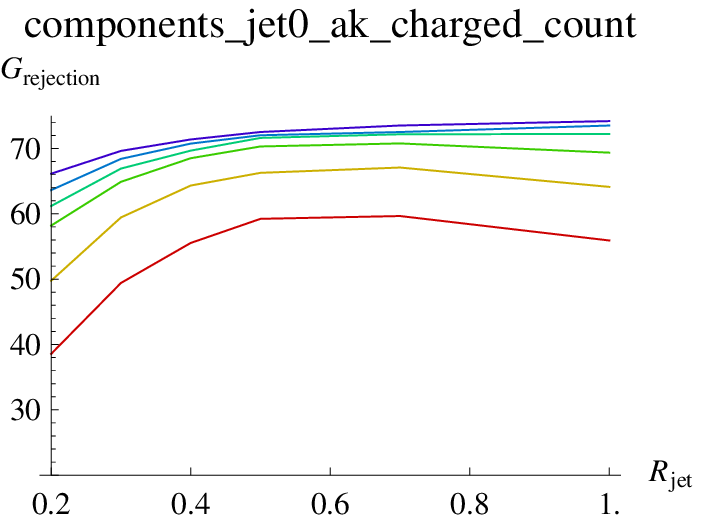}  &
\psfrag{components_jet0}{\small{\qquad Charged Track Count} }
\psfrag{_ak}{ }
\psfrag{_charged}{ }
\psfrag{_count}{ }
\includegraphics[width=0.3\textwidth]{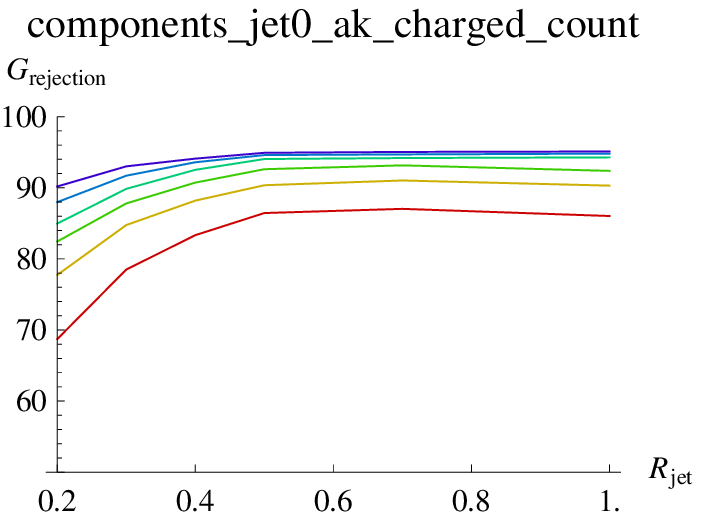}  &
\psfrag{components_jet0}{\small{\qquad Charged Track Count} }
\psfrag{_ak}{ }
\psfrag{_charged}{ }
\psfrag{_count}{ }
\includegraphics[width=0.3\textwidth]{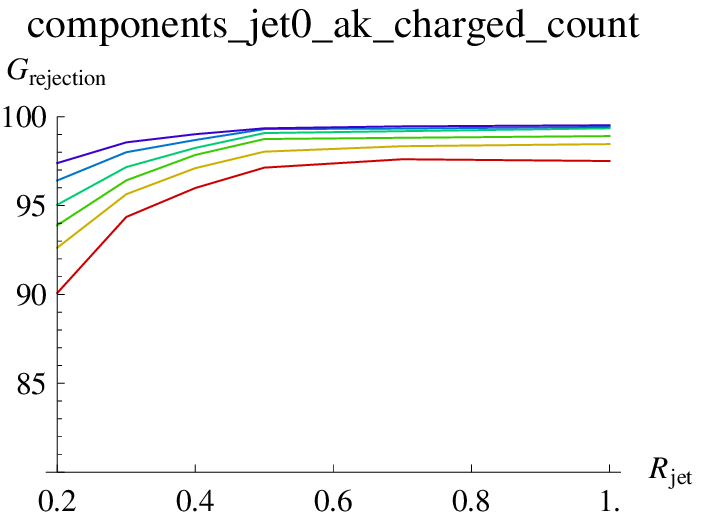}  \\
80\% quark &
50\% quark &
20\% quark
\end{tabular}
\caption{
For a single variable, in this case {\sc Pythia8} \emph{charged track count},
each panel is the gluon rejection for a different quark fraction. A mild 80\% cut
is shown on the left and the harshest 20\%
cut is on the right. Each plot shows the gluon rejection percentage (vertical axis)
as a function of jet size (horizontal axis).
The different lines in each plot correspond to different jet
$p_T$s, with the red (bottom) being
50\,GeV and going by factors of two until the purple (top) at 1600\,GeV.
The vertical scale is different for each plot, but higher rejection is always better.
Jet sizes between $R=0.5$ and $R=0.7$ achieve the best rejection.
Similar to all
count-type variables, higher $p_T$ jets can achieve better
gluon rejection because the shower has more `time' to establish
the different particle counts.
}
\label{fig:components_jet0_ak_charged_count_80_50_20}
\end{center}
\end{figure}

For discrete variables our results can be summarized simply as ``smaller is better''.
Counting all particles gives the best discrimination power, although since neutral particles often cannot
be distinguished, especially in a high-pileup environment, using all particles may not be practical.
If all particles are not available, the next best option is counting charged tracks. We define
charged tracks to mean all
charged particles in the jet above 0.5\,GeV.
%(to make it out of the tracker's magnetic field).
Distributions
of charged track count is shown in
Figure~\ref{fig:components_jet0_ak05_charged_count},
and the gluon rejection for all $p_T$ samples, jet sizes, and
three different quark efficiencies are shown in
Figure~\ref{fig:components_jet0_ak_charged_count_80_50_20}.

\begin{figure}
\begin{center}
\begin{tabular}{cc}
\psfrag{count}{\footnotesize{count} }
\psfrag{arbitrary}{\footnotesize{(arbitrary)} }
\psfrag{components_jet0_ak05_sub_ak010_count}{\footnotesize{ Anti-$k_T$ $R_\mathrm{sub}=0.1$ Subjet Count} }
\includegraphics[width=0.48\textwidth]{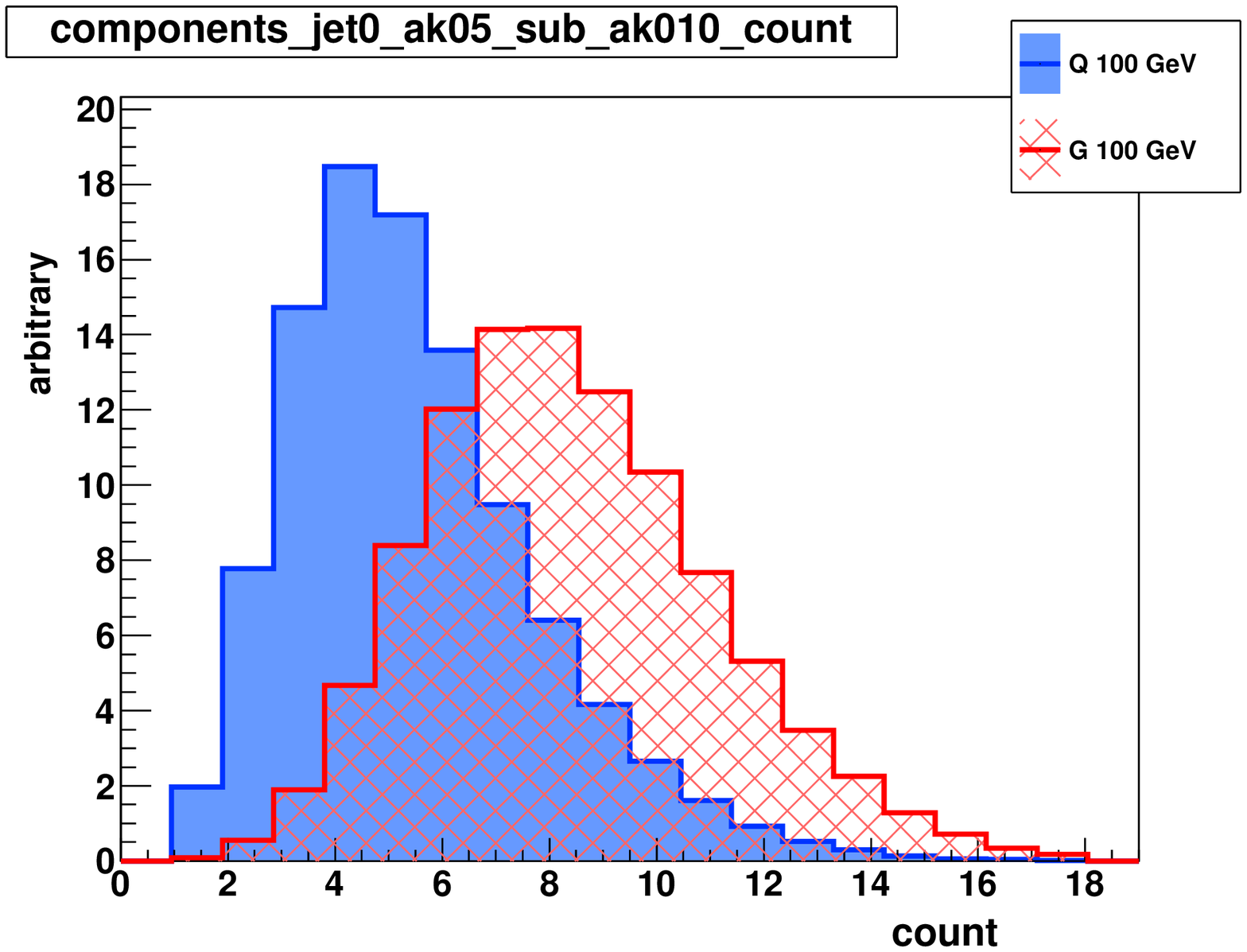}  &
\psfrag{count}{\footnotesize{count} }
\psfrag{arbitrary}{\footnotesize{(arbitrary)} }
\psfrag{components_jet0_ak05_sub_ak010_count}{\footnotesize{ Anti-$k_T$ $R_\mathrm{sub}=0.1$ Subjet Count} }
\includegraphics[width=0.48\textwidth]{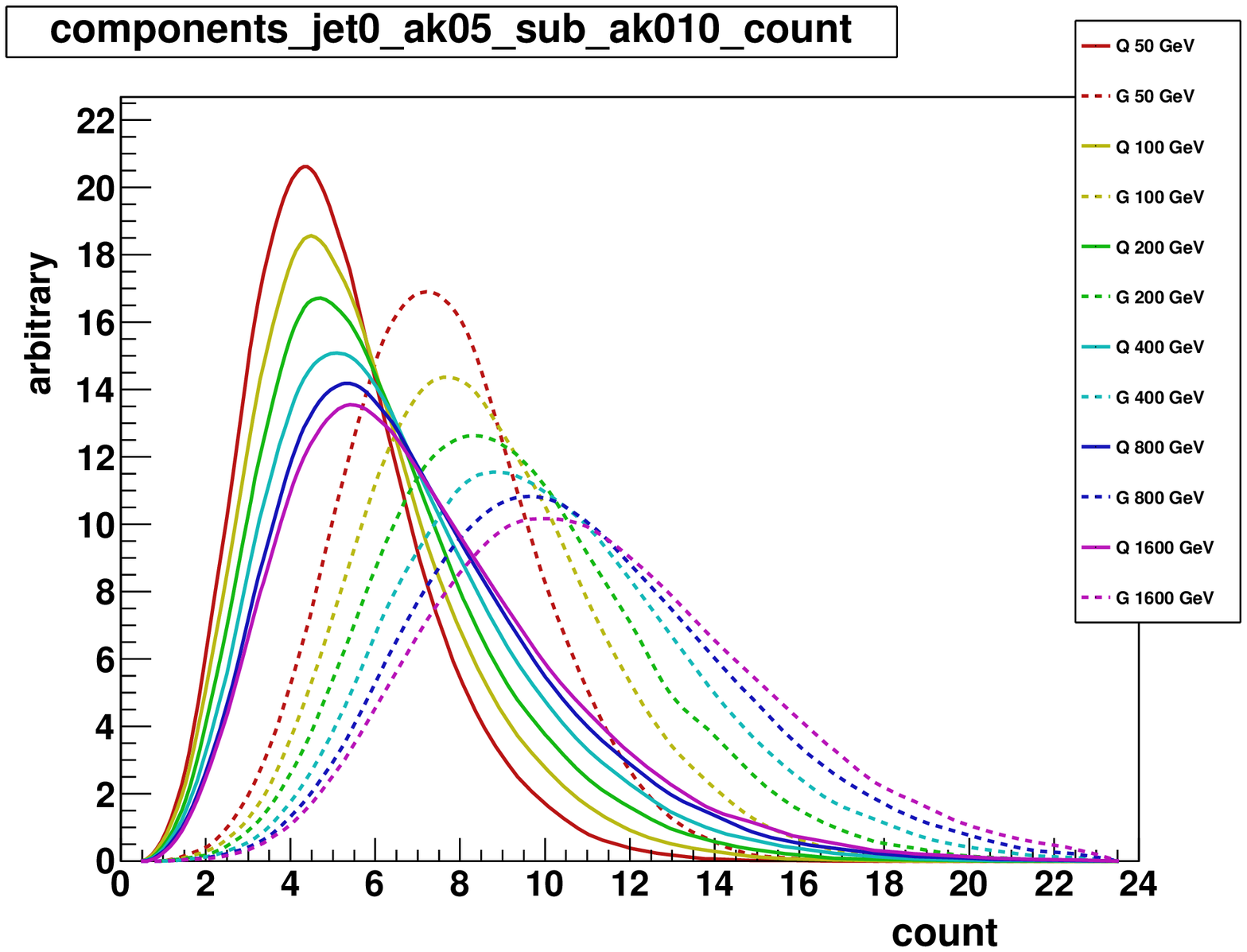}
\end{tabular}
\caption{
Subjet count for {\sc Pythia8} anti-$k_T$ R=0.5 particle jets and subjets using anti-$k_T$ R=0.1, normalized to equal area.
\textbf{Left}: 100\,GeV jets.
\textbf{Right}: All $p_T$s.
}  \label{fig:components_jet0_ak05_sub_ak010}
\end{center}
\end{figure}

\begin{figure}
   \begin{center}
   \begin{tabular}{ccc}
       \psfrag{components_jet0}{Subjet Count}
       \psfrag{_ak}{ }
       \psfrag{_sub}{ }
       \psfrag{__count}{ }
       \includegraphics[width=0.48\textwidth]{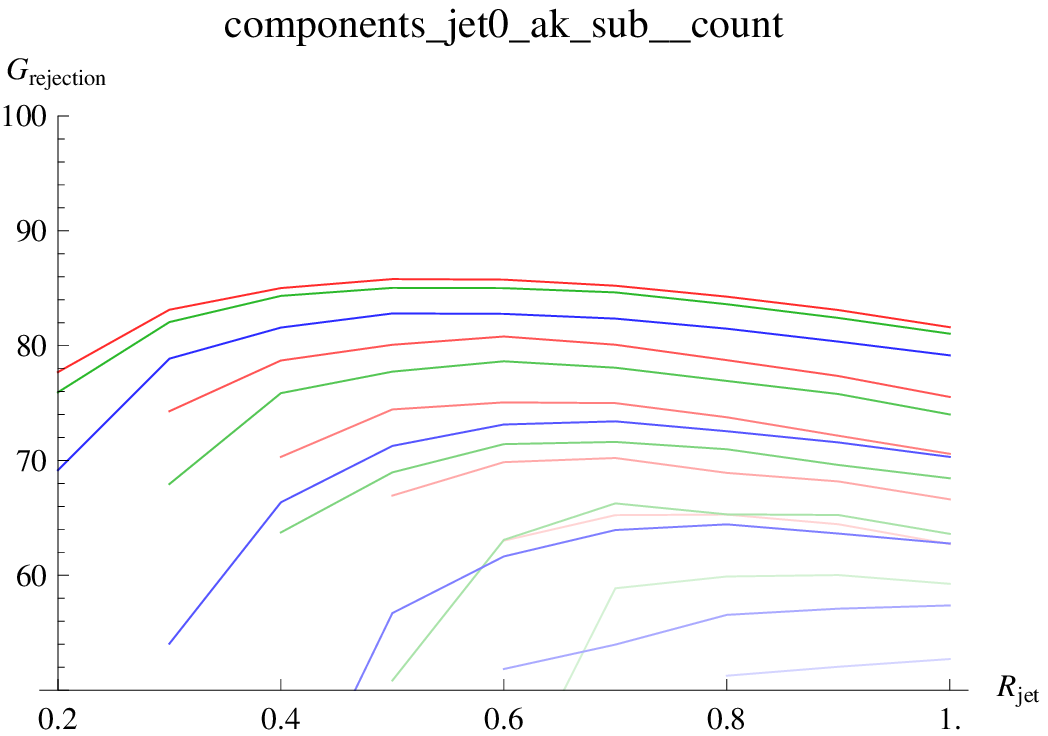} & \qquad &

       \psfrag{subjet_}{1st Subjet's $p_T$ Fraction}
       \psfrag{1}{ }
       \psfrag{stPt_over}{ }
       \psfrag{_jetPt}{ }
       \psfrag{_jet0}{ }
       \psfrag{_ak}{ }
       \psfrag{_sub}{ }
       \psfrag{_}{ }
        \includegraphics[width=0.48\textwidth]{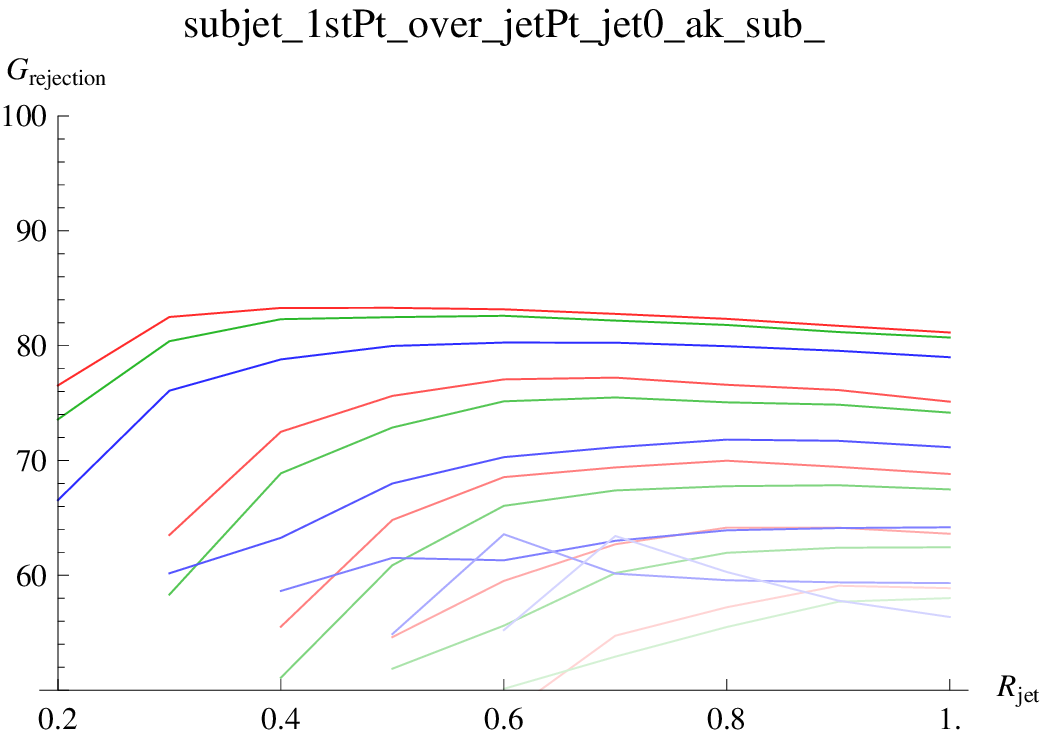} \\

       \psfrag{components_jet0}{Subjet $p_T$ Standard Deviation}
       \psfrag{_ak}{ }
       \psfrag{_sub}{ }
       \psfrag{__stddev}{ }
       \psfrag{_pT}{ }
       \psfrag{_over}{ }
       \psfrag{_jetPt}{ }
       \includegraphics[width=0.48\textwidth]{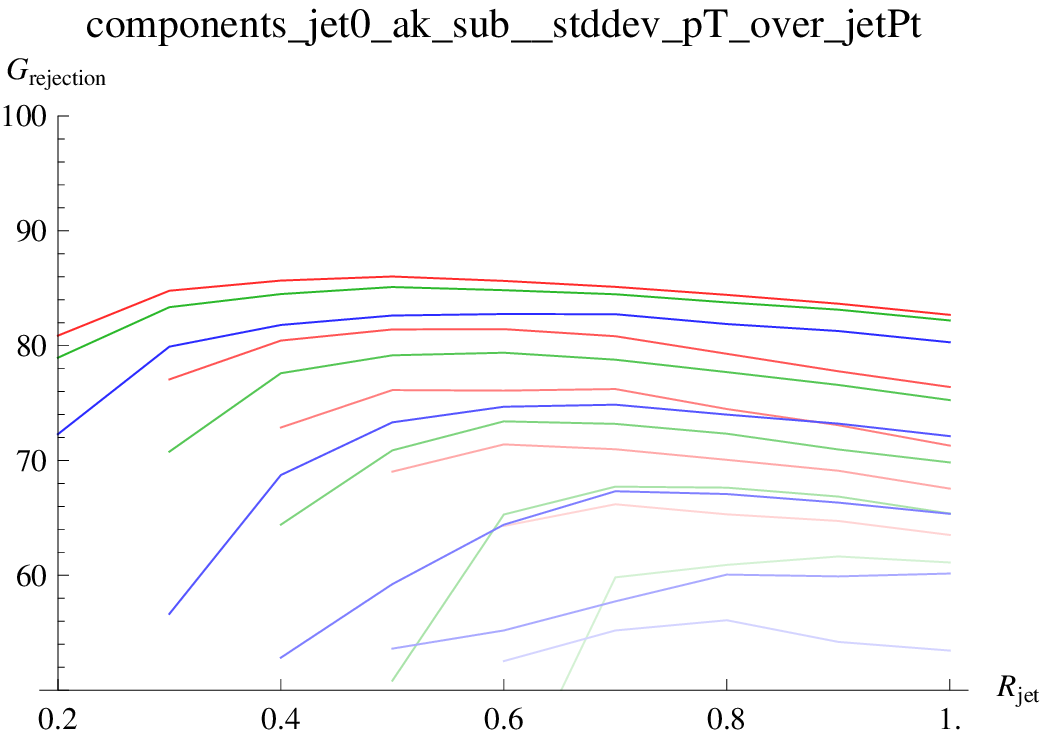}  & \qquad &

       \psfrag{subjet_}{2nd Subjet's $p_T$ Fraction}
       \psfrag{2}{ }
       \psfrag{ndPt_over}{ }
       \psfrag{_jetPt}{ }
       \psfrag{_jet0}{ }
       \psfrag{_ak}{ }
       \psfrag{_sub}{ }
       \psfrag{_}{ }
        \includegraphics[width=0.48\textwidth]{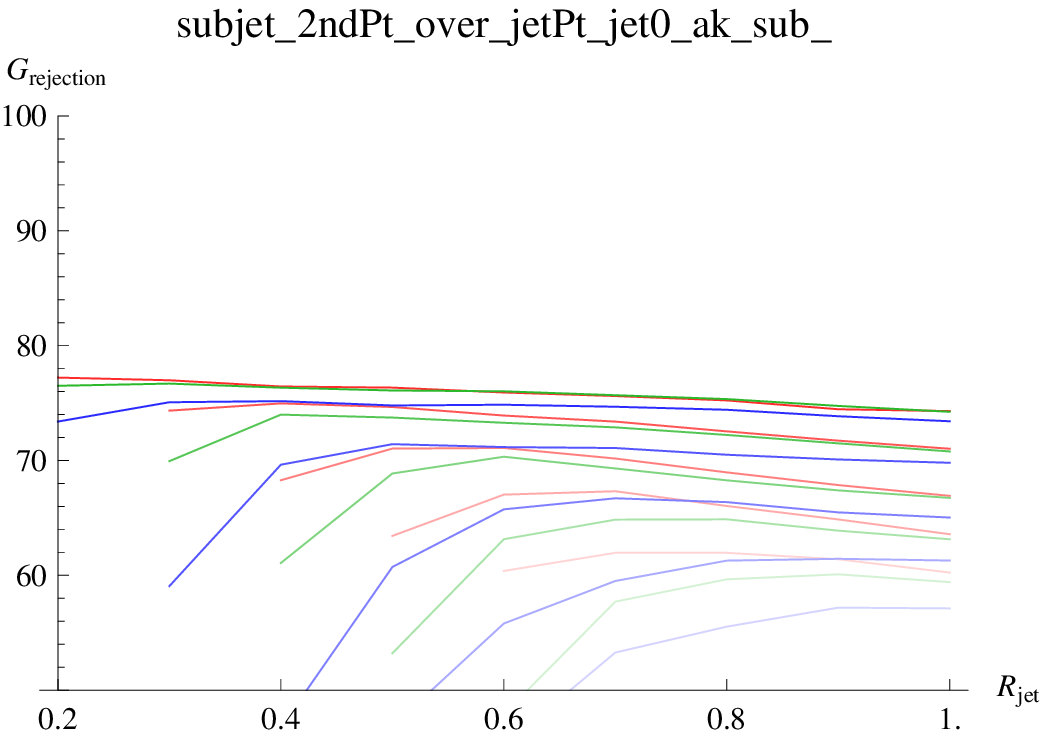}
   \end{tabular}
   \end{center}
\caption{
\label{fig:subjet_sizes}
For subjets, smaller is better.  Gluon rejection at 50\% quark acceptance
is plotted as a function of initial jet size $R_{jet}$.
These scores are averaged over all jet $p_T$ bins from 50\,GeV to 1600\,GeV for {\sc Pythia8} particle jets.
The color corresponds to the subjet algorithm, with anti-$k_T$ in red
being slightly better than CA in green, which is slightly better than $k_T$ in blue.
As for subjet size, the darkest color corresponds to the smallest and best subjet size of $R_\mathrm{sub}=0.1$.
Lightest is the largest and worst subjet size of $R_\mathrm{sub}=R_\mathrm{jet}$.
These trends hold even for subjet variables not plotted.
}
\end{figure}

After charged particle counts, the
smallest subjets do best.
Distributions for anti-$k_T$ R=0.1 subjets are shown in
Figure \ref{fig:components_jet0_ak05_sub_ak010}.
The rejection power of subjets of many types and sizes is shown in
Figure~\ref{fig:subjet_sizes}. The smallest have $R_\mathrm{sub}=0.1$, the
approximate resolution of distinguishable TopoClusters.

Finding
the average particle/track/subjet $p_T$ and normalizing to the
jet $p_T$ gives no more information than the count.  Finally, the
standard deviation of subjet $p_T$s (also normalized by the jet
$p_T$) is useful, but not more than the counts, and also not
as useful when combined with other observables. It is not shown here.

\clearpage

\section{Jet Shapes and Geometric Moments}
\label{sec:Continuous Results}

Unlike counting tracks or taking $p_T$ of the hardest small subjet,
the continuous category requires more detailed definitions. We therefore provide
explanations along with the results in this section.
We first discuss the simplest jet shape, jet mass.
%, which
% which is included here because it
%as we will explain is similar to one of the radial geometric moments.
We next discuss what is traditionally called \emph{the} jet shape,
including its integrated and differential versions.
We then describe some useful variables like angularity and girth which are basically radial moments of
jet shape.
Next we consider more complicated observables like N-subjettiness and
the moments of a two-point function.
Finally we describe 2D moments in the $(\eta,\phi)$ plane like planar flow and pull.

\subsection{Jet Mass}
\label{sec:Jet Mass}

\begin{figure}[b]
\begin{center}
\begin{tabular}{cc}
%\psfrag{components_jet0_ak05_sub_ak010_count}{\footnotesize{ Anti-$k_T$ $R_\mathrm{sub}=0.1$ Subjet Count} }
%\includegraphics[width=0.48\textwidth]{rainbows/components_jet0_ak05_sub_ak010_count_0100}  &
\psfrag{ak05_MDPt}{\footnotesize{ \!\!\!Mass/$p_T$} }
\psfrag{MDPt}{\footnotesize{$m/p_T$} }
\psfrag{arbitrary}{\footnotesize{(arbitrary)} }
\includegraphics[width=0.48\textwidth]{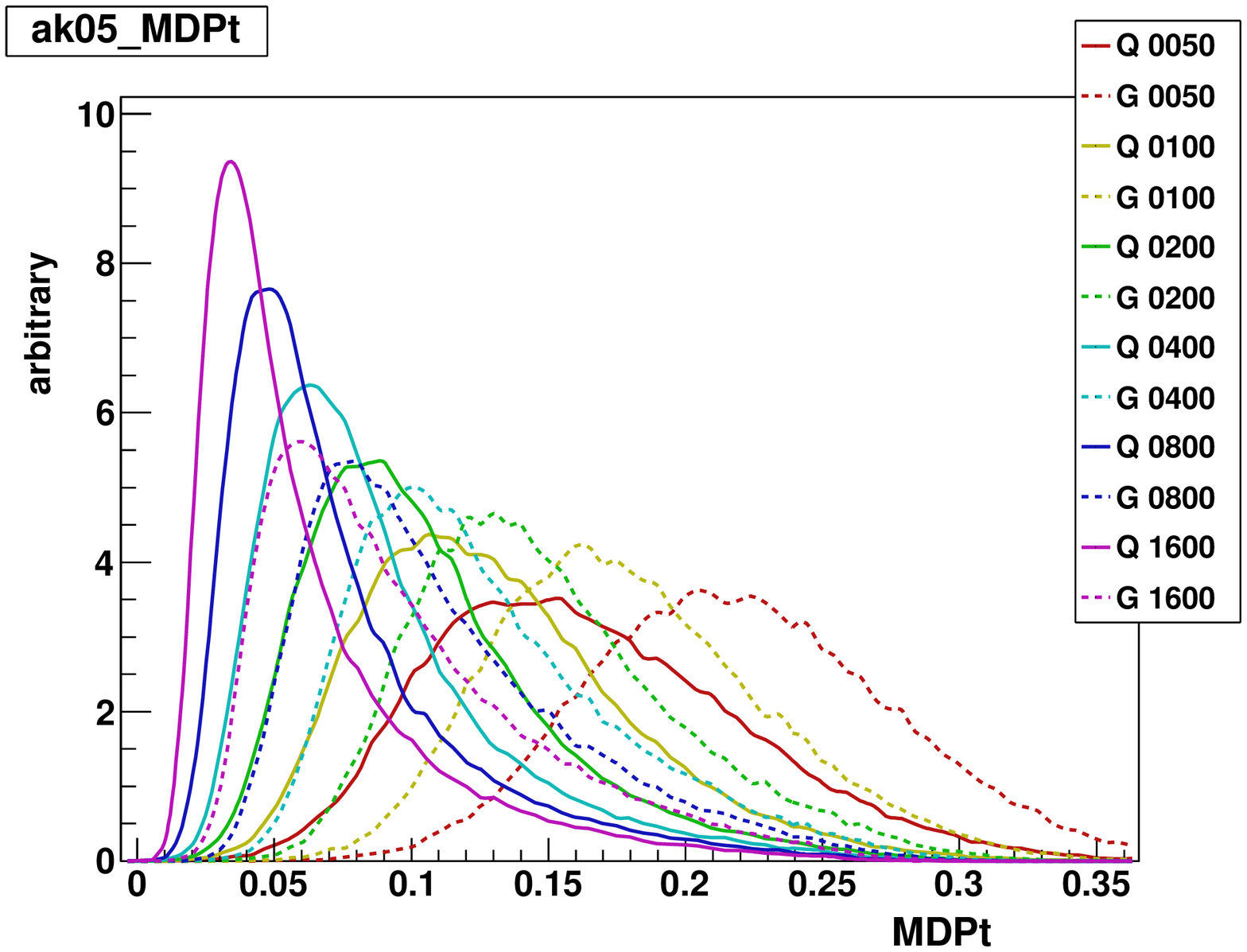}
%\psfrag{ak_mass}{\hspace{-4em}Gluon Rejection for Jet Mass}
%\includegraphics[width=0.48\textwidth]{MathematicaJetsizePt/pt_jetsize_ak_mass_50}
\includegraphics[width=0.48\textwidth]{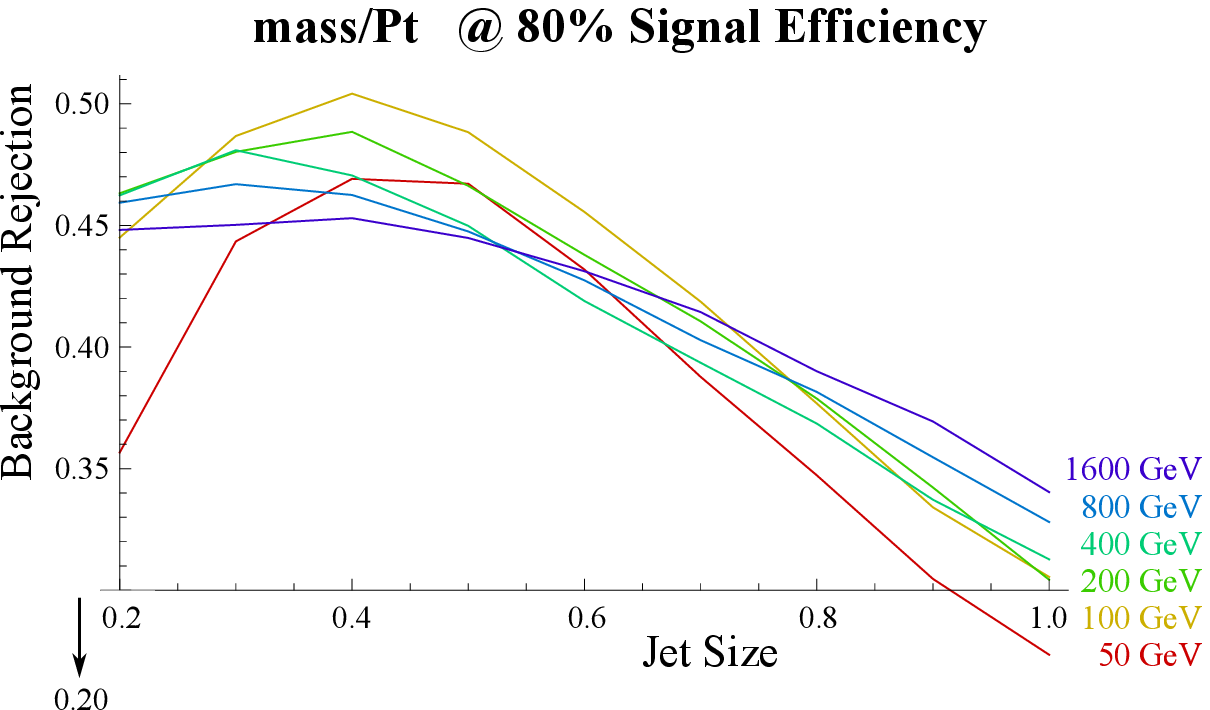}
\end{tabular}
\caption{
Jet mass over jet $p_T$ for {\sc Pythia8} anti-$k_T$ particle jets.
\textbf{Left}: Different jet $p_T$s for R=0.5 jets, normalized to equal area.
\textbf{Right}: Gluon rejection scores as a function of jet size for different $p_T$ samples.
Red is 50\,GeV, yellow is 100\,GeV, and so on, doubling through the spectrum until purple at 1600\,GeV.
The best gluon rejection occurs between R=0.3 and R=0.5.
}  \label{fig:MDPt}
\end{center}
\end{figure}

A jet's 4-vector is obtained by adding up the 4-vectors
of all of the jet constituents.
As long as the constituents
are not collinear, the resulting jet 4-vector will be massive.
This jet mass measures how spread out the constituents of the
jet are.  Distributions of jet mass normalized to jet $p_T$ for different samples, along with their gluon
rejection power is shown in Figure~\ref{fig:MDPt}.
There is already data~\cite{ATLASjetmass} and theoretical calculations~\cite{Dasgupta:2012hg,Chien:2012ur}
of jet mass at the LHC.

\subsection{Traditional Jet Shape}
\label{sec:Traditional Jet Shape}

\begin{figure}[t]
\begin{center}
\begin{tabular}{ccc}
\includegraphics[width=0.2\textwidth]{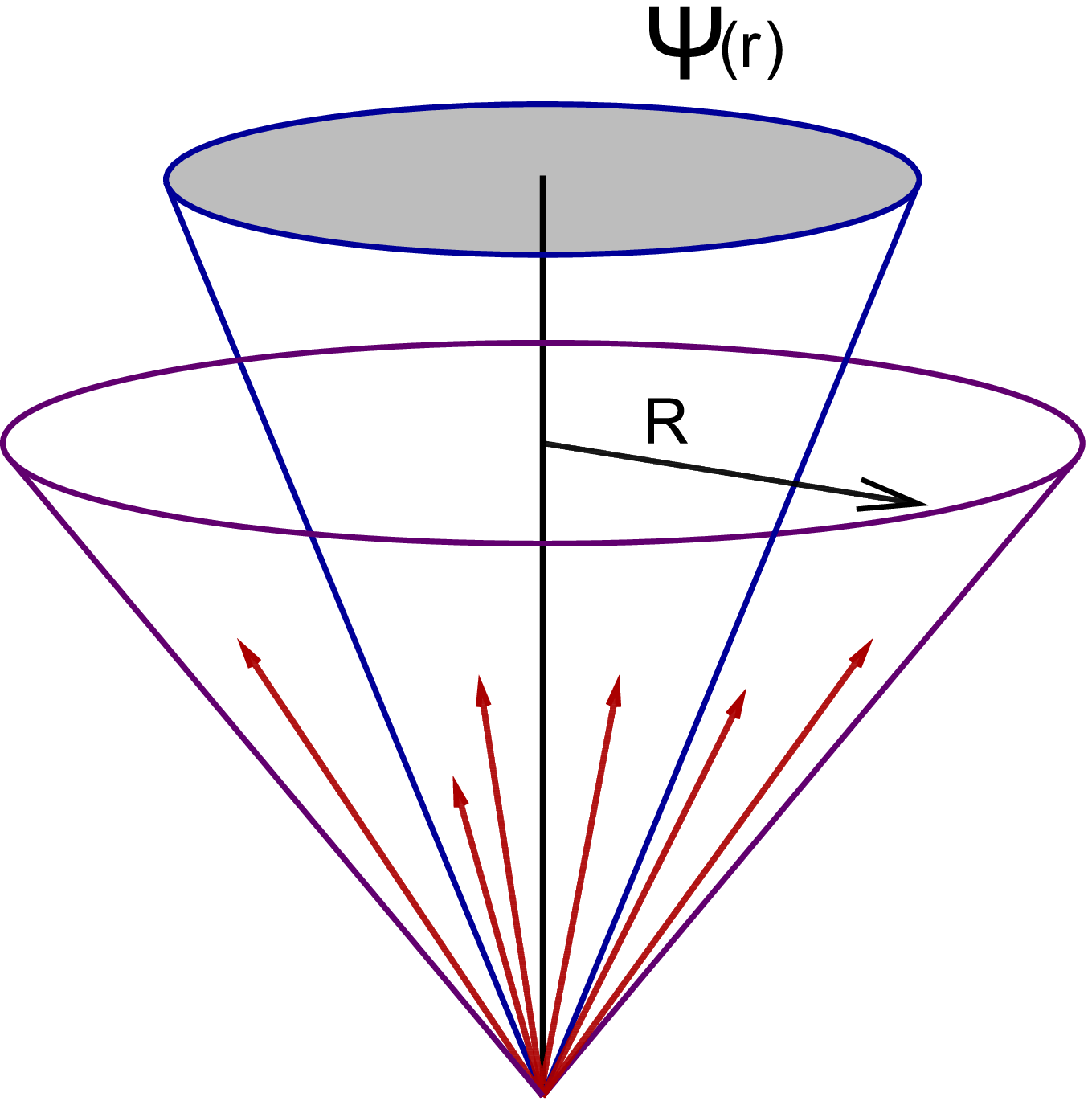} &
\includegraphics[width=0.35\textwidth]{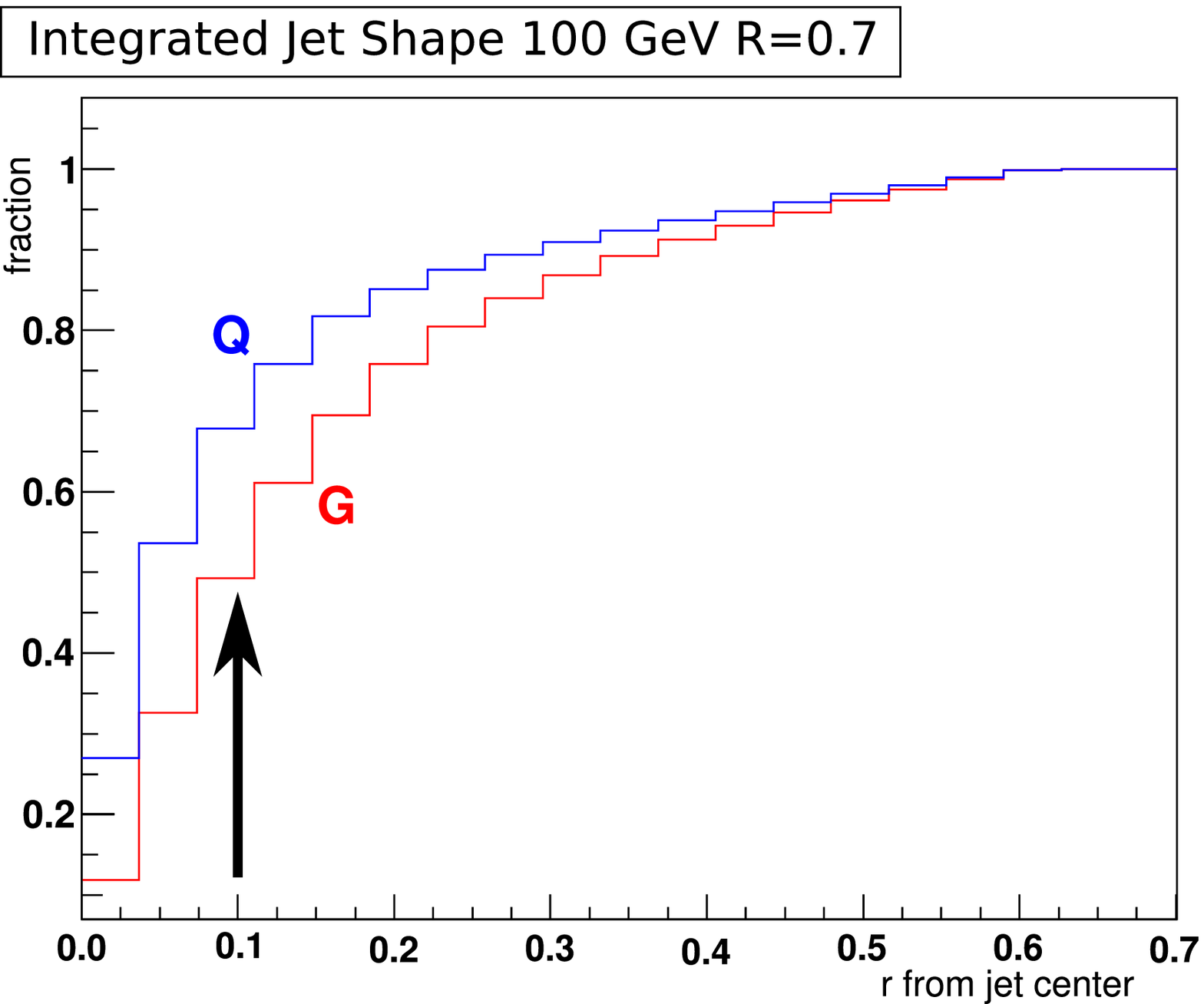} &
%\psfrag{shape_integ_pT_R_jet0_ak07_particles_bin_19_of_38_0089}{\small Integrated Jet Shape out to $r=0.1$}  %  \qquad \qquad for 100 GeV
\includegraphics[width=0.35\textwidth]{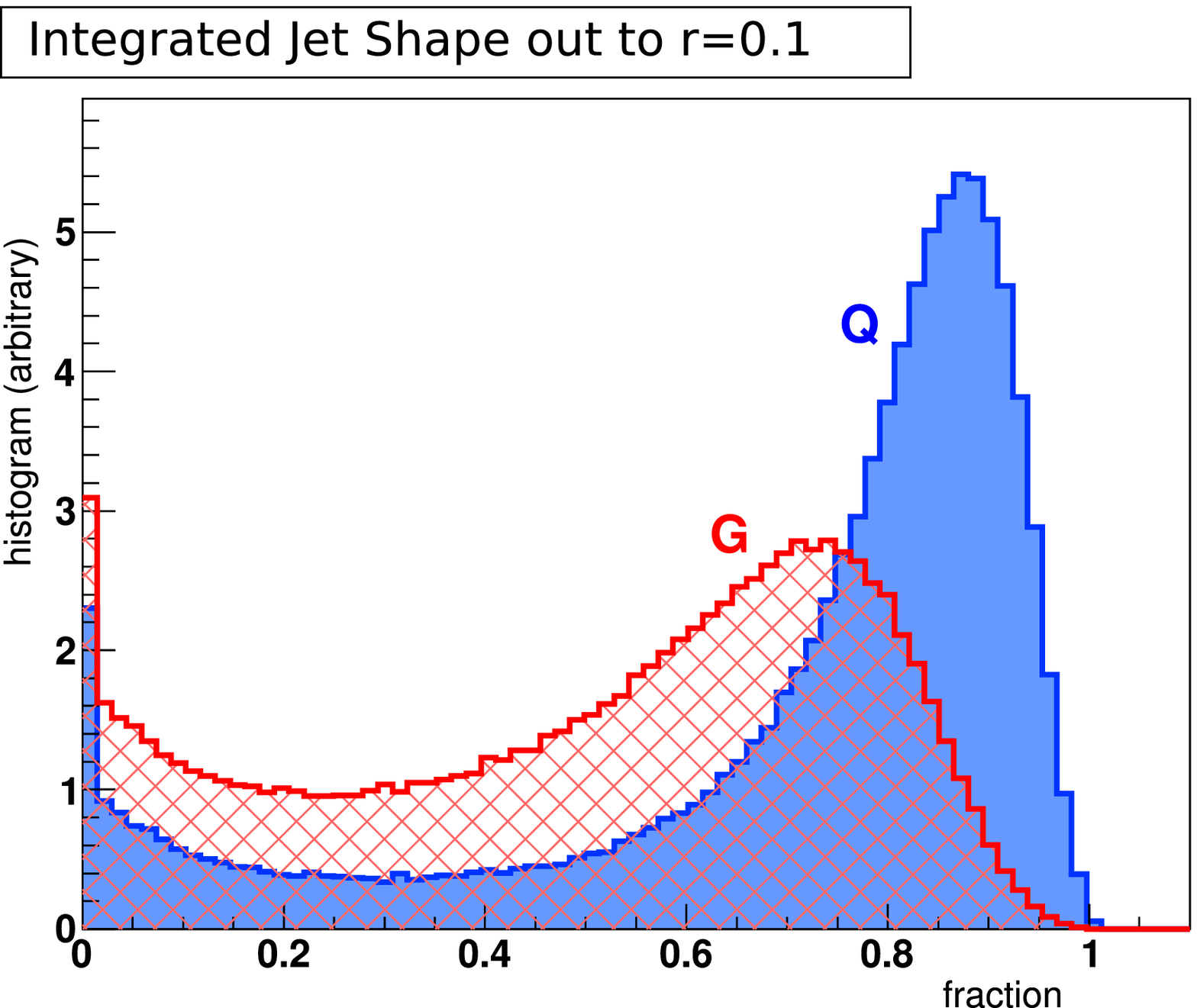}
\end{tabular}
\caption{
The Integrated Jet Shape $\Psi(r)$ is the fraction of the $p_T$ of a jet of cone size $R$
falling within a smaller cone of size $r$, as illustrated in the far left panel.
$\Psi(r=R_\mathrm{jet})=1$ by definition.
In the center is a plot of the integrated jet
shape averaged over all observed jets of a particular type (here our {\sc Pythia8} quark and gluon dijet sample).
%While useful, the integrated jet shape can be misleading.
On the right the \emph{distribution} of $r=0.1$ jet shapes is shown. The mean of these distributions
gives $\Psi(0.1)$ for quarks and gluons.
The distribution is clearly not a simple Gaussian centered around the average value, indicating that
much information is discarded in considering only the integrated jet shape.
%Even plotting the $\pm 1 \sigma$ spread
%around the average would not convey the interesting structure of
%the distribution.
The rise at low $r$ is due to jets where the
parton underwent a semi-hard splitting leading to little $p_T$
deposited along the jet axis.
} \label{fig:jetshape}
\end{center}
\end{figure}

The Jet Shape is an example of an IRC safe observable
that is commonly used in jet-property studies.
Each jet has its own integrated jet shape $\Psi(r)$, which
measures the fraction of the jet's total $p_T$ that falls
within $r$ of the jet axis. This is illustrated
in Figure~\ref{fig:jetshape} and defined more precisely as
\begin{equation}
\textrm{Integrated Jet Shape: } \Psi(r) = \int_0^r \frac{p_T(r')}{p_T^{jet}} dr' \ .
\end{equation}
An important distinction must be made between this definition,
which is different function for each jet, and what is commonly
plotted as `jet shape,' which is averaged over all jets seen by
a detector (with some cuts.)  In Figure~\ref{fig:jetshape}, this
averaged integrated jet shape is the left plot, whereas the
distribution of integrated jet shapes out to a single radius of
$r=0.1$ is the right plot.  The distribution is clearly
\emph{not} a Gaussian centered around the average value.  Given
$\Psi(0.1)$ for a particular jet that you want to classify,
it's more useful to know the full distribution for quarks and
gluons than just the two average values.  Historic measurements
and calculations are for the average rather than the full
distribution.  The same is true for jet masses: often average
masses are calculated and measured for different $p_T$s rather
than mass distributions.

%Figure \ref{fig:jetshape} illustrates the definition of the Integrated Jet Shape,
%shows the familiar plot of averages for different inner-cone sizes,
%and finally displays the \emph{distribution} for a particular size.
%The distribution illustrates
%why the familiar averages are not particularly useful for a tagger.
%Knowing the full distribution allows one to establish
%a quark acceptance / gluon rejection estimate for every possible cut.
%Moreover, the full multidimensional distribution over several different
%inner-cone sizes takes advantage of different correlations within quark jets
%as compared to gluon jets.

%\includegraphics[width=0.4\textwidth]{intshapes_plot_1_crop}  % Can't find file

Measurements at CDF agreed well \cite{Acosta:2005ix} with
{\sc Pythia} Tune A and {\sc Herwig} out to $p_T^{jet}=380$\,GeV.
At higher $p_T$, shapes got narrower, which is consistent with the mix of quark and gluon jets
evolving from 27\% quark at 50\,GeV to 80\% quark at 350\,GeV.
Early ATLAS data also agrees moderately well \cite{Feng:2010wh} with simulations.
When used event-by-event, often a particular annulus was chosen
to be integrated over, for example $0.2 < r < 0.7$ in the CMS
jet shape briefs \cite{Kurt:2008zz} and \cite{Kurt:2008zza}.
At the Tevatron, CDF chose $0.3 < r < 0.7$ \cite{Acosta:2005ix}.
% Here distributions are often shown, but
This particular choices were not optimized for distinguishing quarks from
gluons. % (though we find later it's pretty good.)

\subsection{Radial Geometric Moments \label{sec:radial_geometric_moments}}

We refer to any geometric moment that is linear in $p_T$ and independent of angle around the jet axis
as a \emph{radial moment}.   Linearity in $p_T$ is required for IR/collinear safety.
Specifically, the $p_T$ in each radial bin is weighted by a kernel $f(r)$ and summed up to
form the moment $M_f$:
\begin{equation}
\textrm{Radial moment using kernel $f(r)$} \qquad
M_f = \sum_{i \in \mathrm{jet}} \frac{p^i_T}{p^{jet}_T} f(r_i)
\end{equation}
Distances $r$ of each particle or cell from the jet center are calculated on the
(rapidity,phi) cylinder.
The jet center is taken as the $(y,\phi)$ of the jet's 4-vector,
but the $p_T$-weighted centroid is almost identical. It is
important to use rapidity rather than pseudorapidity for the
jet location because the jet is massive.
A radial moment sums a function of these distances, weighted by
$p_T$, then normalized to the total $p_T$ of the jet.
Energies and angles, rather than $p_T$s and $r$'s give similar results,
but are less appropriate to hadron colliders.

The integrated jet shape $\Psi(0.1)$ corresponds to the
moment where $f(r)$ is 1 out to $r=0.1$ and 0 beyond.  The
differential jet shape $\psi(0.3)$ corresponds to a kernel that
is 1 in a small window around $r=0.3$. One series of kernels are
powers of $r$: $r$, $r^2$, $r^3$, $\cdots$.  These most closely
correspond to the traditional geometric notion of `moments.'
%Other kernels are possible and may be more useful in different
%contexts.
Radial moments like these are interesting
because it may be possible to calculate them accurately in QCD,
see for example~\cite{Ellis:2010rw}.

An orthonormal set of kernel functions fully characterizes the
radial distribution of $p_T$ for a \emph{single jet}, but even
knowing the 1D distributions for an infinite set of orthogonal
functions would not give complete information about the
underlying high-dimensional distribution with all correlations preserved.
In other words, knowing this series for a particular jet would allow a full
reconstruction of where the $p_T$ in \emph{that} jet goes, but the
same isn't true for the 1D distributions.

%By recording two moments, you loose the information that a particular jet's moment $M_A$
%had value $a$ and \emph{simultaneously} moment $M_B$ had value
%$b$ rather than any other the other possible values for $b$.

\subsection{Linear Radial Geometric Moment: Girth, Width, or Jet Broadening \label{sec:girth}}

\begin{figure}
\begin{center}
\begin{tabular}{cc}
%\psfrag{components_jet0_ak05_sub_ak010_count}{\footnotesize{ Anti-$k_T$ $R_\mathrm{sub}=0.1$ Subjet Count} }
%\includegraphics[width=0.48\textwidth]{rainbows/components_jet0_ak05_sub_ak010_count_0100}  &
\psfrag{radial_moment_ak05_1_Pt_4v_y}{\footnotesize{ \!\!\!Girth for R=0.5} }
\psfrag{girth}{\footnotesize{girth} }
\psfrag{arbitrary}{\footnotesize{(arbitrary)} }
\includegraphics[width=0.48\textwidth]{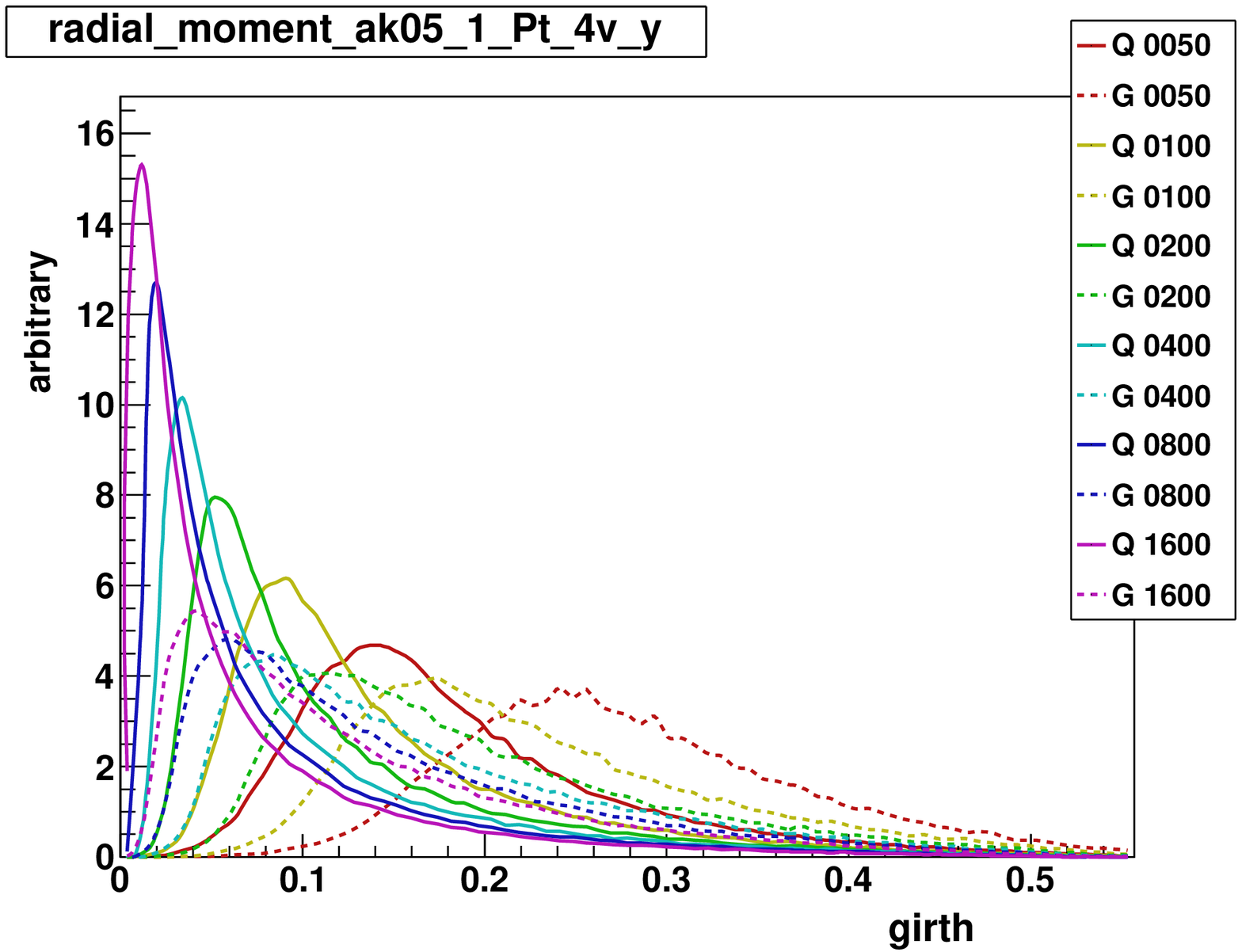}
\psfrag{radial_moment}{Gluon Rejection for Girth}
\psfrag{_ak}{ }
\psfrag{_}{ }
\psfrag{1}{ }
\psfrag{_Pt}{ }
\psfrag{_}{ }
\psfrag{4}{ }
\psfrag{v_y}{ }
\includegraphics[width=0.48\textwidth]{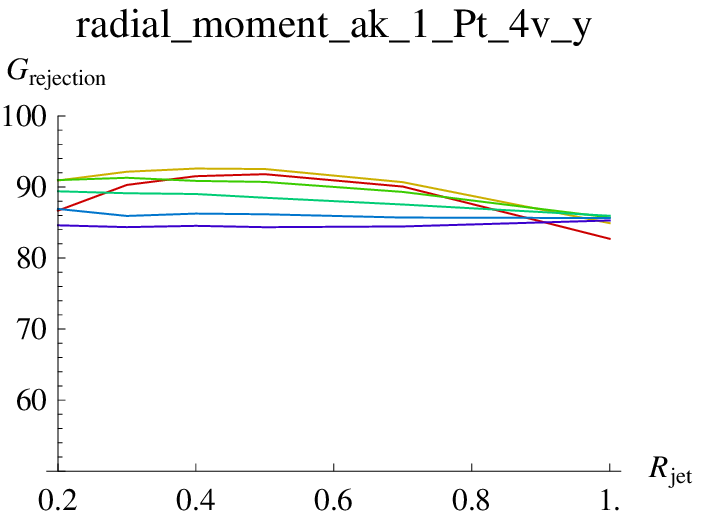}
\end{tabular}
\caption{
Girth (also called width, or the linear radial moment) for {\sc Pythia8} anti-$k_T$ particle jets.
\textbf{Left}: Different jet $p_T$s for R=0.5 jets, normalized to equal area.
\textbf{Right}: Gluon rejection scores as a function of jet size for different $p_T$ samples.
Red is 50\,GeV, yellow is 100\,GeV, and so on, doubling through the spectrum until purple at 1600\,GeV.
For the lower $p_T$ samples, the best gluon rejection occurs around R=0.5.
}  \label{fig:girth}
\end{center}
\end{figure}

The linear radial moment, or \emph{girth},
is a special case of
a generic radial moment with $f(r)=r$.
For discrete constituents, it is
defined as %~\cite{Gallicchio:2010sw}
\begin{equation}
\textrm{Girth}:\qquad  g =  \sum_{i \in \mathrm{jet}} \frac{ p_T^i }{ p_T^{jet} } r_i \ .
\label{eqn:girth}
\end{equation}
The girth distribution is shown in
Figure~\ref{fig:girth}.

ATLAS calls this variable \emph{width}.
This is a hadron-collider version of a popular LEP variable
called \emph{jet broadening}.
Jet Broadening, as measured at ALEPH \cite{Barate:1998cp} and
OPAL \cite{Abreu:1998ve}, leads to distributions very similar
to the linear moment, simply because the small-angle
approximation of $k_T \approx p_T \, r$ is valid.
At LEP, jet broadening was given by
\begin{equation}
B_\mathrm{jet} = \frac{ \sum_i | \vec p_i \cross \hat n_\mathrm{jet} | } { \sum_i | \vec p_i | }
   = \frac{ \sum_i | \vec k_Ti  | } { \sum_i | \vec p_i | } \ .
\end{equation}

We examined higher-power radial moments, and found them less useful
for quark/gluon discrimination.  CMS \cite{Gavrilov:2010zz} has examined the second radial
moment. For small narrow, nearly transverse jets, the second moment
is equivalent to the jet mass.
Higher-powered moments have the disadvantage of being most sensitive to
the edges of the jet, where the most contamination lies.
%The CMS paper does address the magnetic field and some differences
%between PYTHIA and HERWIG++.

On the other hand, we have found that a very good discriminator uses a lower power,
specifically the square root of the distance:
\begin{equation}
M_{1/2} = \sum_{i \in \mathrm{jet}} \frac{ p_T^i }{ p_T^{jet} } \sqrt{r_i} \, .
\label{eqn:sqrt_profile}
\end{equation}

\subsection{Jet Angularities \label{sec:jet_angularities}}

\begin{figure}[t]
\begin{center}
\includegraphics[width=0.5\textwidth]{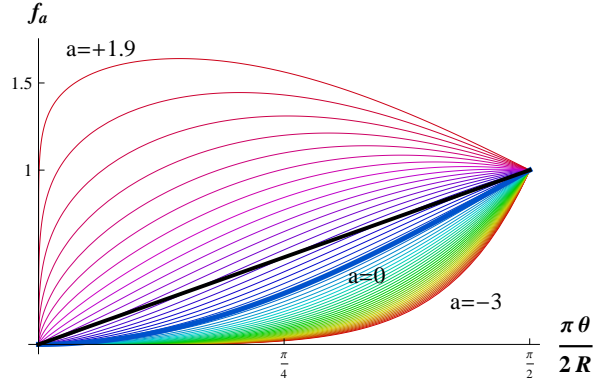}
\caption{
Profiles $f_a(\tilde{\theta})$ for different choices of the angularity $a$ parameter spaced at 0.1 intervals
(in rainbow) and linear radial-moment ``girth'' (in black).
%I'd be embarrassed to point this out if it wasn't a confusion
%at a workshop talk I gave, but
These \emph{profile} shapes
have nothing to do with the shapes of the \emph{distributions} resulting
from integrating these moments over jets and forming a histogram of the results.
} \label{fig:angularities}
\end{center}
\end{figure}

\begin{figure}
\begin{center}
\begin{tabular}{ccc}
\includegraphics[width=0.3\textwidth]{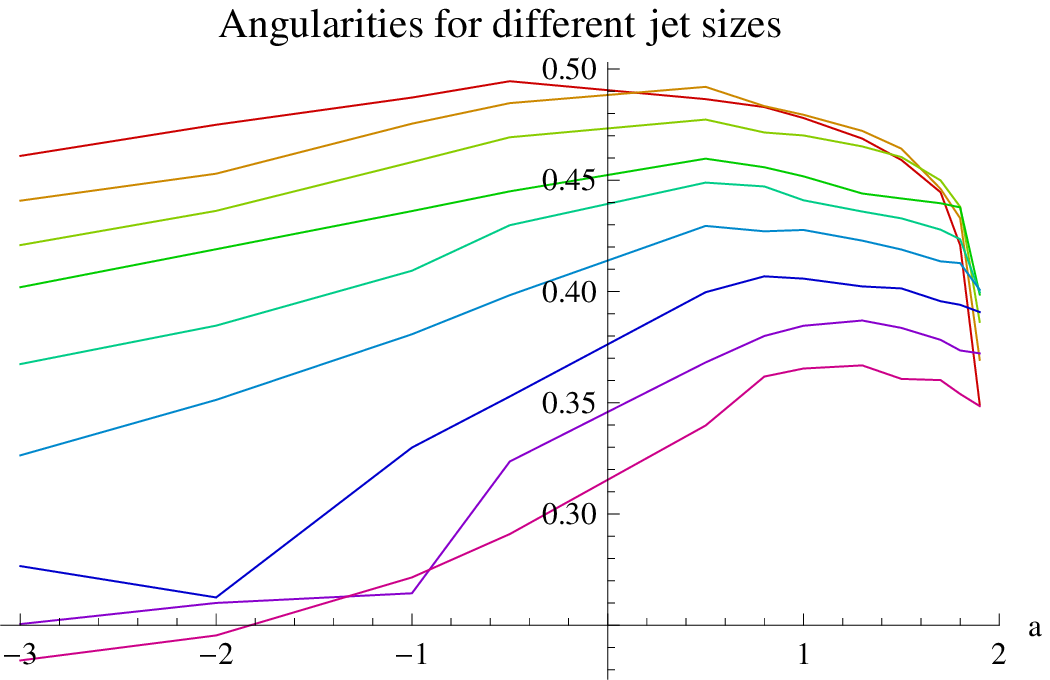}  &
\includegraphics[width=0.3\textwidth]{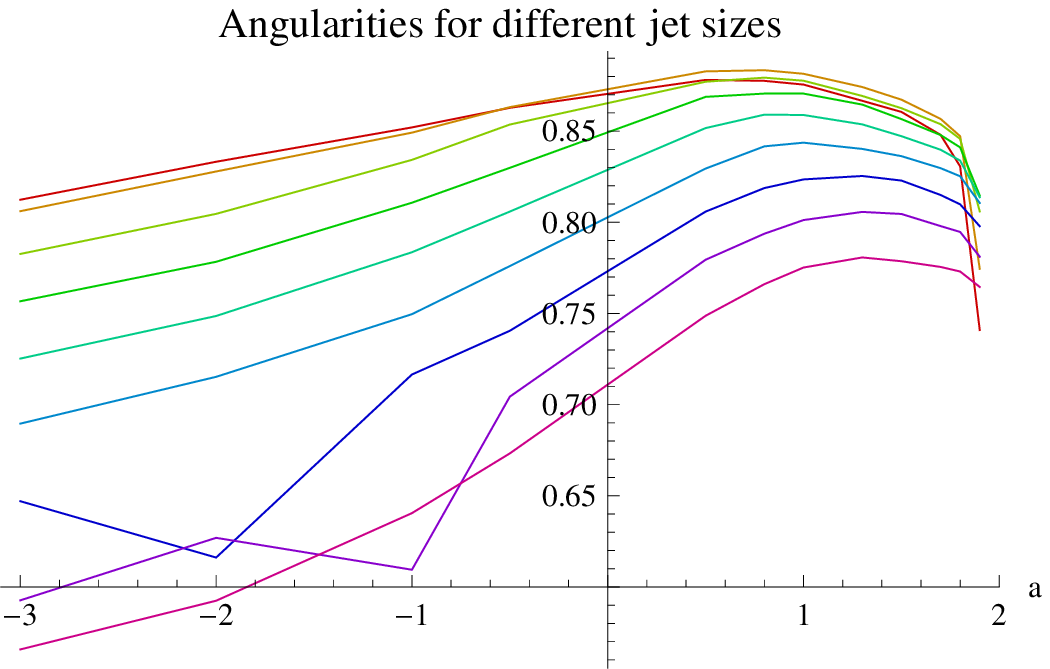}  &
\includegraphics[width=0.3\textwidth]{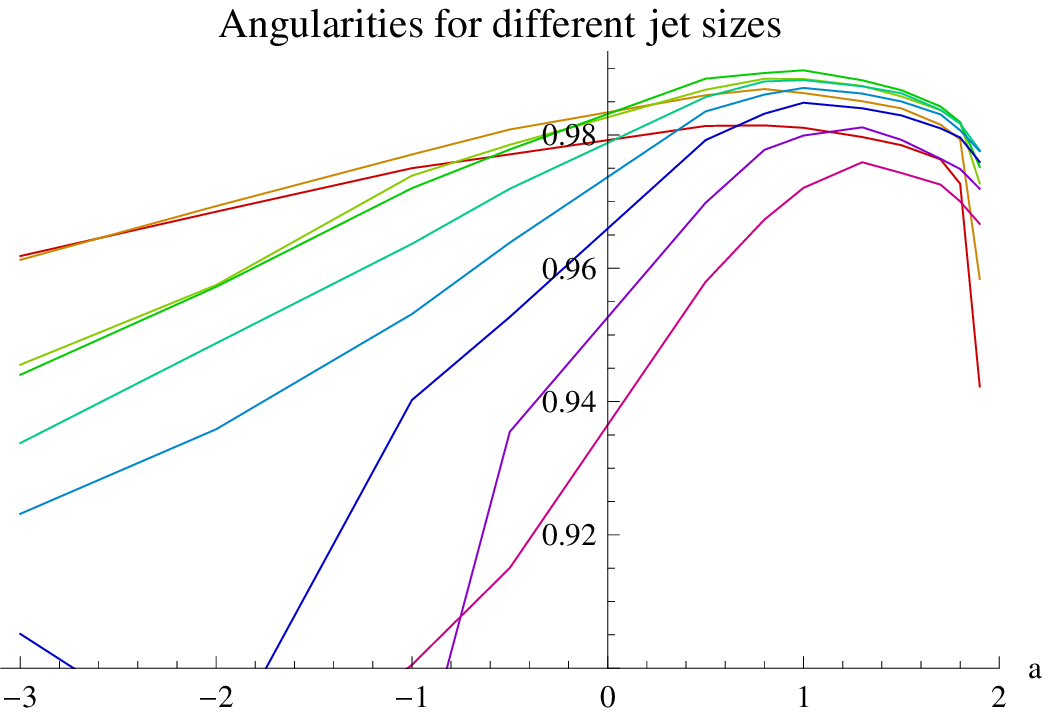}  \\
80\% quark &
50\% quark &
20\% quark
\end{tabular}
\caption{
Gluon rejection power for angularities as a function of angularity parameter $a$.
Each line represents growing {\sc Pythia8} particle jet size from $R$=0.2 in red up to $R$=1.0 in purple.
Here the scores for all $p_T$s are averaged.
The best angularities perform
slightly better than masses, but worse than track and subjet counts.
}
\label{fig:angularities_results}
\end{center}
\end{figure}

Jet Angularities are also radial moments, but their ``radial
distances'' are rescaled into the angular coordinates
appropriate for $e^+e^-$ event shapes.
%are the angles that the particles or cells make with the jet axis: $\theta_i$.
They are defined by~\cite{Almeida:2008yp} as
\begin{equation}
\textrm{Jet Angularities}:~~~~ A_a =
%\frac{1}{ m_{jet} }
\sum_{i \in \mathrm{jet}}  E_i \ f_a( \tilde{\theta} ) \,,
\end{equation}
with
\begin{equation}
f_a(\tilde{\theta})
   =  \sin^a \! \tilde{\theta} \,
        \left( 1 - \cos \tilde{\theta} \right)^{1-a}
\qquad \mathrm{and} \qquad
\tilde{\theta} = \frac{\pi |r_i|}{2R}
\end{equation}
for $a<2$. The kernel function $f_a(\theta)$ is inspired by
full event-shape angularities~\cite{Berger:2003iw}, but
squished so that the edge of a jet at $|r_i|\!=\!R$ is mapped
to $\pi/2$. Profiles for different choices of the $a$ parameter
are shown in Figure~\ref{fig:angularities}. Note that the
energies $E_i$ are used in the definition, instead of the
$p_T$s popular with hadron colliders.
Also, angularities are often normalized by
the jet mass, but this is not the most useful for our purposes.
A given angularity has two parameters ($R_\mathrm{jet}$ and
$a$) in addition to any discrete choices like normalization
(none, jet mass, jet $p_T$, jet $E$) or angle used
($\tilde{\theta}$ as defined, or geometric $\theta$.)
Gluon rejections for different choices of $a$ are shown
in Figure~\ref{fig:angularities_results}.

\subsection{Optimal Kernel for Radial Moment \label{sec:optimal_kernel}}

Rather than sticking to powers of $r$,
sines and cosines (like angularity), or another
orthonormal basis, we looked for
the kernel $f(r)$ that gives the best discrimination power
between quarks and gluons for each $p_T$.
Because the goal is to find the best \emph{function},
the optimization problem is technically infinite dimensional.
But through reasonable smoothness criteria, it can be reduced
to adjusting a few control-points of a spline or coefficients
of an orthonormal basis.
%There is also plenty of redundancy.
Since adding a constant doesn't change the discrimination power,
we chose our kernels to have $f(0)=0$. This means that the energy
deposits near the crowded and noisy jet center count least.
Multiplying by a constant (even a negative one)
also does not affect discrimination power, so we normalized our
trial profiles so their maximum value was +1.
%The trickiest part in testing the kernels is to pick a measure of discrimination power
We evaluated the ROC curve at three different quark
efficiencies, 20\%, 50\%, and 80\%.

\begin{figure}[t]
\begin{center}
\includegraphics[width=0.45\textwidth]{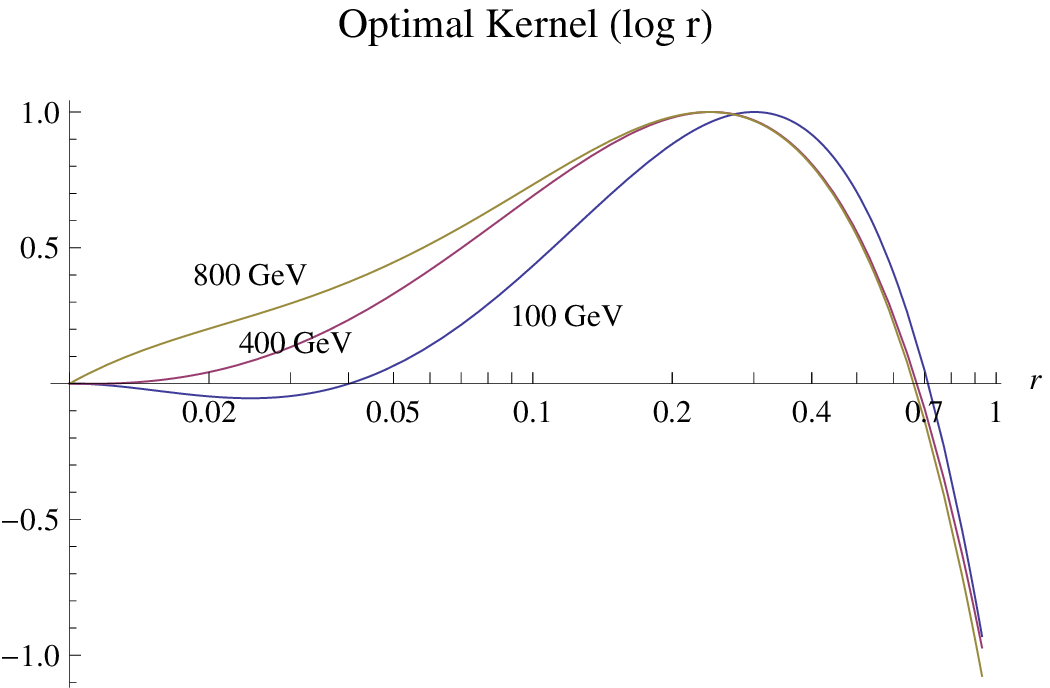}
\includegraphics[width=0.45\textwidth]{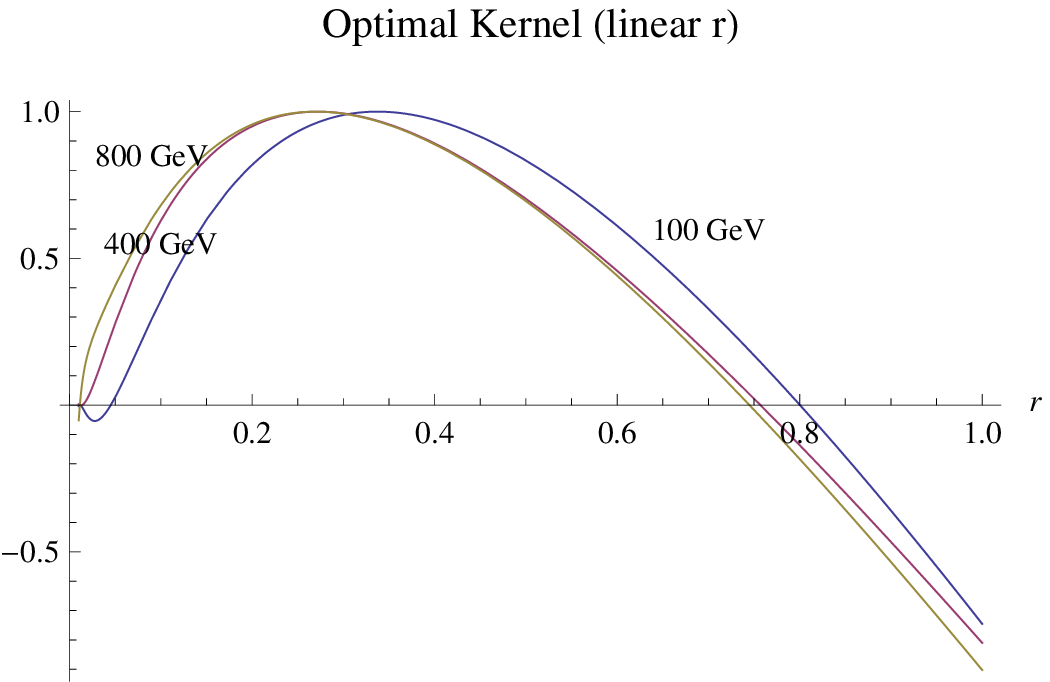}
\caption{Profiles for the optimal kernels found for various jet sizes.
Kernels for the higher-$p_T$ jets give a higher weight to $p_T$ near the jet axis.}
\label{fig:optimal_kernels}
\end{center}
\end{figure}

The best kernels we found had rejection scores that were not significantly
higher than those for girth (equation \ref{eqn:girth})
or the square-root profile (equation \ref{eqn:sqrt_profile}).
For this reason, we won't go into much more detail.
Some general trends did appear.  By construction, all
kernels started out at zero at the center of the jet and rose
to +1 at some distance away.  In the best kernels, this happened
around $r=0.4$ for low-$pT$ jets, 0.3 for
100\,GeV jets and 0.24 for 400 and 800\,GeV jets.
Beyond this, it mattered less what happened,
but the best kernels did fall toward the edge of the jet.
Examples of such kernels are shown in Figure~\ref{fig:optimal_kernels}.

%When the jet
%shape team on CDF takes their $0.3<r<0.7$ annuli, they are
%capturing the region of high discrimination, but with
%artificially sharp cutoffs.  Their cone-size was 0.7, and it
%doesn't matter if you ask which fraction of the $p_T$ was
%inside or outside of 0.3.

%It might be worth pursuing this program in data using
%purified samples of quark and gluon jets.
%Trial kernel profiles need to be constructed, then each jet in the
%sample needs to be integrated against it, which is slow
%and involves keeping detailed information about the location
%and energy within each jet.
%Once a histograms is built up for each trial kernel
%using the quark and gluon samples, the ROC curve needs to
%be computed, and a discrimination power score needs to be
%assigned.

\subsection{N-subjettiness \label{sec:N-subjettiness}}

N-subjettiness \cite{Thaler:2010tr} is a family of jet shapes
that attempt to characterize the degree to which a jet has exactly $N$ subjets.
$N$ is one of the input parameters, and is commonly taken to be 1, 2 or 3.
$N$-subjettiness finds exactly $N$ axes within the jet and associates each particle or $p_T$ deposit to
the nearest axis. These are the $N$ subjets.
The $N$-subjettiness score $\tau_N$ is sum of $p_T$-weighted radial moments for each subjet.
In this moment, each bit of $p_T$ is multiplied by its distance
to the subjet axis $\Delta R$ raised to a power $\beta$, which must be positive.
Specifically, this is
\begin{equation}
\tau_{N \! , \, \beta} = \frac{1}{d_0} \sum_{J=0}^N \ \ \sum_{k \in \mathrm{subjet}_J} p_{T,k} \, (\Delta R_{J,k})^\beta \ ,
\end{equation}
where $d_0$ is a normalization involving the jet size $R_0$
to keep $\tau_{N \! , \, \beta}$ between zero and one:
\begin{equation}
d_0 = R_0^\beta \sum_{k \in \mathrm{jet}} p_{T,k} \ .
\end{equation}
There are three parameters: $N$, the exponent $\beta$, and the method
of choosing axes.  A simple way of choosing $N$ axes is to undo
a $k_T$ or Cambridge-Aachen clustering exactly $N$ steps.  A more effective method
is to choose the axes that minimize the score $\tau$.
It's clear from the definition that the \emph{more} a jet looks like it contains
exactly $N$ well-separated and individually well-collimated subjets,
the \emph{lower} the $\tau$ score is.

\begin{figure}[h]
\begin{center}
\includegraphics[width=0.66\textwidth]{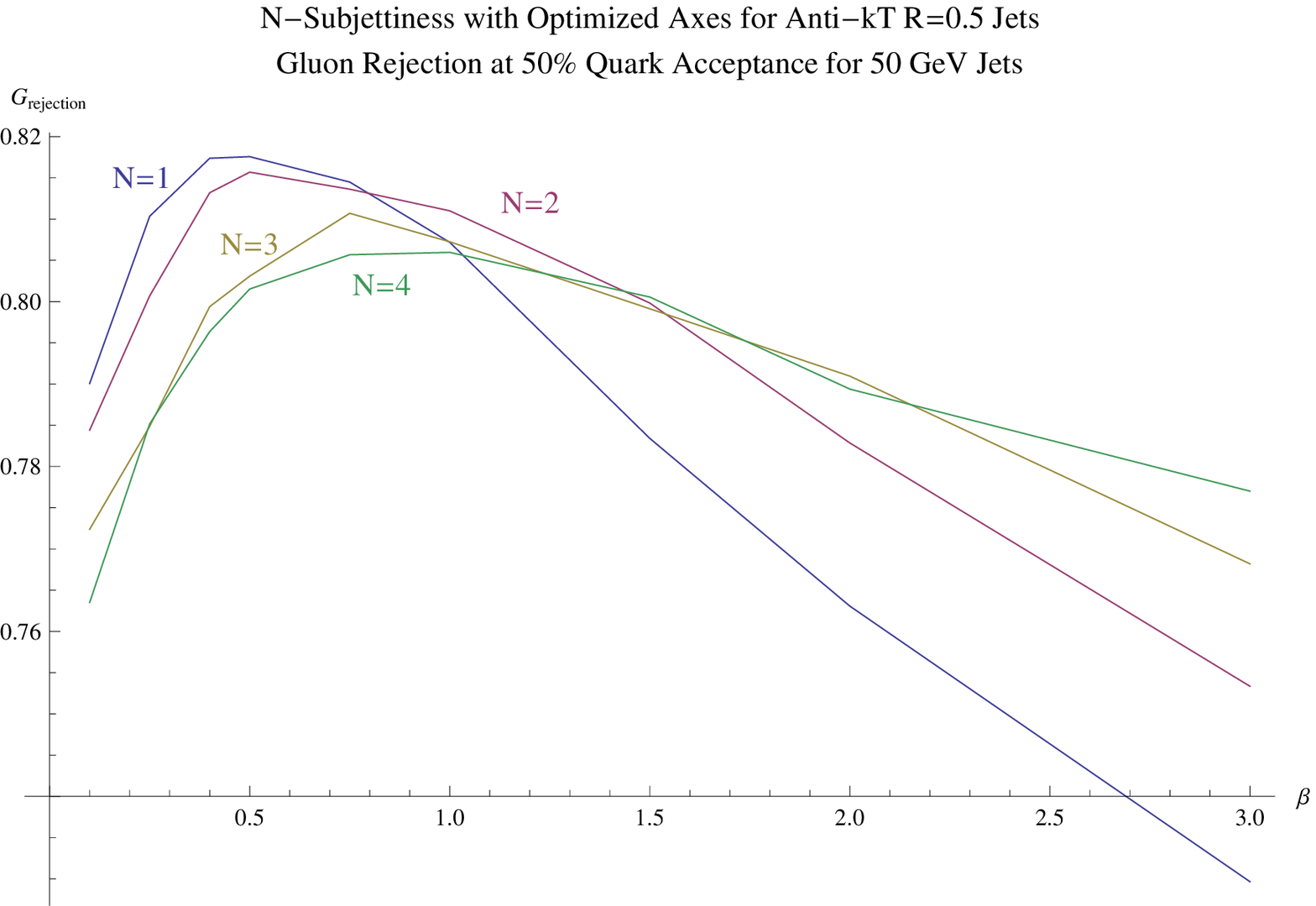}  % same information
\includegraphics[width=0.66\textwidth]{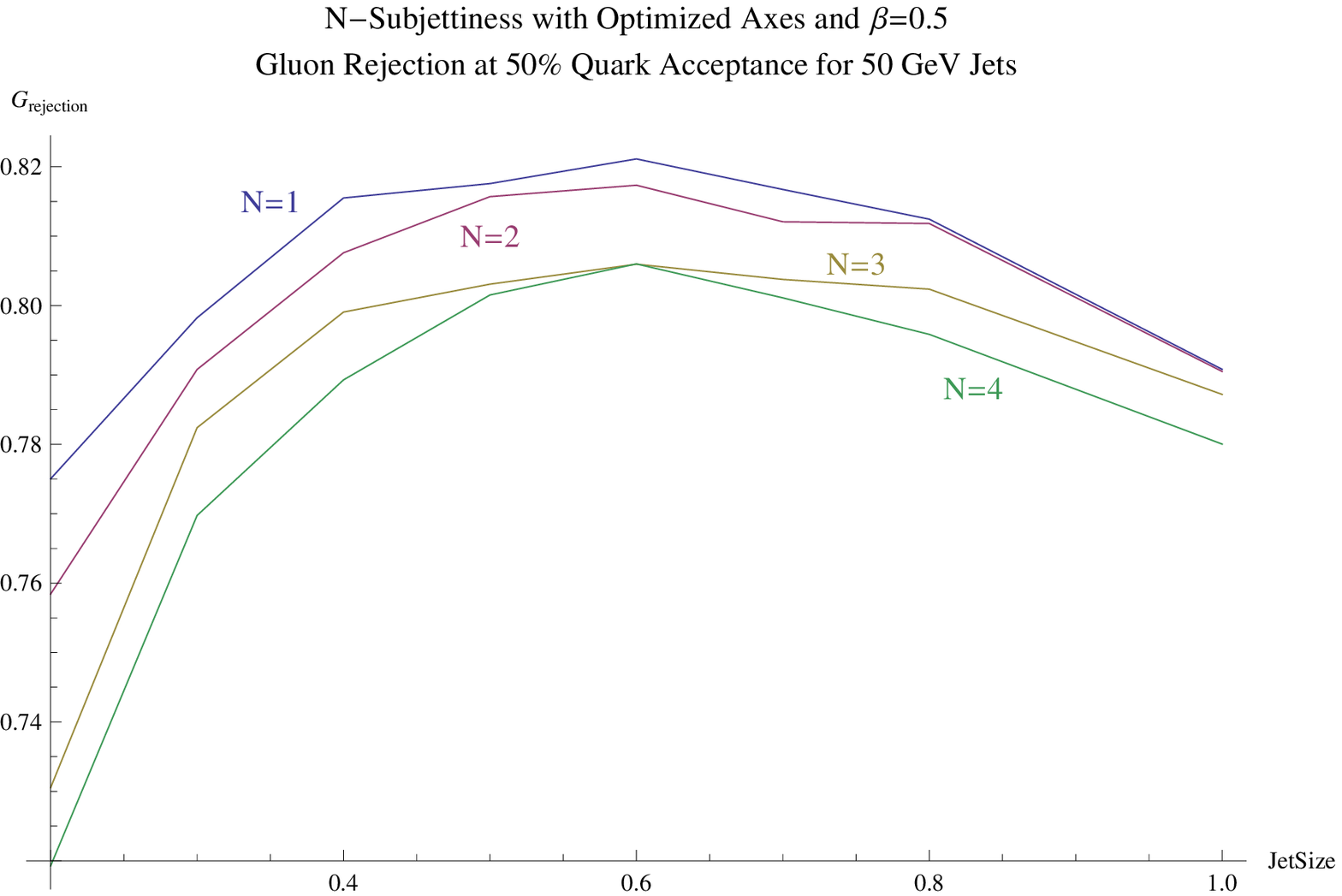}
\caption{
N-subjettiness gluon rejection as the parameters $N$, $\beta$, and jet-size are varied.
This is for 50\,GeV particle jets simulated in {\sc Herwig++}.
The best $\beta$s are between 1/2 and 1.  As usual, the best jet sizes are
between 0.5 and 0.7.  These trends hold for {\sc Pythia8} and for higher $p_T$ jets.
} \label{fig:nsubjet}
\end{center}
\end{figure}

In the gluon-rejection plots in Figure~\ref{fig:nsubjet}, only charged tracks were included at ATLAS's request.
If a jet contains only $N$ charged tracks (with $p_T >  1$\,GeV here), the
axes will coincide with these tracks and the score will be zero.
It will also be zero if there are fewer than $N$ tracks.
This explains the excess in the zero-bin of some $N$-subjettiness distributions (not shown).
$N$-subjettiness is also appealing
because it can be calculated accurately in QCD, at least in some contexts~\cite{Feige:2012vc}.

% Adjust width to fit on same page

\clearpage

\subsection{Two-Point Moment \label{sec:Two-Point_Moment}}

\begin{figure}[h]
\begin{center}
\vspace{1in}
\includegraphics[width=0.80\textwidth]{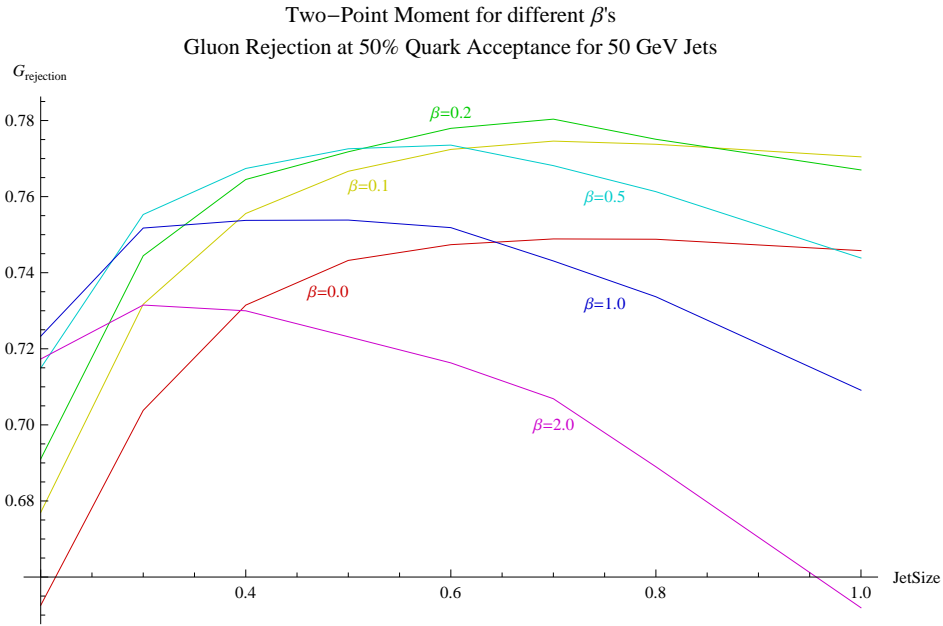}
\caption{
Two-Point Moment gluon rejection as the jet size and $\beta$ parameter are varied.
This is for 50\,GeV charged-track jets simulated in {\sc Herwig++}.
The best $\beta$'s are small, around 1/4.  As usual, the best jet sizes are
between 0.5 and 0.7.  These trends hold for all-particle jets, {\sc Pythia8}, and for higher $p_T$.
} \label{fig:twopointmomentBetas}
\end{center}
\end{figure}

A two-point moment is a sum over every pair of
constituents (energy deposits or tracks).  It is a sum of
the product of $p_T$s of each pair, times their
separation $\Delta R$ raised to a power $\beta$.
It is normalized by the jet $p_T^2$ to make
it dimensionless and less sensitive to the jet $p_T$ itself.
\begin{equation}
T_\beta = \frac{1}{(p_T^\mathrm{jet})^2}
    \sum_{i \in \mathrm{jet}} \sum_{j \in \mathrm{jet}} p_T^i \,  p_T^j \, \Delta R ^ \beta
\end{equation}
It is a moment of the two-point function,
which would be a function of $\Delta R$.
As long as $\beta > 0$, this is IRC safe.
This is meant to capture an average separation
between constituents.
In Figure~\ref{fig:twopointmomentBetas},
the gluon rejection is shown as a function
of the jet size for different values of $\beta$.

\clearpage

\subsection{Two-Dimensional Geometric Moments \label{sec:other_2d_moments}}

\begin{figure}[h]
   \[
      \textrm{Covariance Tensor: \ }
      \mathbf{C} =  \sum_{i \in \mathrm{jet}} \frac{ p_T^{i} }{ p_T^{jet} }
      \left(
      \begin{array}{cc}
         \Delta \eta_i \Delta  \eta_i & \Delta \eta_i \Delta \phi_i \\
         \Delta \phi_i \Delta  \eta_i & \Delta \phi_i \Delta \phi_i
      \end{array}
      \right)
   \]
   \begin{center}
    \begin{tabular}{m{3in}m{.1in}m{3in}}
       \includegraphics[width=0.5\textwidth]{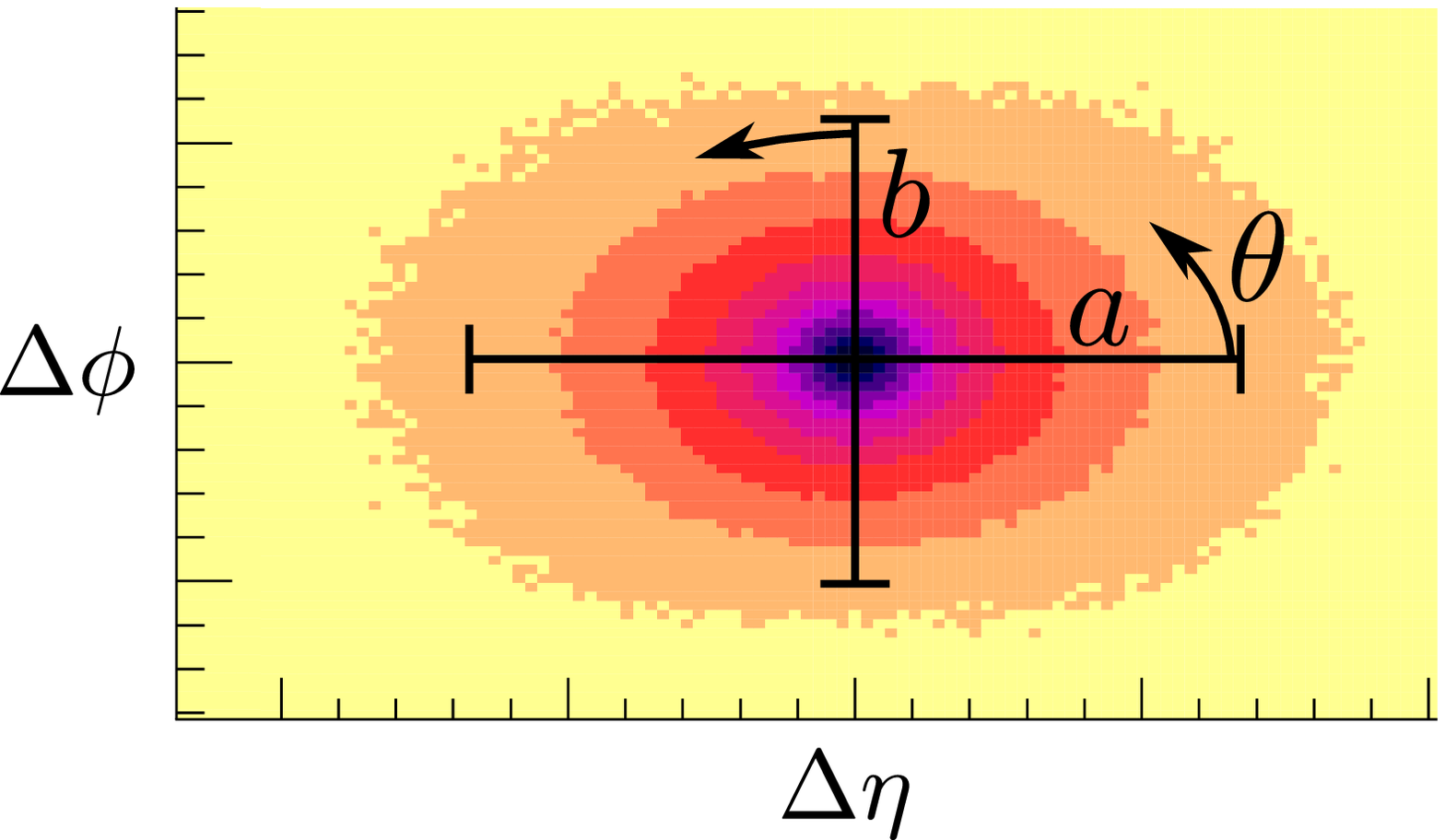}  &
       \quad &
       \begin{tabular}{rl}
         \multicolumn{2}{l}{\bf{Combination of Eigenvalues}}  \\
         Eigenvalues:      & $a > b$ \\
         Quadratic Moment: & $g = \sqrt{a^2 + b^2}$ \\
         Determinant:      & $det = a \cdot b$ \\
         Ratio:            & $\rho = b/a$ \\
         Eccentricity:     & $\epsilon = \sqrt{a^2 - b^2}$  \\
         Planar Flow:      & $p\!f = \frac{4 a b}{(a+b)^2}$   \\
         Orientation:      & $\theta$ \\
       \end{tabular}
    \end{tabular}
   \end{center}
\caption{
The Covariance Tensor and its eigenvalues and eigenvectors.
%(Moment of Inertia)
} \label{fig:covariance_tensor}
\end{figure}

The radial moments above ignored how the $p_T$ was distributed
\emph{around} the jet axis.  Motivated by the moment-of-inertia
and covariance tensors, a second order 2D geometric moment
tensor can be formed as shown in Figure~\ref{fig:covariance_tensor}. 
Combinations of its
eigenvalues and eigenvectors (like Planar Flow) have been used used
to distinguish boosted objects.

None of these variables turn out not be particularly useful for quark/gluon discrimination,
so no distributions are shown here.
Whether a quark emits a gluon or a gluon splits, the the 2-body
kinematics are similar.  Since it's this leading emission that dominates
the subsequent shower, it is understandable that these shapes might not
differ significantly between quarks and gluons.

%In general, an $R=1.0$ fat jet has $\pi/(0.1 \cdot 0.1) \gtrsim
%300$ course calorimeter cells and leads to a
%300-\emph{dimensional} distribution, a problem familiar from
%image analysis and classification.  While what's his face's
%Zernike transform is an ortho-normal set on a disk, so what.
%The Zernikie transform, or the amusingly named `Seven Moments
%of Hu' might be good at telling a photo of bricks from clouds,
%but they're not used for facial recognition, nor are the
%automatically useful here.  In fact, the Zerniki basis
%especially bad because it peaks at the edges.

%%%%%%%%%%%%%%%%%%%%%%%%%%%%%%%%%%%%%%%%%%%%%%%%%%%%  PULL %%%%%%%%%%%%%%%%%%%%%%%%%%%%%%%%%%%%%%%%%%%%%%%

\clearpage

\subsection{Pull \label{sec:pull}}
\begin{figure}[b]
\begin{center}
%\begin{tabular}{cc}
%%{\blue Quark Jets}
%Quark Jets
%&
%%{\red Gluon Jets}
%Gluon Jets
%\\
%    \includegraphics[trim=10cm 4cm 10cm 11cm,clip,width=0.3\textwidth,height=0.3333\textwidth]{acc_highres_ellip3M_dEtadPhi_raw_qqZbb_011}
%&
%    \includegraphics[trim=10cm 4cm 10cm 11cm,clip,width=0.3\textwidth,height=0.3333\textwidth]{acc_highres_ellip3M_dEtadPhi_raw_ggZgg_011}
%\end{tabular}
\includegraphics{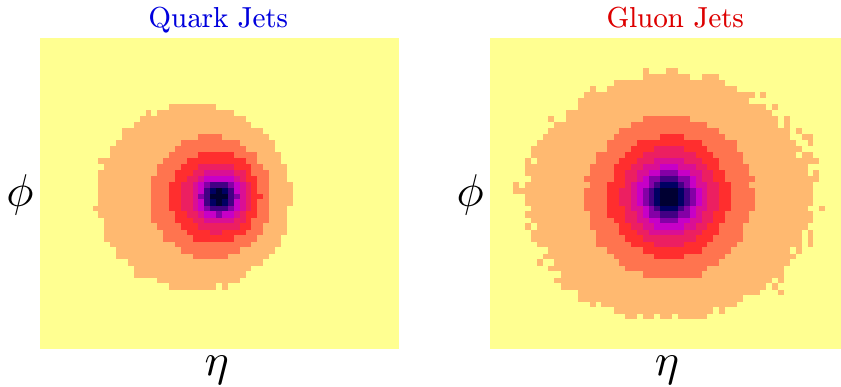}
\end{center}
\caption{
Distribution of radiation in quark and gluon jets accumulated over 3 million back-to-back 100\,GeV dijet events with fixed parton kinematics.
The color shows the average showered $p_T$ density in
$(\eta,\phi)$ for an ensemble of events with fixed parton-level kinematics.
(Contours are stepped in factors of two, which somewhat obscures
the nearly identical jet $p_T$ in both cases.)
}
\label{fig:acc_highres3M_raw_crop}
\end{figure}

In figure \ref{fig:acc_highres3M_raw_crop}, we show accumulated $p_T$ for
the same quark and gluon parton showered  millions of times.
In the large $N_C$ approximation
where these concepts apply, quarks have a single color
connection and gluons have one color and one anti-color
connection.
In this particular event, the quark was color-connected with the
beam remnant that went off to the left toward $\eta=-\infty$.
The gluon was connected to both outgoing beams.

Pull tries to quantify the color connections.
It was introduced in Ref.~\cite{Gallicchio:2010sw}, and then
immediately used in the D\O\ search for $\zh$ with $Z\to \nu
\bar{\nu}$ \cite{D0pull}.
The {\bf pull vector} of a jet is designed to point toward
the jet or beam that its color-connected to. The pull vector is
a $p_T$ weighted moment that tends to point toward the
color-connected partner of the jet's initiating quark.
If the jet was initiated by a gluon, it is color-connected
two two different places, so we might expect less pull. The pull
vector is defined as
\begin{equation}
\textrm{Pull Vector } \vec t = \sum_{i \in \mathrm{jet}} \frac{ p_T^i \, |r_i|  }{ p_T^{jet} }\, \vec r_i
\qquad
\textrm{where } \vec r_i \equiv (y_i - y_\mathrm{jet}, \phi_i - \phi_\mathrm{jet}) \,.
\end{equation}
If the factor of $|r_i|$ were removed, this would be the jet's
$p_T$-weighted centroid. Unlike other moments, the pull vector is
explicitly designed to \emph{not} be rotationally invariant.
The most effective way to use the pull vector in the Higgs
study was to calculate a {\bf pull angle}, which is the angle
between the pull vector and the direction where it `should'
point if the jet was color-connected to some other object.
We did not find pull angle very useful in distinguishing quarks from gluons.

\clearpage

\section{Combining Variables}
\begin{figure}[b]
   \begin{center}
      \begin{tabular}{ccc}
         %\blue{Quark} &
         %\red{Gluon} \\
         \psfrag{Signal}{\qquad \qquad \bf{Quark}}
         \psfrag{components_jet0_ak05_charged_count}{\footnotesize{Charged Count}}
         \psfrag{components_jet0_ak07_charged_count}{\footnotesize{Charged Count}}
         \psfrag{radial_moment_ak05_1_Pt_4v_y}{\footnotesize{Girth}}
         \psfrag{radial_moment_ak04_1_Pt_4v_y}{\footnotesize{Girth}}
         \includegraphics[width=0.32\textwidth]{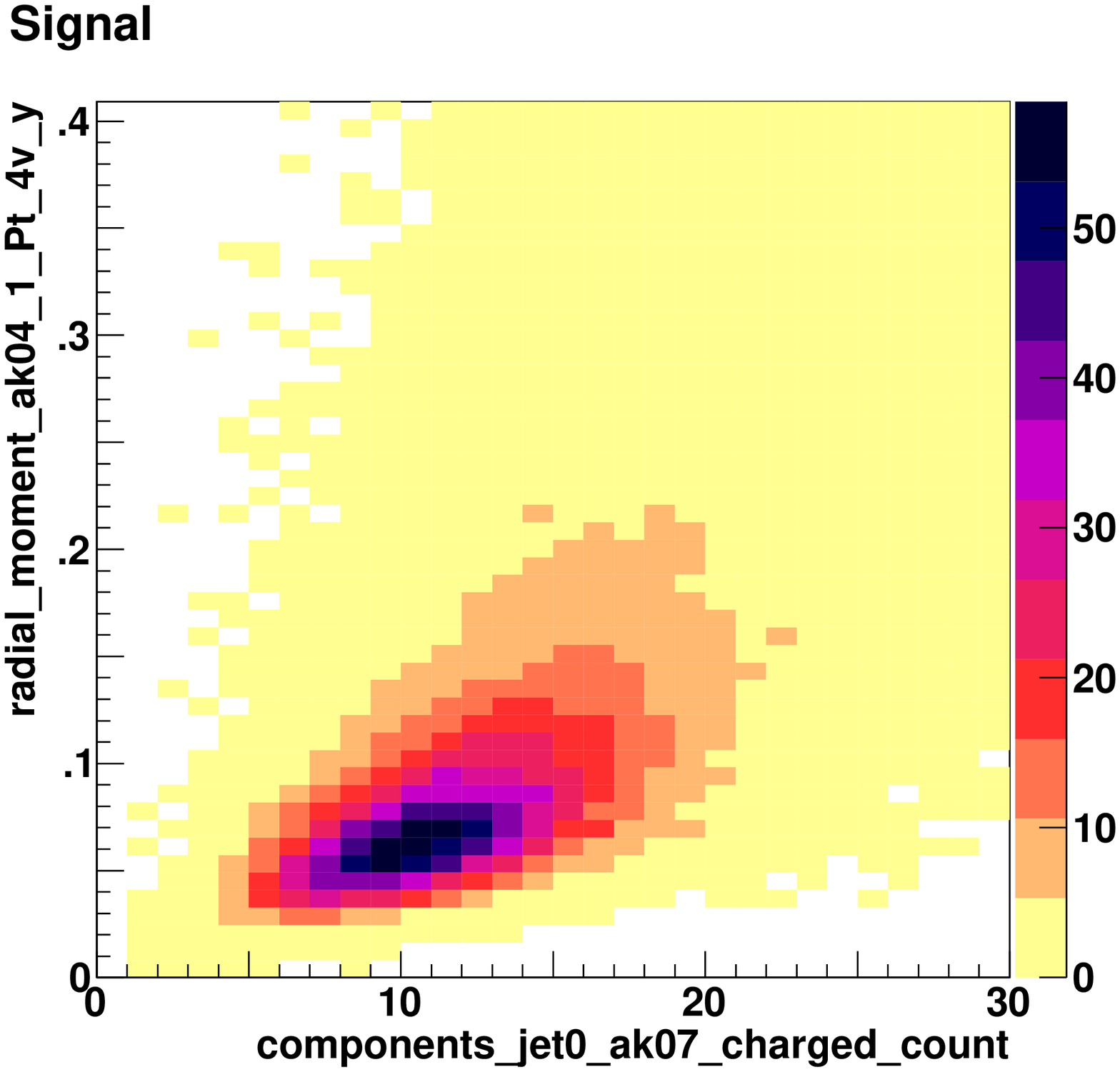}
        &
         \psfrag{Background}{\qquad \qquad \bf{Gluon}}
         \psfrag{components_jet0_ak05_charged_count}{\footnotesize{Charged Count}}
         \psfrag{components_jet0_ak07_charged_count}{\footnotesize{Charged Count}}
         \psfrag{radial_moment_ak05_1_Pt_4v_y}{\footnotesize{Girth}}
         \psfrag{radial_moment_ak04_1_Pt_4v_y}{\footnotesize{Girth}}
         \includegraphics[width=0.32\textwidth]{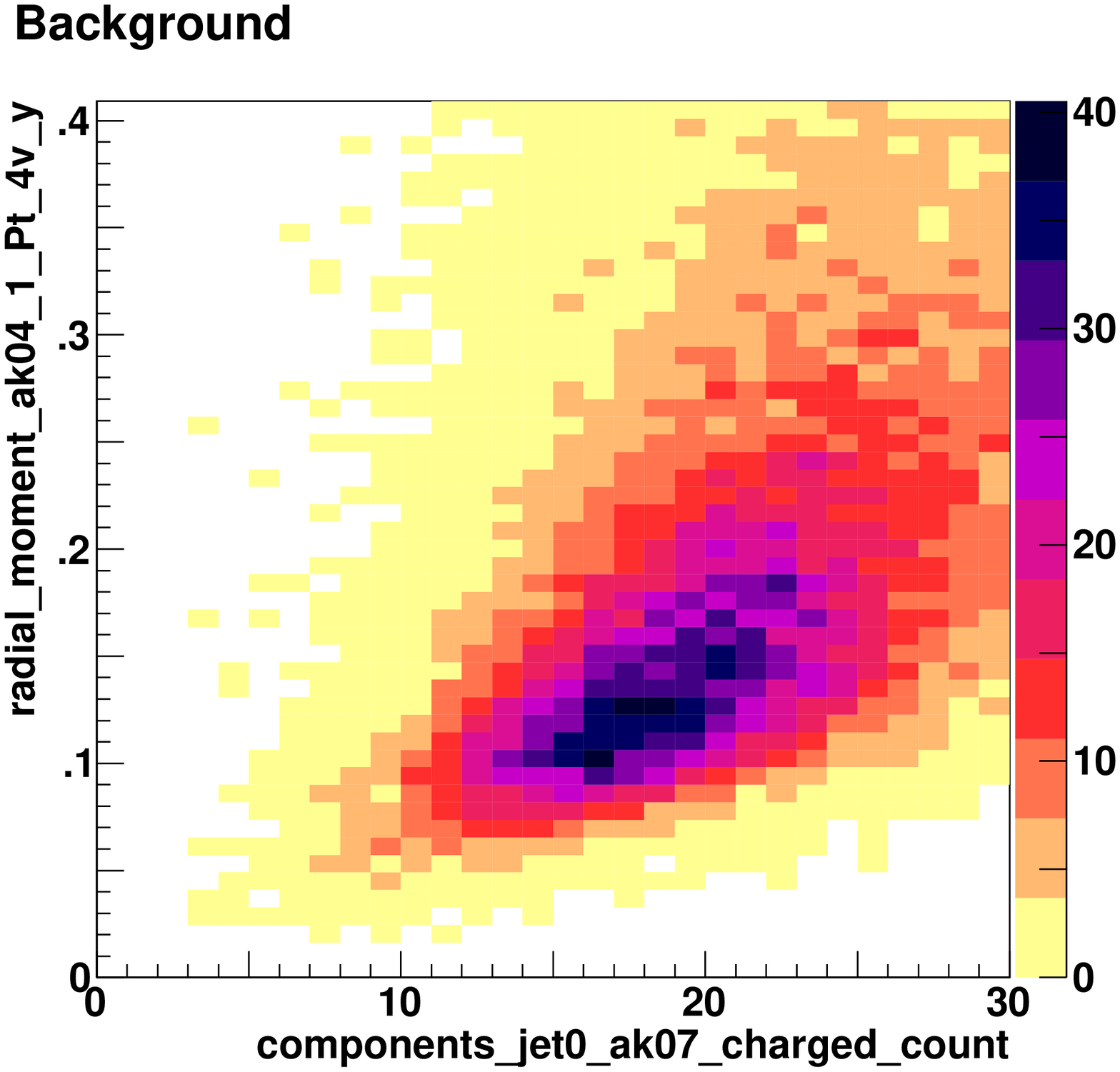}
         &
         \psfrag{components_jet0_ak05_charged_count_shape_optimal_pT_R_jet0_ak04_particles}{\qquad \scriptsize{Log(Q/G)}}  % title
         \psfrag{components_jet0_ak05_charged_count}{\footnotesize{Charged Count}}
         \psfrag{components_jet0_ak07_charged_count}{\footnotesize{Charged Count}}
         \psfrag{radial_moment_ak05_1_Pt_4v_y}{\footnotesize{Radial Moment}}
         \psfrag{radial_moment_ak04_1_Pt_4v_y}{\footnotesize{Girth}}
        \psfrag{Likelihood}{\quad  \bf{Likelihood: $q/(q+g)$}}
        \psfrag{components_jet0_ak05_charged_count_shape_optimal_pT_R_jet0_ak04_particles}{\qquad \scriptsize{Log(Q/G)}}  % title
      \includegraphics[width=0.32\textwidth]{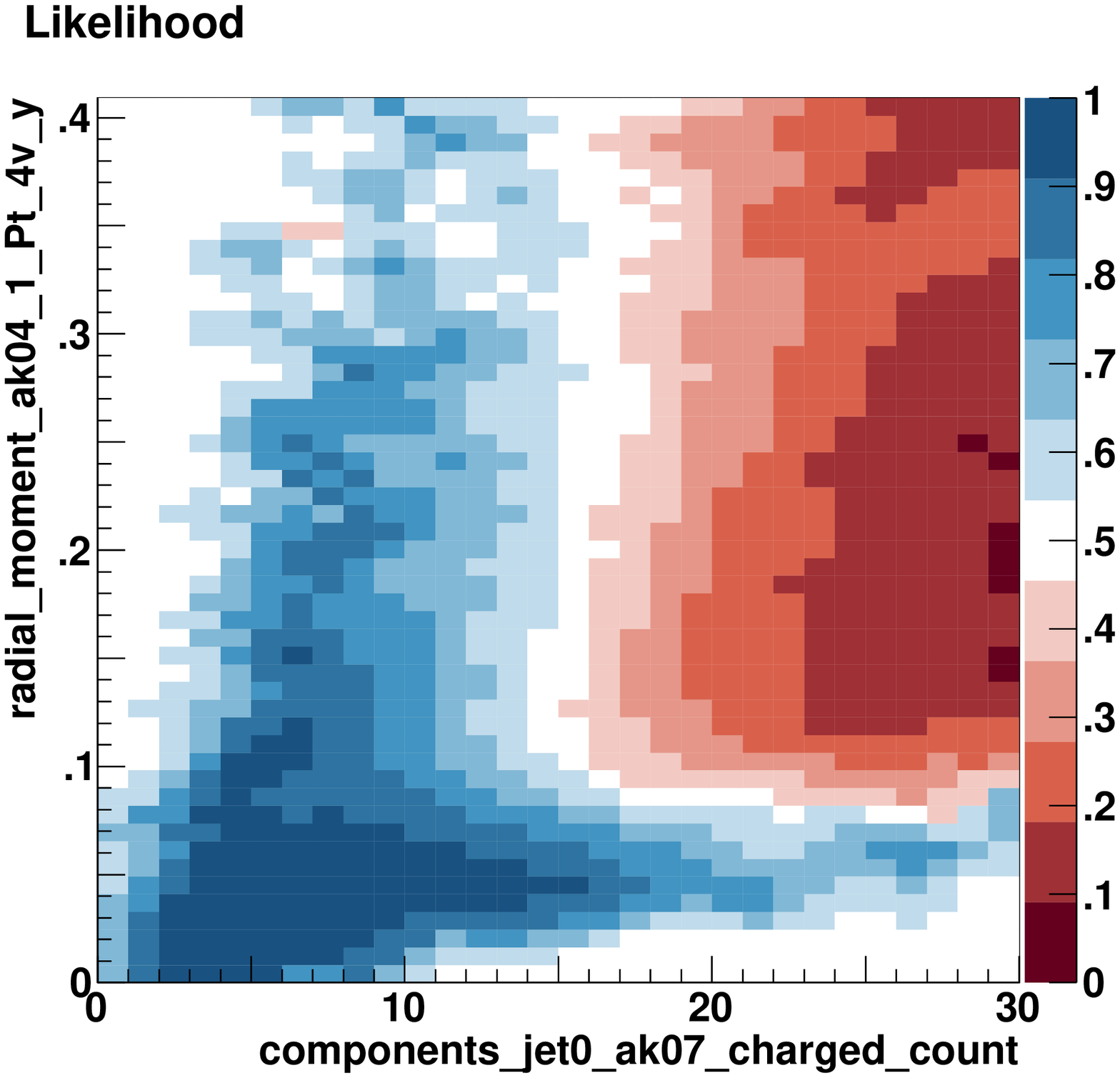}
     \end{tabular}
   \end{center}
\caption{
Combining Variables:  2D distributions are shown for a powerful pair
of variables.  The Likelihood can be formed by combining
these histograms bin-by-bin as $q/(q+g)$, where $q$ and $g$ are the 
fraction of events in the appropriate bin of the 
quark and gluon histogram, respectively.
The blue regions mean that an event with that pair
of values is more likely to be quark.
A cut on the likelihood correspond to a cut along one of the contours,
and this can be proven to be the optimal cut for that signal efficiency.
These plots are for {\sc Pythia8} 200\,GeV particle jets.
}
\label{fig:combining}
\end{figure}

A multivariate tagger can make the best use of
several variables at the same time.
In Figure~\ref{fig:combining}, the 2D distributions
of a good pair of variables is plotted for the quark
and gluon samples.  To find the best cut contours,
one method is to combining these histograms
into a likelihood histogram.
This is done bin-by-bin by reading the values
of the quark and gluon histograms and computing $q/(q+g)$.
If particular values are measured for each of the two
variables, this likelihood is proportional to
the probability that it is a quark jet.
The constant of proportionality will depend on the
prior distribution of quarks and gluons in your sample via Bayes' Theorem,
but does not affect the contours.

A cut on on this likelihood score corresponds to a cut along
some contour in the 2D plane. Each such cut gives some efficiency for
keeping quark jets and some other efficiency for rejecting gluon jets.
Cutting on the likelihood is optimal in the sense that it maximizes
gluon rejection for every given quark acceptance~\cite{tmva}.
Some ways of visualizing the effects of cuts and multivariate improvements
were discussed in~\cite{Gallicchio:2010dq,Cui:2010km}.

%To compute the likelihood bin-by-bin as described requires
%a huge number of events. 
To populate a 2D histograms 
%with sufficient events
such that each bin has a statistically meaningful number is difficult 
without an enormous number of events.
%usinga limited sample.
For more than 2 variables, it is practically impossible to populate
the higher-dimensional histograms with any accuracy.
For example, for 5 variables, even if each variable had only 10 coarse divisions,
there would still be $10^5$ bins to populate.
This is where multivariate techniques like
Boosted Decision Trees are useful~\cite{tmva}.
Using a limited number of
training events, these techniques assign a score to each point.
With a large enough training sample, this score is in 1-1 correspondence with the
likelihood.

\section{Comparing Variables}
A pair of variables that always performs at or near the top in our multivariate rankings is
charged track count combined with girth (also known as the linear radial moment jet width).
This pair was shown in Figure~\ref{fig:combining}.
In a computationally intensive search, we also came up with the best
group of five variables, which differed for each $p_T$ window and
ranking method (gluon rejection at 80\%, 50\%, or 20\%.)

\begin{figure}
\begin{center}
\psfrag{combo005_ak02MDPt_componentsjet0ak04particlesbroadening_componentsjet0ak04subak010av_641d1850de95b29a}{\footnotesize{best group of 5}}
\psfrag{combo002_componentsjet0ak07chargedcount_radialmomentak041Pt4vy}{\footnotesize{charged mult \& girth}}
\psfrag{components_jet0_ak05_charged_count_T_radial_moment_ak05_1_Pt_4v_y}{\footnotesize{charged mult * girth}}
\psfrag{components_jet0_ak07_charged_count}{\footnotesize{charged mult R=0.5}}
\psfrag{components_jet0_ak05_sub_ak010_count}{\footnotesize{subjet mult R$_\mathrm{sub}$=0.1}}
\psfrag{components_jet0_ak07_sub_ak010_count}{\footnotesize{subjet mult R$_\mathrm{sub}$\!=0.1}}
\psfrag{ak03_MDPt}{\footnotesize{mass/Pt R=0.3}}
\psfrag{radial_moment_ak05_2cos_Pt_4v_y}{\footnotesize{cos moment R=0.5}}
\psfrag{angularity_ak05_p100_Pt_4v_y_Rmax_overPt}{\footnotesize{ang a=1 R=0.5}}
\psfrag{radial_moment_ak05_1_Pt_4v_y}{\footnotesize{girth R=0.5}}
\psfrag{subjet_1stPt_over_jetPt_jet0_ak05_sub_ak010}{\footnotesize{1st subjet $p_T$}}
\psfrag{decluster_kT_min_jet0_ak05_kt010}{\footnotesize{decluster $k_T$}}
\psfrag{shape_optimal_pT_R_jet0_ak10_particles}{\footnotesize{optimal R=1.0}}
\psfrag{pull_Eta_ak03_0_E_4v_y}{\footnotesize{pull $\eta$ R=0.3}}
\psfrag{pull_R_ak03_2_Pt_4v_y}{\footnotesize{$|$pull$|$ R=0.3}}
\psfrag{planar_flow_covariance_jet0_ak03_sub_ak010_kTE}{\footnotesize{planar flow R=0.3}}
\psfrag{components_jet0_ak07_charged_count, radial_moment_ak05_1_Pt_4v_y}{\footnotesize{best pair}}
%\psfrag{LHC 200 : Background Rejection}{}
%\psfrag{}{\footnotesize{count R=0.5}}
\psfrag{group of length 5}{\footnotesize{best group of 5}}
\psfrag{components_jet0_ak07_charged_count, radial_moment_ak04_1_Pt_4v_y}{\footnotesize{best pair (charge ct \& girth)}}
\psfrag{shape_optimal_pT_R_jet0_ak02_particles}{\footnotesize{optimal kernel at 80\%}}
\psfrag{subjet_1stPt_over_jetPt_jet0_ak05_sub_ak010}{\footnotesize{1st subjet R=0.5}}
\psfrag{subjet_2ndPt_over_1stPt_jet0_ak05_sub_ak010}{\footnotesize{2nd subjet R=0.5}}
\psfrag{subjet_3rdPt_over_jetPt_jet0_ak05_sub_ak010}{\footnotesize{3rd subjet R=0.5}}
\psfrag{components_jet0_ak05_sub_ak010_avg_kT}{\footnotesize{avg $k_T$ of R$_\mathrm{sub}$\!=0.1}}
\psfrag{decluster_kT_jet0_ak05_kt010}{\footnotesize{decluster\,$k_T$\,R$_\mathrm{sub}$\!=0.1}}
\psfrag{shape_integ_pT_R_jet0_ak07_particles_bin_20_of_38_0100}{\footnotesize{jet shape $\Psi(0.1)$}}
\psfrag{Signal eff}{\hspace{-1in}\Large{Quark Jet Acceptance}}
\psfrag{LHC 100 : Background Rejection}{\Large{Gluon Rejection}}
\psfrag{LHC 100 : Bkg Rejection}{\Large{Gluon Rejection}}
\psfrag{LHC 100 : Sig Minus Bkg}{$\epsilon_S - \epsilon_B$}
\includegraphics[width=\textwidth]{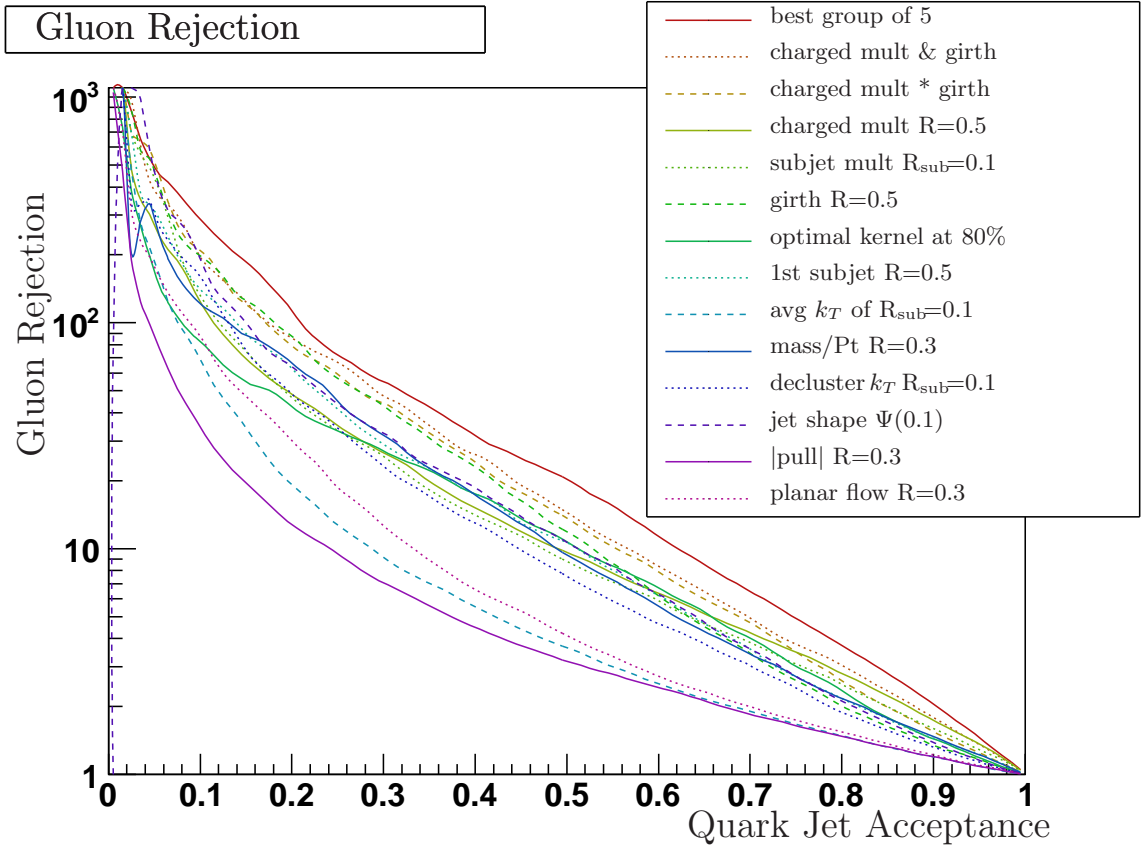}
\caption{
ROC curves for 100\,GeV {\sc Pythia8} jets for selected variables.
These curves show the background (gluon jet) rejection efficiency ($1/\eB$) 
as a function of the signal (quark jet) acceptance efficiency ($\eS$).
}
\label{fig:roc_variables}
\end{center}
\end{figure}

The ROC curves (gluon rejection versus
quark acceptance) for interesting variables
are shown in Figure~\ref{fig:roc_variables}
for $p_T$=100\,GeV jets.
Unlike previous curves, the rejection is plotted
on a logarithmic axis.  A 1\% background acceptance
corresponds to a background rejection of $10^2$.
The best curve corresponds to the best
group of five variables, but the best pair
(charged track multiplicity \& girth) is not far behind.
Simply taking the product of these two creates a single variable that
does better than each individually, but only for
for harder cuts (lower quark acceptances.)
The good discrete observables like counting charged tracks
or counting small subjets do best at high quark acceptance
(and as we saw before, high jet $p_T$).
The good continuous observables like girth tend to do
best at lower quark acceptance and lower jet $p_T$.
Jet mass tends to be somewhere in the middle, and 2D
geometric moments like pull and planar flow are never particularly
powerful.

%For high (80\%) quark efficiency, the charged track counts are
%best. For the low (20\%) efficiencies that also optimize the
%significance improvements, radial moments, especially optimal
%kernels, are best. (Similar observables like jet mass or $a=+1$
%angularity also perform well here.) This remains true for all
%jet $p_T$. At higher $p_T$, the charged track count becomes
%competitive even at 20\%. Together, these are consistently the
%best pair of variables everywhere with the possible exception
%of 50\,GeV 20\% where two optimal shapes or mass and angularity
%combine to do well.

\begin{figure}
\begin{center}
\psfrag{components_jet0_ak05_charged_count}{\footnotesize{charge count}}
\psfrag{combo005_ak02MDPt_componentsjet0ak04particlesbroadening_componentsjet0ak04subak010av_641d1850de95b29a}{\footnotesize{best group of 5}}
\psfrag{combo002_componentsjet0ak07chargedcount_radialmomentak041Pt4vy}{\footnotesize{charged mult \& girth}}
\psfrag{components_jet0_ak05_charged_count_T_radial_moment_ak05_1_Pt_4v_y}{\footnotesize{charged mult * girth}}
\psfrag{components_jet0_ak07_charged_count}{\footnotesize{charged mult R=0.5}}
\psfrag{components_jet0_ak05_sub_ak010_count}{\footnotesize{subjet mult R$_\mathrm{sub}$=0.1}}
\psfrag{components_jet0_ak07_sub_ak010_count}{\footnotesize{subjet mult R$_\mathrm{sub}$\!=0.1}}
\psfrag{ak03_MDPt}{\footnotesize{mass/Pt R=0.3}}
\psfrag{radial_moment_ak05_2cos_Pt_4v_y}{\footnotesize{cos moment R=0.5}}
\psfrag{angularity_ak05_p100_Pt_4v_y_Rmax_overPt}{\footnotesize{ang a=1 R=0.5}}
\psfrag{radial_moment_ak05_1_Pt_4v_y}{\footnotesize{girth R=0.5}}
\psfrag{subjet_1stPt_over_jetPt_jet0_ak05_sub_ak010}{\footnotesize{1st subjet $p_T$}}
\psfrag{decluster_kT_min_jet0_ak05_kt010}{\footnotesize{decluster $k_T$}}
\psfrag{shape_optimal_pT_R_jet0_ak10_particles}{\footnotesize{optimal R=1.0}}
\psfrag{pull_Eta_ak03_0_E_4v_y}{\footnotesize{pull $\eta$ R=0.3}}
\psfrag{pull_R_ak03_2_Pt_4v_y}{\footnotesize{$|$pull$|$ R=0.3}}
\psfrag{planar_flow_covariance_jet0_ak03_sub_ak010_kTE}{\footnotesize{planar flow R=0.3}}
\psfrag{components_jet0_ak07_charged_count, radial_moment_ak05_1_Pt_4v_y}{\footnotesize{best pair}}
%\psfrag{LHC 200 : Background Rejection}{}
%\psfrag{}{\footnotesize{count R=0.5}}
\psfrag{group of length 5}{\footnotesize{best group of 5}}
\psfrag{components_jet0_ak07_charged_count, radial_moment_ak04_1_Pt_4v_y}{\footnotesize{best pair (charge ct \& girth)}}
\psfrag{shape_optimal_pT_R_jet0_ak02_particles}{\footnotesize{optimal kernel at 80\%}}
\psfrag{subjet_1stPt_over_jetPt_jet0_ak05_sub_ak010}{\footnotesize{1st subjet R=0.5}}
\psfrag{subjet_2ndPt_over_1stPt_jet0_ak05_sub_ak010}{\footnotesize{2nd subjet R=0.5}}
\psfrag{subjet_3rdPt_over_jetPt_jet0_ak05_sub_ak010}{\footnotesize{3rd subjet R=0.5}}
\psfrag{components_jet0_ak05_sub_ak010_avg_kT}{\footnotesize{avg $k_T$ of R$_\mathrm{sub}$\!=0.1}}
\psfrag{decluster_kT_jet0_ak05_kt010}{\footnotesize{decluster\,$k_T$\,R$_\mathrm{sub}$\!=0.1}}
\psfrag{shape_integ_pT_R_jet0_ak07_particles_bin_20_of_38_0100}{\footnotesize{jet shape $\Psi(0.1)$}}
\psfrag{Signal eff}{\hspace{-1in}\large{Quark Jet Acceptance}}
\psfrag{LHC 100 : Significance}{\large{Significance Improvement}}
\includegraphics[width=\textwidth]{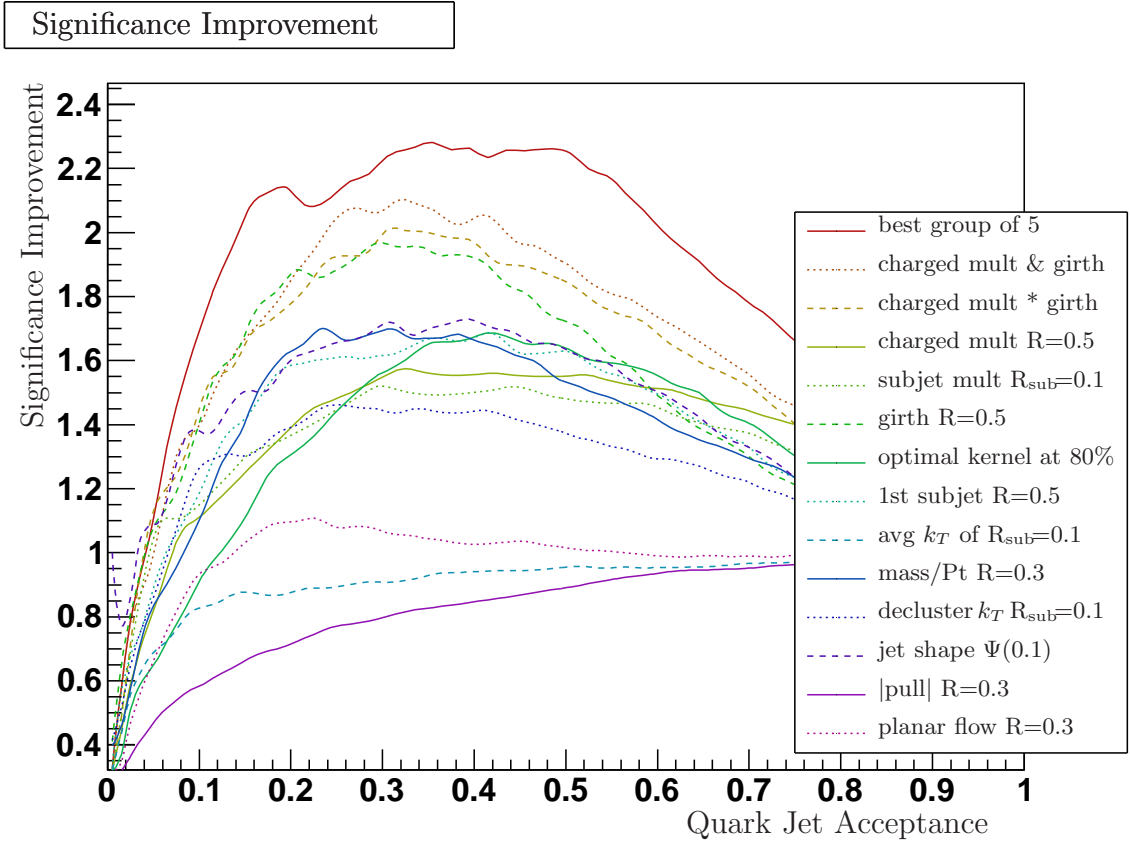}
\caption{
Significance Improvement Curves for 100\,GeV {\sc Pythia8} jets for selected variables.
These curves show the significance improvement $\eS/\sqrt{\eB}$ as a function of $\eS$.
}
\label{fig:sic_variables}
\end{center}
\end{figure}

The best variable depends on the desired signal
acceptance operating point, which depends on the application.  For example,
one might try to maximize the significance ($\sigma \sim S/\sqrt{B}$) of a
small signal above a large background.  An advantage of maximizing the significance
is that, each operating point translates into
an improvement factor (which should be greater than one if the
cut is useful), independent of the initial significance.  This improvement factor is also independent of
integrated luminosity and the signal and background counts
themselves. To see this, note that cutting on a variable changes the significance
by
\begin{equation}
\sigma \equiv \frac{S}{\sqrt{B}}
\qquad \ra{\text{cut}} \qquad
\frac{\eS S}{\sqrt{ \eB B}}
= \left(\frac{\eS}{\sqrt{\eB}}\right) \sigma \,.
\end{equation}
With a simple 1-1 transformation, a ROC curve can be
turned into a significance improvement curve (SIC)~\cite{Gallicchio:2010dq}.
Samples are shown in Figure~\ref{fig:sic_variables} for the 100\,GeV sample.

For all variables, the cuts that optimize quark/gluon
significance improvement tend to be quite harsh, leaving only
$\sim$20\% of the quark sample. For rare signals with few
events, looser cuts might be required to see any events at all.
In cases where the background to a quark jet new physics signal
is not 100\% gluon-jets, looser cuts end up giving the optimal
significance improvement. QCD backgrounds at low $p_T$ are only
around 80\% gluon.  This is discussed further in Section~\ref{sec:mixed background}.

\begin{figure}
   \begin{center}
      \psfrag{LHC 50 : Background Rejection}{\Large{Gluon Background Rejection}}
      \psfrag{Signal eff}{\hspace{-1in}\Large{Quark Signal Efficiency}}
      \includegraphics[width=\textwidth]{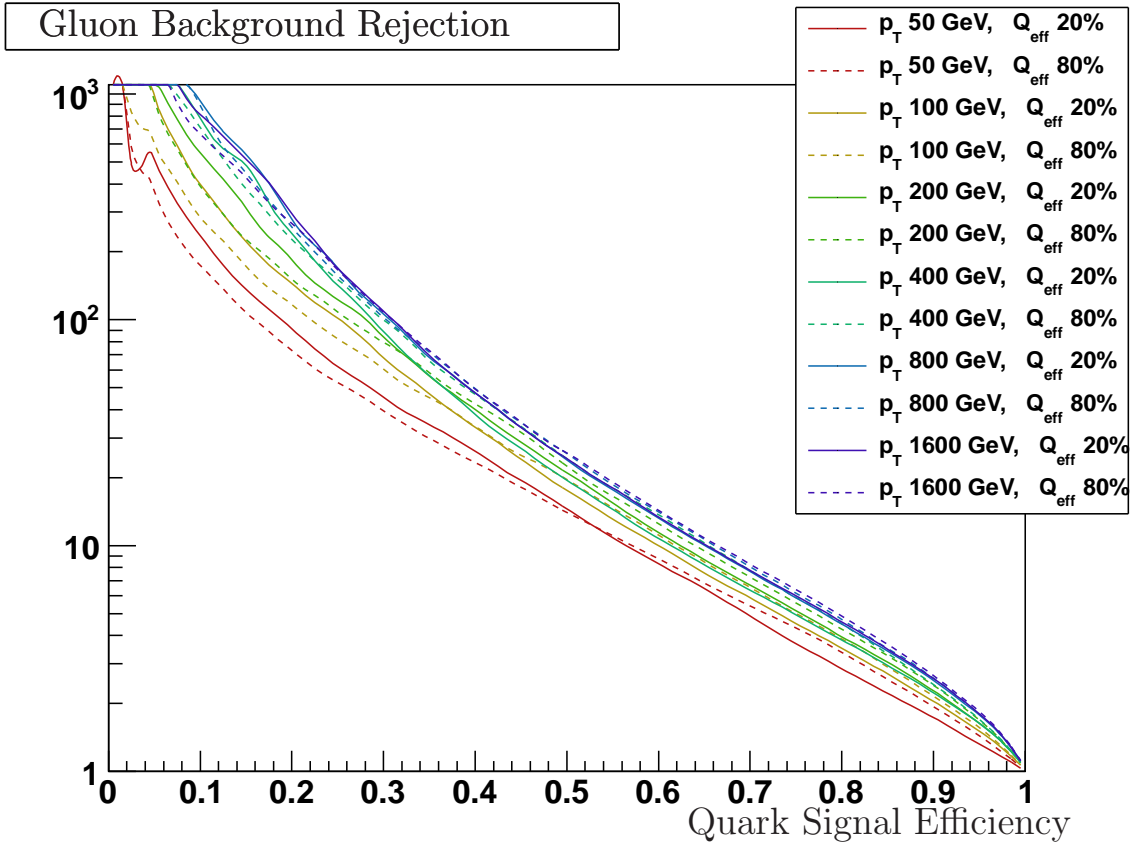}
   \end{center}
\caption{
Gluon Background Rejection for the best groups of five {\sc Pythia8} variables for each $p_T$.
The dashed lines are the groups that maximize gluon rejection at 80\% quark efficiency.
The solid lines maximize significance improvement, which the next figure
shows happens around $\sim 20\%$ quark efficiency.
Higher $p_T$ jets lead to greater rejection power, mostly because the
charged track count is a more powerful observable.
}
\label{fig:rocs_for_pts}
\end{figure}

\begin{figure}
   \begin{center}
      \psfrag{LHC 50 : Significance}{\!\!\!\!\Large{Significance Improvement}}
      \psfrag{Signal eff}{\hspace{-1in}\Large{Quark Signal Efficiency}}
      \includegraphics[width=\textwidth]{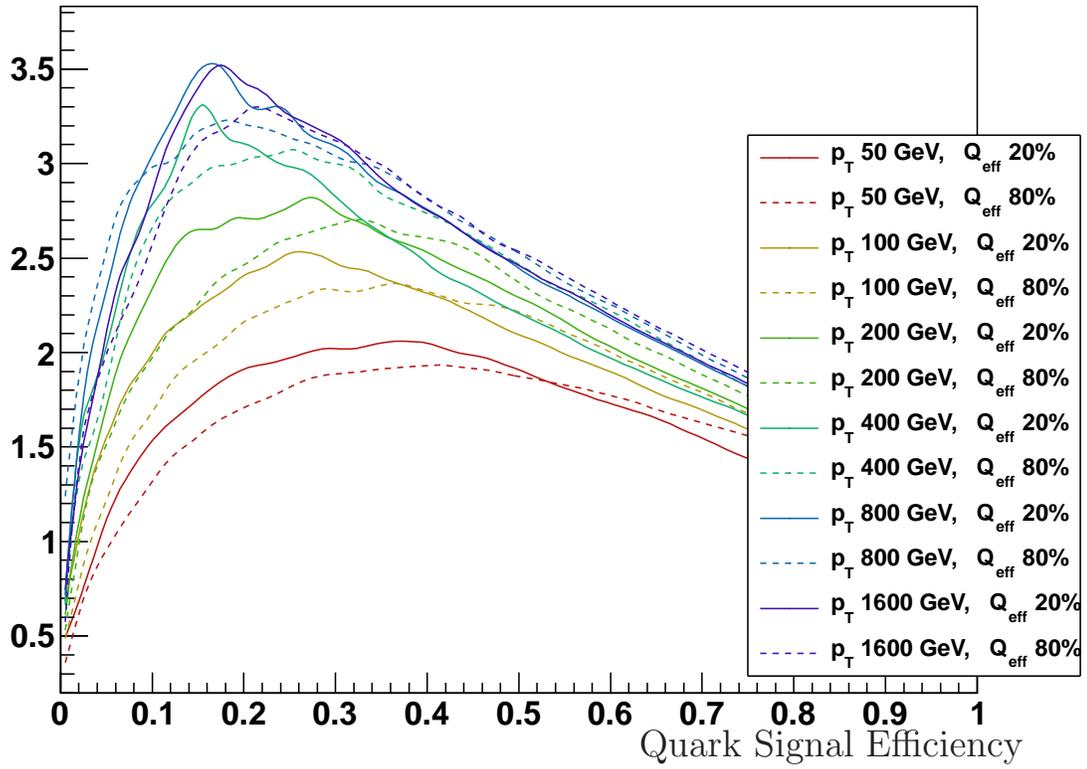}
   \end{center}
\caption{
Significance Improvement curves for best groups of five variables.
This contains the same information as the previous figure,
but shows that small differences in gluon rejection at low efficiencies
lead to large differences in significance improvement.
}
\label{fig:sics_for_pts}
\end{figure}

Gluon rejection curves for the best group of five variables
for each of the $p_T$ samples are shown in Figure~\ref{fig:rocs_for_pts}.
There is one best group of five when optimizing rejection at 20\% quark
acceptance, and a different group of five when optimizing at 80\%.
Transformations of these into Significance Improvement Curves is
shown in Figure~\ref{fig:sics_for_pts}.
Exact scores depend on whether {\sc Pythia8} or {\sc Herwig++} is used
and whether all particles are used or just charged tracks.  This
is discussed further in Section~\ref{sec:Herwig vs Pythia}. 
Results are shown at the end, in Table~\ref{tab:herpy}.

%Also, keeping 50\% or 80\% of the quark jets while throwing
%away the most gluon-like jets can be useful in other contexts.
%The best choice of variables (and parameters like jet size)
%depend on the desired signal efficiency. Charged particle count
%tends to be good at high efficiency, whereas radial moments
%tend to be good at lower efficiency when a more pure quark
%sample is required or where significance is maximized.

%Does this hold up for different jet $p_T$s? By picking a
%better jet algorithm or size $R$, can we do better?  We examine
%this by picking three reference quark signal efficiencies:
%20\%, 50\%, and 80\%.  For each signal efficiency, we show how
%the background rejection changes for the different $p_T$
%samples and different jet sizes.  The \emph{charged particle
%count} is shown for different $R_\mathrm{jet}$ and $p_T$ in
%Figure~\ref{fig:components_jet0_ak_charged_count_80_50_20}.

%For each variable, each jet size, and each $p_T$, and each
%Monte Carlo, we could show:
%%
%\begin{itemize}
%  \item Distributions for quarks and gluons
%  \item ROC curve from cutting on the distributions
%  \item Measure of power: separation, or rejection at 20\%,
%      50\% or 80\%.
%\end{itemize}
%%
%We limit ourselves to the most powerful, most familiar, and
%most surprising.

%\begin{figure}
%   \begin{center}
%      \psfrag{LHC 200 : Significance}{\!\!Significance Improvement \qquad for 200\,GeV}
%      \psfrag{Signal eff}{\hspace{-1in}Quark Signal Efficiency}
%      \includegraphics[width=0.75\textwidth]{qvsg_dijet6_BDT100_all_ppgg_LHC7jetwin_jetpt200_hard_signif_best005_ROCsSignif}
%   \end{center}
%\caption{
%200\,GeV groups of variables from singles
%(blue) pairs (cyan) up to groups of five (red),
%optimizing significance improvement.
%}
%\end{figure}

%\begin{figure}
%   \begin{center}
%      \psfrag{LHC 50 : Significance}{\!\!\!Significance Improvement}
%      \psfrag{Signal eff}{\hspace{-1in}Quark Signal Efficiency}
%      \includegraphics[width=0.75\textwidth]{qvsg_dijet6_qvsgPts_combo5_ROCsSignif}
%   \end{center}
%\caption{
%Final Result Significance Improvement for the best groups of 5 variables.
%}
%\end{figure}

%\begin{figure}
%   \begin{center}
%      \includegraphics[width=0.75\textwidth]{BTagAtlas}
%   \end{center}
%\caption{
%Atlas $B$-tagging rejection curves.  Gluon rejection is not as good
%as $B$-tagging rejection of light quarks,
%but better than rejection of charm, and similar rejection of $\tau$.
%Of course these curves are from data while ours do not even
%include a detector simulation.
%(Note the horizontal scale here cuts off at 0.3
%efficiency.)
%}
%\end{figure}

\clearpage

\section{Using Impure Samples to Verify Underlying Pure Distributions}

\begin{figure}[b]
\begin{center}
%\vspace{-0.25in}
\includegraphics[width=0.75\textwidth]{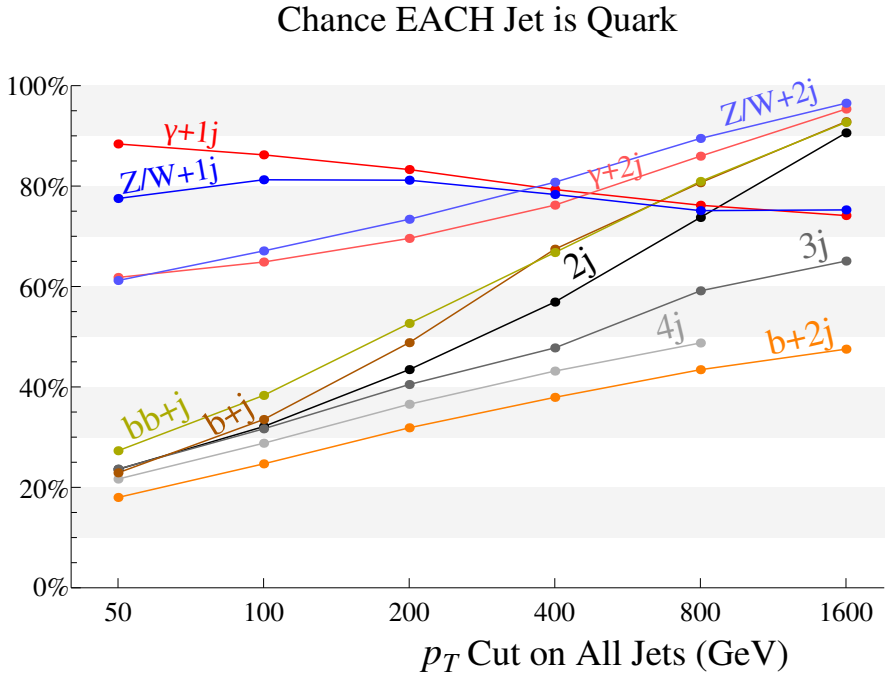}
\vspace{-0.1in}
\caption{
The chance that a given jet is a
light quark jet rather than a gluon jet.
(This ratio does not include bottom or charm.)
The $W$ and $Z$ were nearly identical and combined on this plot,
but they are slightly different from the photon, mostly due to the $\gamma$
and lepton cuts, which were each at 20\,GeV.
%This has to do with the $\gamma$
%and lepton cuts.
%rather than the underlying
%physics of producing massive, as
%opposed to massless, electroweak bosons.
\label{fig:Chance_EACH_Jet_is_Quark}
}
\end{center}
\end{figure}

When training a multivariate method to distinguish quarks from
gluons, it would be ideal to have huge samples of pure quarks
and pure gluons.  The quark fractions of several samples are
shown in Figure \ref{fig:Chance_EACH_Jet_is_Quark}.
For low jet $p_T$, none of the samples are more 90\% pure.
By making cuts, this can be increased at the cost of having
fewer training events.  For example, in $\gamma+2$jet events,
when the softer jet is close to the photon, it is very likely
a quark jet.  Similar cuts can be made to purify gluons from
multi-jet samples.  This was discussed in reference \cite{Gallicchio:2011xc}.

Ideally, one would like to combine information from high cross section, low-purity samples
with low cross section, high-purity samples.
One approach to combining information from different samples would be to first verify the jet property predictions
Monte Carlo generators.
If the Monte Carlo generators are sufficiently accurate,
huge numbers of simulated events can then be used to train
multivariate classifiers. The distributions of each jet property and correlations between them
can be checked against data.
For example, the jet mass distribution
of a simulated 90/10 mix should match a 90\% pure sample.
If two observables provide most of the
discrimination power, their 2D histogram can be compared to the
linear combination of pure quark and gluon jets produced by the Monte
Carlo.  Low statistics, high purity samples will constrain some
combination of measurements, while high statistics mixed
samples will constrain a different combination.

More generally, each jet can be assigned a likelihood of being
quark or gluon based on its kinematics (nearness to a photon for example)
rather than its intra-jet properties.
If it passes some kinematic
threshold, the jet and its likelihood can be used to train a
classifier.  More certain probabilities can be assigned higher
weights.  Although the multivariate techniques popular
in high energy physics and implemented in the
TMVA package~\cite{tmva} require a separate signal and background samples,
within each sample events can be assigned different weights.

An alternative data-driven approach is to fit for the underlying
pure distributions in an impure sample where the quark/gluon fractions
are assumed to be known.  For example, in a small window around 50\,GeV,
dijets with certain kinematic cuts are 23\% quarks. Jets in a $\gamma+$jet sample are 88\% quarks.
If the jet mass distribution is measured for these two 50\,GeV samples, the underlying
quark and gluon distributions can found by solving
a simple 2x2 linear equation bin-by-bin.  These pure underlying
quark and gluon mass distributions can then be re-weighted and compared
to a sample with a different known quark/gluon fraction.  If there is
consistency across samples, it would justify the use of
quark and gluon jet discrimination.  ATLAS calls this the template method \cite{Aad:2011he, ATLAS-CONF-2012-138}.
In this approach, bottom and charm jets are accounted for by trusting the Monte
Carlo simulation for both their distribution and the value of their small composition fraction.

Fitting to pure or mixed samples assumes a universality to quark and gluon properties, which may only
partly hold. Selection effects might induce atypical distributions, for example if harsh cuts
keep only the kinematic tails of distributions.  For example, jet
mass might look different at very high $\eta$.
One sample may be busier than another, with many other
nearby jets leaking into the ones you are interested in.
Jets in a training sample may
have different color-connections than jets in the sample you
ultimately wish to tag.  Color-singlet quark pairs from a $W$ might look
different than beam-connected quarks in $\gamma+$jet.
%%he
%radiation pattern for the $\epsilon_{abc}$ color-tensor for the SUSY R-parity
%violating $udd$ operator is not even present in the standard model.
%(Though it is well-approximated by recent versions of Pythia8 \cite{Desai:2011su}.)
%Many of the same issues arise in other substructure studies,
Many of these issues are not particular to quark/gluon tagging and
should be kept in mind for any substructure study.

\section{Choosing the Operating Point for a Mixed Background \label{sec:mixed background}}

The cut that most improves significance depends on the
quark/gluon composition of the real signal and background.
So far we have considered the signal to be only quark jets
and the background to be only gluon jets.
QCD has around 20\% quark jets at low $p_T$ and more for higher $p_T$.
Other common backgrounds were
shown in Figure \ref{fig:Chance_EACH_Jet_is_Quark}.

Once a tagger is trained, you need to pick an operating point.
For example, you could pick the fraction of quark jets you are willing keep:
the quark efficiency $\epsilon_q$.
This translates into a particular cut on the observables, either
directly or via a cut on a multivariate output.
The ROC curve shows the gluon efficiency $\epsilon_g$ (the fraction of gluon jets
that get past the best cut) for a given $\epsilon_q$.

If the signal is not pure quark and the background is not pure gluon,
a cut with these quark and gluon efficiencies will translate into
signal and background efficiencies in a way that depends on: the fraction
of signal made of quarks $s_q$, signal made of gluons $s_g$,
background made of quarks $b_q$, and background made of gluons $b_g$.
In this case,
\begin{equation}
\epsilon_s = s_q \epsilon_q + s_g \epsilon_g
\qquad
\mathrm{and}
\qquad
\epsilon_b = b_q \epsilon_q + b_g \epsilon_g \ .
\label{eqn:mixed_composition}
\end{equation}

\begin{figure}
\begin{center}
%\vspace{-0.25in}
\includegraphics[width=0.49\textwidth]{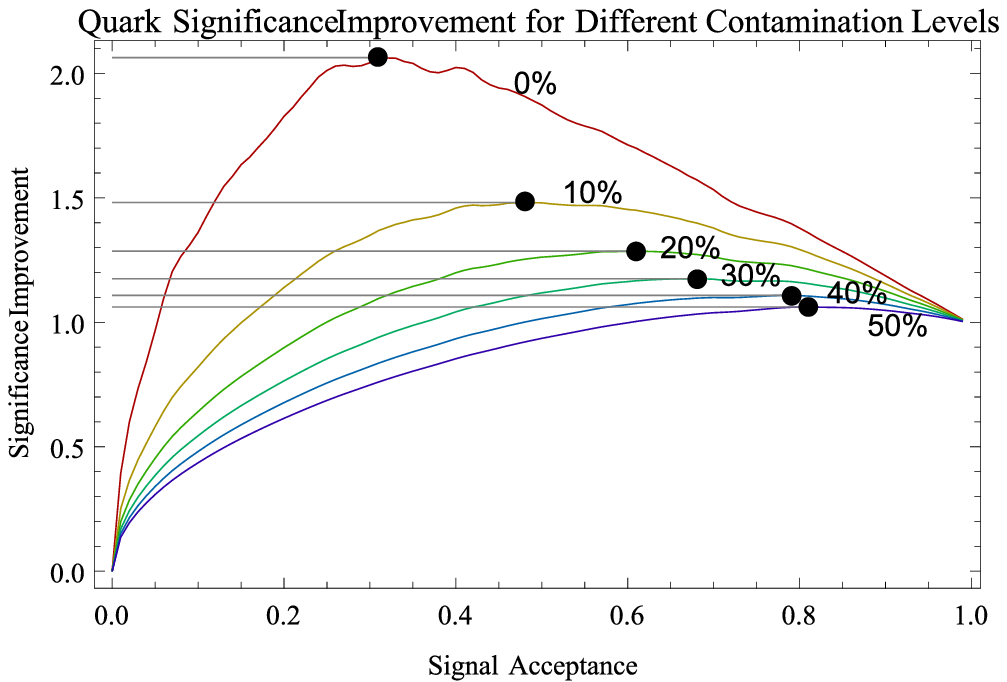}
\includegraphics[width=0.49\textwidth]{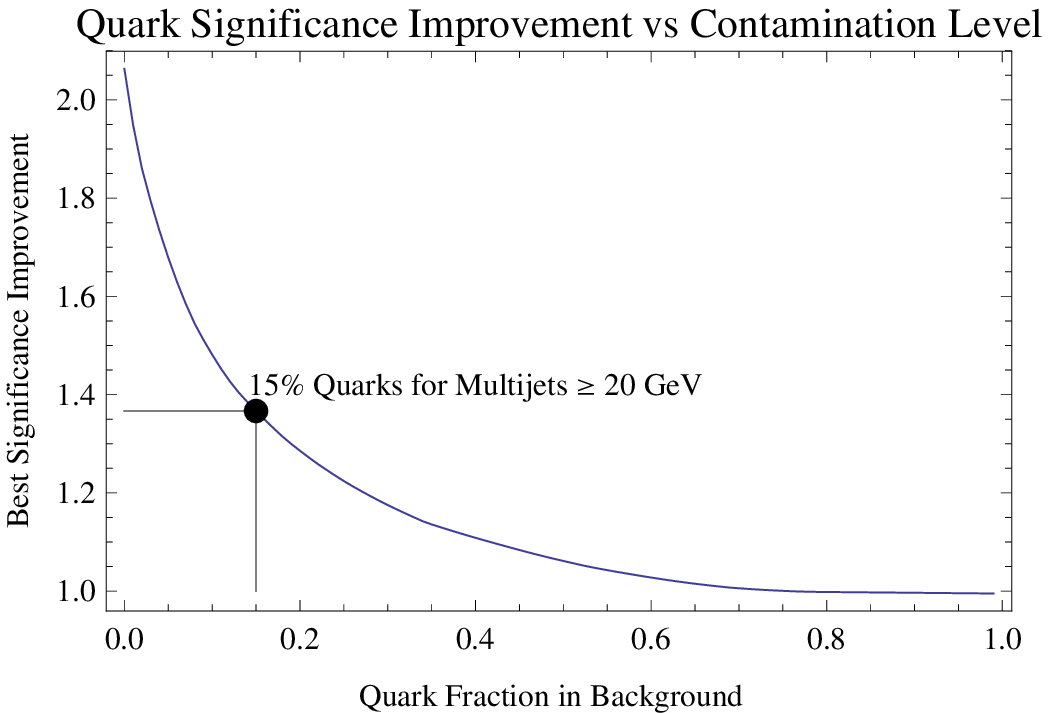}
\vspace{-0.1in}
\caption{
On the left is an illustration of how the significance improvement
curve changes when the background is not pure gluon jets, but contains
the indicated fraction of quark jets.
The points show the maximum significance improvement.
On the right, these maxima are plotted a function of the quark
fraction in the background.  The QCD background for jets
with $p_T \ge 20$\,GeV is approximately 15\% quark.  For higher $p_T$,
the QCD quark fraction goes up and the maximum significance decreases.
\label{fig:SignifQuarkVsContamination}
}
\end{center}
\end{figure}

Now suppose you start with $S$ signal events and $B$
background events. A cut with  signal efficiency $\epsilon_s$
and background efficiency $\epsilon_b$ changes the statistical significance
in a simple way as before:
\begin{equation}
\sigma \ = \  \frac{S}{\sqrt{B}}
\qquad
\rightarrow
\qquad
\frac{S \epsilon_s}{\sqrt{B \epsilon_b}}
\ = \  \sigma \frac{\epsilon_s}{\sqrt{\epsilon_b}}
\label{eqn:sic_formula}
\end{equation}
If the signal started with some significance $\sigma$,
the cut will improve it by a factor $\epsilon_s/\sqrt{\epsilon_b}$,
which is called the ``Significance Improvement'' in the plots below.
It depends not only on the performance of the tagger, but on the
quark/gluon makeup of your signal and background and where one chooses
to operate.  A ROC curve for quark/gluon discrimination
($\epsilon_g$ vs $\epsilon_q$) can be easily transformed into a
Significance Improvement Curve ($\epsilon_s/\sqrt{\epsilon_b}$ vs $\epsilon_s$)
using equations \ref{eqn:mixed_composition} and \ref{eqn:sic_formula}.
The first plot in Figure \ref{fig:SignifQuarkVsContamination}, several such curves
are shown for signals that are 100\% quark, but backgrounds
that are mixed.  The curve labeled 0\% has a
background that is purely gluon jets and corresponds to the curves shown previously.
Even when the background has only
10\% quark jets, the maximum achievable significance improvement has fallen
quite a bit.  The second plot in this figure shows those maxima as a
continuous function of the background's quark fraction.
A typical low-$p_T$ QCD background has 15\% quarks and is indicated.
Clearly the tagger will be most useful when the signal and background are
both quite pure.

The same analysis can be performed for the published $B$-tagger ROC curves.
Generic low-$p_T$ QCD backgrounds have around 2\% $B$-jets. For this value,
the significance improvement peaks at around 60\% $B$-acceptance.
This is the typical operating point of these taggers.

\clearpage

\section{Comparing {\sc Herwig++} to {\sc Pythia8} \label{sec:Herwig vs Pythia}}

Recently, ATLAS presented some quark/gluon measurements \cite{ATLAS-CONF-2012-138} of
charged track count and linear radial moment, which they call jet width.
They compared their data to {\sc Herwig++} and to two {\sc Pythia8} tunes.
They found that both simulations described the quark jet properties better than gluon jet
properties.

\begin{figure}[b]
\begin{center}
\psfrag{components_ak07_charged_count}{\textbf{Charged Track Count ($n_\mathrm{trk}$)}}
\psfrag{radial_moment_ak05_charged_1_Pt_4v_y}{\textbf{Linear Radial Moment (jet width)}}
\psfrag{ak05_charged_MDPt}{\textbf{mass/$p_T$}}
\psfrag{nsub_beta025_axes3_tau1_ak06_charged}{\textbf{1-subjettiness, optimized axes $\beta=1/4$}}
\includegraphics[width=0.49\textwidth]{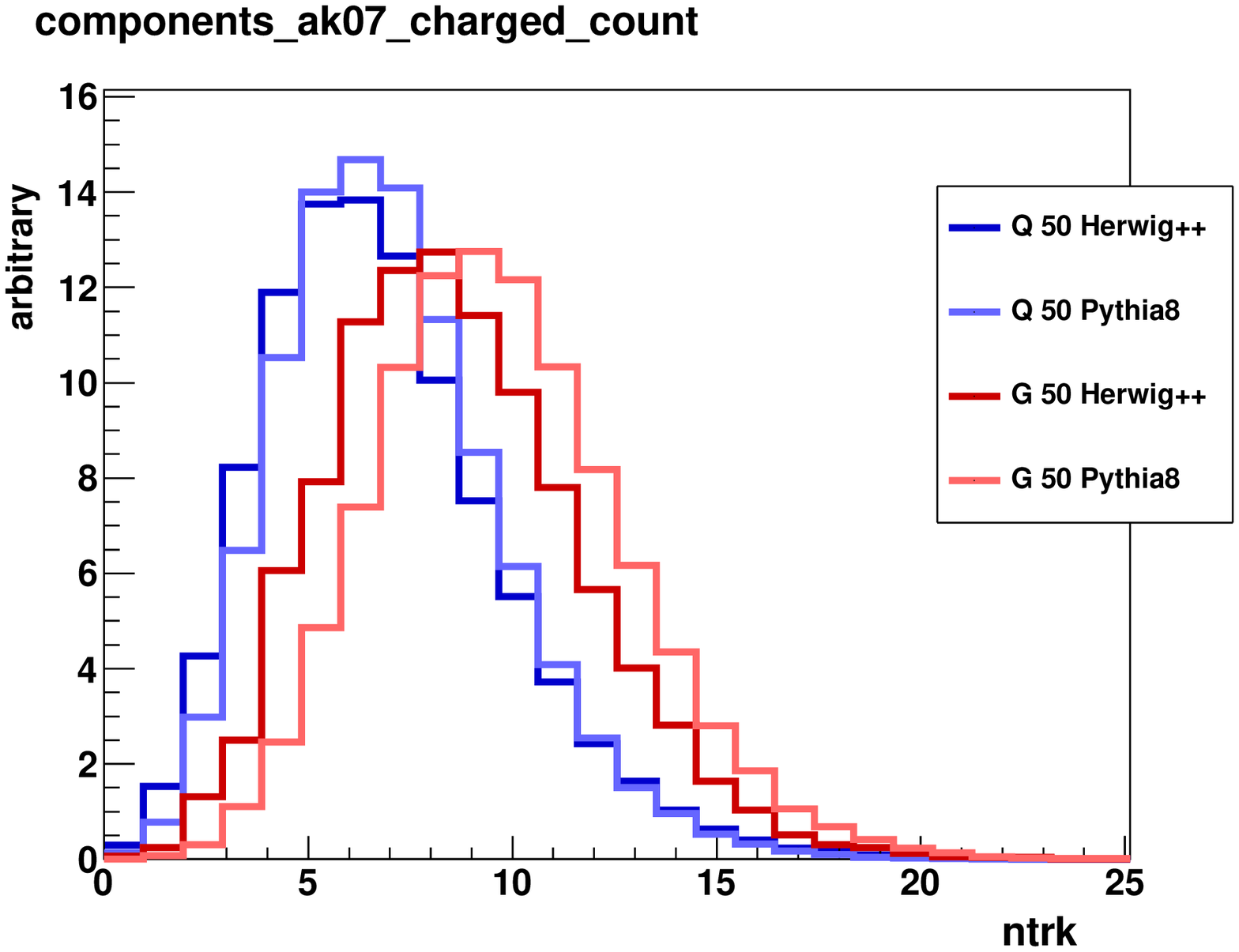}
\includegraphics[width=0.49\textwidth]{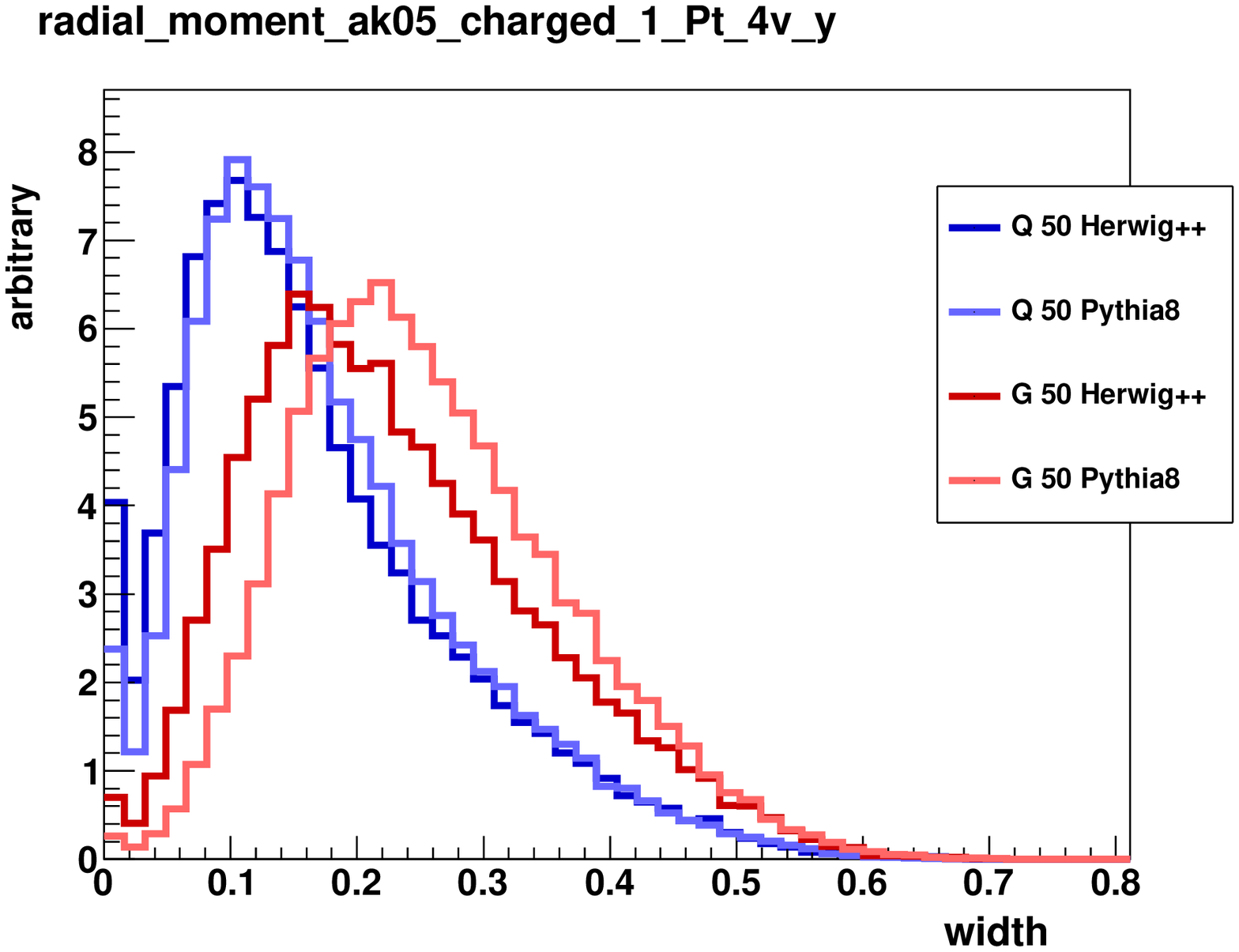}
\includegraphics[width=0.49\textwidth]{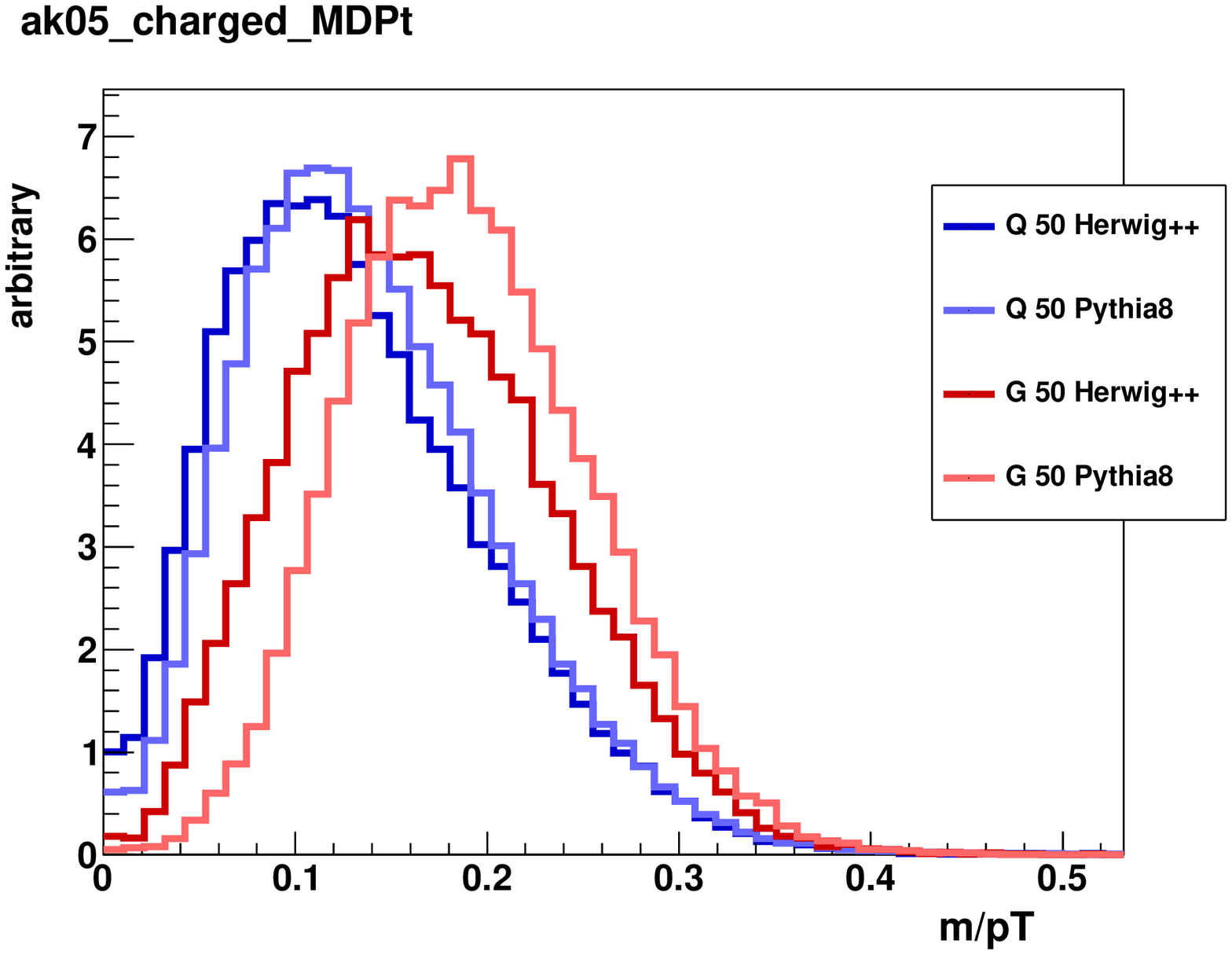}
\includegraphics[width=0.49\textwidth]{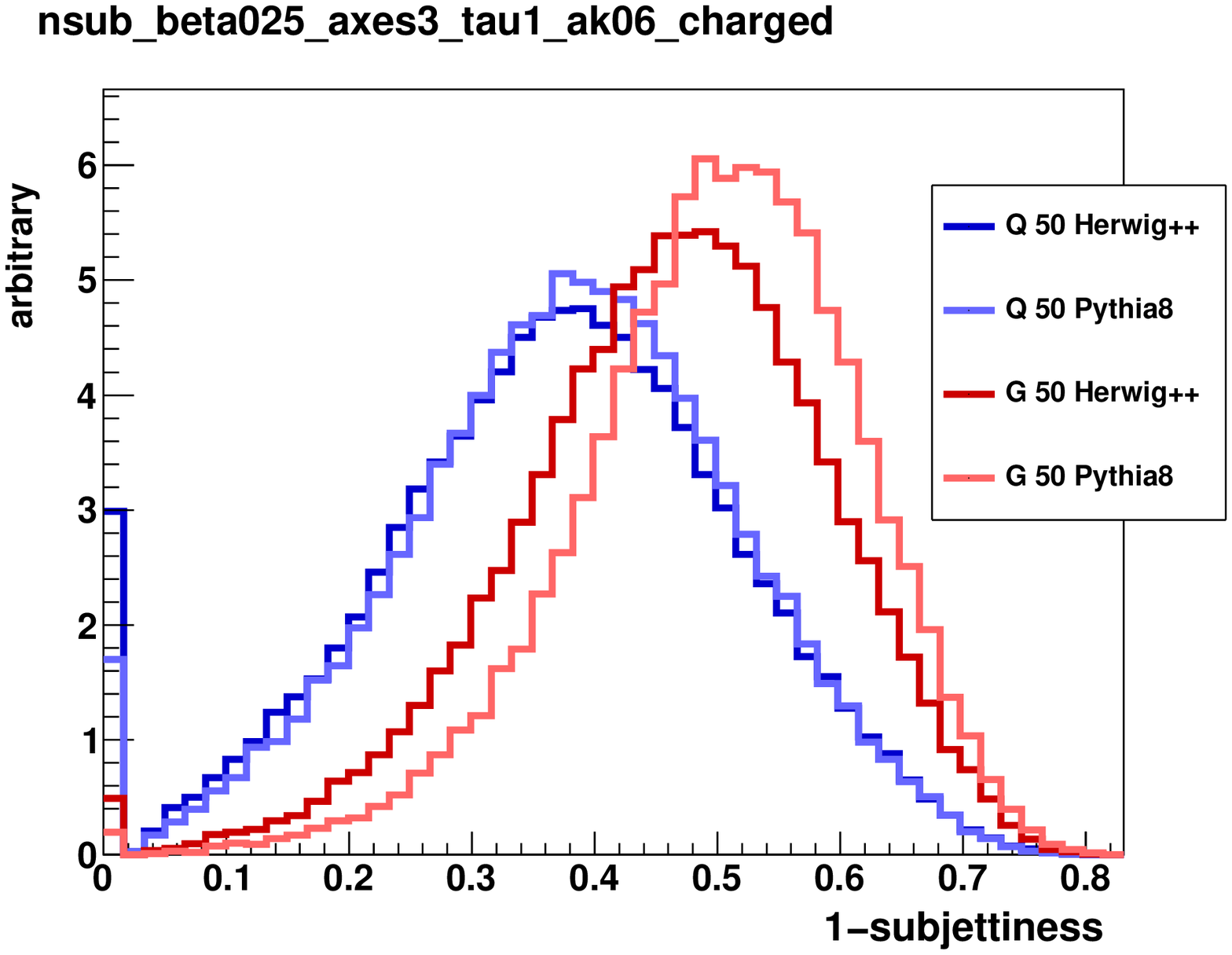}
\end{center}
\caption{
Distributions of charged track count and linear radial moment
(here calculated using only the charged tracks within the jet) for 50\,GeV jets.
Quark samples are blue and Gluon samples red.
{\sc Pythia 8.165} is the lighter shade and {\sc Herwig++ 2.5.2} is in the darker shade.
}
\label{fig:herwig_vs_pythia}
\end{figure}

 Our simulation, shown in Figure \ref{fig:herwig_vs_pythia},
indicates that various simulations agree with each other for quark jets.
We find that the distributions for gluon jets in {\sc Pythia 8.165} are consistently
farther away from the quark distributions than they are in {\sc Herwig++ 2.5.2}.
ATLAS found that {\sc Herwig++} agreed with data better for charged tracks,
but {\sc Pythia8} agreed better for width.
In both the data and {\sc Herwig++}, quarks and gluons
look more similar to each other than they do in {\sc Pythia8} (which was used for most of the plots in this paper).
Gluon rejection is consistently around  10\% worse for {\sc Herwig++} than for {\sc Pythia8}.
As a function of
things like jet-size or radial moment power, or the number of subjets,
the difference between {\sc Herwig++} and {\sc Pythia8} is just an overall shift.
This means that all of the single-variable trends described in the bulk of this paper still hold.
We find less advantage in combining variables with
multivariate techniques for jets simulated using {\sc Herwig++} than with {\sc Pythia8}.
%For 50 GeV jets, gluon rejection for 50\% quark acceptance starts at
%77.3\% for one variable, and moves to
%78.3\% for two, and 79.2\% for four.
%For 100 GeV, it's only slightly better and goes from 80.3\% to 82.8\%.

%Herwig++ 2.5.2 as compared to Pythia 8.165

ATLAS also found that various jet grooming techniques not only helped
with pileup, but the groomed jet mass was better simulated
than the ungroomed mass.
% Further studies need to be done
%on the variables appropriate for quark/gluon discrimination.
To reject pileup, ATLAS also
%prefers to calculate their jet moments
used only the charged tracks, thus ignoring their calorimeter
for everything except the overall jet $p_T$. To account for this, in our comparison of
{\sc Pythia8} to {\sc Herwig++}, we use only charged tracks. A summary and comparison of various simulations using all
and just charged tracks is shown in Table~\ref{tab:herpy}.

Combining variables sometimes helps, but mostly for harder jets.
At 50\% quark efficiency, combining radial moment with track count
gives an additional 0.4\% to 1.9\% gluon rejection, depending on
jet $p_T$ and generator.  For {\sc Herwig++}, this improvement increases
with jet $p_T$, but for {\sc Pythia8}, it reaches its maximum at 200\,GeV.
These small shifts should be compared to gluon efficiency,
not gluon rejection (one minus efficiency).  For {\sc Pythia8} 200\,GeV
jets, going from 9\% efficiency down to 7\% efficiency has a
sizable effect on significance improvement. It directly
translates into a 20\% better $S/B$.  But {\sc Herwig}'s efficiency
here starts at the higher 19\% and only goes down
to 17.3\% -- a more modest 9\% improvement in $S/B$.
The improvements for 50 and 100\,GeV jets are smaller than for 200\, GeV jets.

\clearpage

\section{Conclusions \label{sec:conc}}
In this paper, we have performed a comprehensive Monte-Carlo-based multivariate study of how quark and gluons can be distinguished on an event-by-event basis. We considered thousands of variables, which can generally be classified in two broad classes: discrete variables that count the number of tracks in a jet, and continuous variables, such as jet shapes. A general conclusion is that discrete variables help more at high jet $p_T$ and high quark efficiency (loose cuts), whereas continuous variables help more at lower jet $p_T$ and lower quark efficiency (tight cuts).  Overall discrimination also works better at higher $p_T$. 

We find, not surprisingly, that the more information that is used, the better the discrimination: counting smaller subjets work better than larger subjets; using all particles to calculate jet moments rather than just charged tracks works better. For charged track count, even using larger R=0.7 jets helps. 
The best two-variable discriminant often involves one variable from each class. Including three or more variables in a multivariate discriminant shows only small improvements over two variables. 
The marginal improvement of combining variables is also larger at high $p_T$ than at low $p_T$.

We have found that there is a consistent difference in how variables perform when {\sc Pythia8}
or {\sc Herwig++} is used for the simulation. {\sc Pythia8} simulations show more differences between quarks and gluons, but unfortunately, early studies with data
indicate that {\sc Herwig++} more accurately predicts quark/gluon discrimination power. 
We find that single variables can reject about
90\% of gluon jets, keeping 50\% of quark jets if {\sc Pythia8} can be trusted, but only around 80\% of gluon jets with {\sc Herwig++}. Some quantitative results are shown in Table \ref{tab:herpy}.
Since there are significant differences between different event generators, one use of quark and gluon tagging would be to tune these generators to match data more accurately. Such improved tunings could have important implications for a number of substructure based analyses.

In conclusion, quark/gluon discrimination is difficult, but not impossible. 85\% gluon rejection at 50\% quark acceptance seems feasible, however further input from experiment is needed.

\noindent
\newcommand{\st}{$^*$\!\!}
\begin{table}[t]
\begin{tabular}{|l|c|c|c|c||c|c|c|c|} \hline
\bf{Gluon Efficiency \% at} & \multicolumn{4}{c||}{\bf{50\,GeV}}  & \multicolumn{4}{c|}{\bf{200\,GeV}} \\
%\cline{2-9}
\bf{50\% Quark Acceptance} & \multicolumn{2}{c|}{Particles} &  \multicolumn{2}{c||}{Tracks} & \multicolumn{2}{c|}{Particles} &  \multicolumn{2}{c|}{Tracks} \\
%\cline{2-9}
% & P8 & H\footnotesize{++}\!\! & P8 & H\footnotesize{++}\!\! & P8 & H\footnotesize{++}\!\! & P8 & H\footnotesize{++}\!\! \\ \hline
 & P8 & \!H++\!\! & P8 & \!H++\!\! & P8 & \!H++\!\! & P8 & \!H++\!\! \\ \hline

2-Point Moment $\beta$=1/5                                   &   8.7\st  &  17.8\st  &  13.7\st  &  22.8\st  &   8.3  &  15.9  &  13.2  &  19.6  \\
1-Subjettiness $\beta$=1/2                                   &   9.3  &  18.5  &  14.2  &  22.9  &   7.6  &  16.2  &  12.3  &  19.4  \\
2-Subjettiness $\beta$=1/2                                   &   9.2  &  18.6  &  13.9  &  23.6  &   6.8  &  15.7\st  &   9.8  &  18.7\st  \\
3-Subjettiness $\beta$=1                                     &   9.1  &  19.3  &  14.6  &  24.4  &   5.9\st  &  16.7  &   8.6\st  &  19.5  \\
Radial Moment $\beta$=1 (Girth)                              &  10.3  &  20.5  &  16.1  &  24.9  &  11.2  &  18.9  &  15.3  &  21.9  \\
Angularity $a=+1$                                            &  10.3  &  20.0  &  15.8  &  24.5  &  12.0  &  19.3  &  14.0  &  21.6  \\
Det of Covariance Matrix                                     &  11.2  &  21.2  &  18.1  &  27.0  &   9.4  &  20.9  &  13.5  &  24.6  \\
Track Spread: $\sqrt{<p_T^2>}/p_T^\mathrm{jet}$              &  16.5  &  25.3  &  16.5  &  25.3  &   9.3  &  20.1  &   9.3  &  20.1  \\
Track Count                                                  &  17.7  &  26.4  &  17.7  &  26.4  &   8.9  &  21.0  &   8.9  &  21.0  \\
Decluster with $k_T$, $\Delta R$                             &  15.8  &  24.5  &  20.1  &  28.4  &  13.9  &  20.1  &  16.9  &  23.4  \\
Jet $m/p_T$ for R=0.3 subjet                                 &  13.1  &  25.9  &  16.3  &  27.7  &  11.9  &  24.2  &  14.8  &  26.2  \\
Planar Flow                                                  &  28.7  &  34.4  &  28.7  &  34.4  &  39.6  &  42.9  &  39.6  &  42.9  \\
Pull Magnitude                                               &  37.0  &  39.0  &  32.9  &  35.6  &  30.6  &  30.2  &  29.6  &  30.6  \\
\hline
Track Count \& Girth                                          &   9.9  &  20.1  &  13.4  &  23.2  &   7.1  &  17.3  &   7.7\st  &  18.7  \\
R=0.3 $m/p_T$ \& R=0.7 2-Pt $\beta$=1/5                   &   7.9\st  &  17.7  &  12.2\st  &  22.1  &   5.7  &  14.4\st  &   8.5  &  17.9  \\
1-Subj $\beta$=1/2 \& R=0.7 2-Pt $\beta$=1/5              &   8.5  &  17.3\st  &  12.9  &  22.1  &   6.0  &  14.6  &   8.6  &  17.7\st  \\
Girth \& R=0.7 2-pt $\beta$=1/10                          &  12.6  &  21.9  &  12.6  &  21.9\st  &   9.2  &  18.0  &   9.2  &  18.0  \\
1-Subj $\beta$=1/2 \& 3-Subj $\beta$=1                       &   8.9  &  18.0  &  14.0  &  23.2  &   5.6\st  &  15.0  &   8.4  &  18.4  \\
%Best Pair                                                     &   7.9  &  17.3  &  12.2  &  21.9  &  5.6  &  14.4  &   7.7  &  17.7  \\
\hline
Best Group of 3                                               &   7.5  &  17.0  &  11.0  &  20.9  &   4.7  &  14.0  &   6.9  &  16.6  \\
Best Group of 4                                               &   7.1  &  16.7  &  10.6  &  20.5  &   4.5  &  13.7  &   6.2  &  16.3  \\
Best Group of 5                                               &   6.9  &  16.4  &  10.4  &  20.0  &   4.3  &  13.3  &   6.1  &  15.9  \\
\hline
\end{tabular}
\caption{\label{tab:herpy}
Comparison of gluon efficiencies at the 50\% quark 
acceptance working point.
All of the single variables use R=0.5 jets, whereas combinations sometimes include R=0.7 jets.
Gluon efficiencies, rather than gluon rejections (one minus efficiencies), are 
shown because a fractional improvement here is the same fractional improvement in $S/B$.
Divided by two, it is also the fractional improvement in $S/\sqrt{B}$.
These scores have $\pm$0.5\% statistical errors, 
but they are correlated --- the differences between 
variables has smaller spread, 
as does the improvement when combining variables.
Because of the large 
number of variables and parameters, and the larger
number of possible combinations of these, there is 
definitely a look-elsewhere-type effect when choosing 
the top pair.
Many pairs statistically tied for the top spot in each category,
so five pairs were chosen as representative.
Their scores are marked with asterisks, as are the best
individual variables in each category.
The best groups of 3, 4, and 5 start to show diminishing returns.
}
\end{table}

\section*{Acknowledgments}
The authors would like to thank D. Mateos, D. Miller,  A. Schwartzman, M. Silva and M. Swiatlowski for useful discussions.
This work was supported in part by the Department of Energy
under grant DE-SC003916. Computations for this paper were
performed on the Odyssey cluster supported by the FAS Research
Computing Group at Harvard University.

\end{document}